\documentclass[aps,prx,reprint]{revtex4-1}
\usepackage{blindtext}
\usepackage{graphicx}
\usepackage{subcaption}
\usepackage{xcolor}
\usepackage[fleqn]{amsmath}
\usepackage[font=small,skip=3pt]{caption}
\usepackage{cleveref}
\usepackage{makecell}
\begin{document}
\title{Three-Dimensional Inhomogeneity of Electron-Temperature-Gradient Turbulence in the Edge of Tokamak Plasmas}
\author{J. F. Parisi$^{1,2}$}
\email{jparisi@pppl.gov}
\author{F. I. Parra$^{1}$}
\author{C. M. Roach$^2$}
\author{M. R. Hardman$^3$}
\author{A. A. Schekochihin$^3$}
\author{I. G. Abel$^4$}
\author{N. Aiba$^5$}
\author{J. Ball$^6$}
\author{M. Barnes$^3$}
\author{B. Chapman-Oplopoiou$^2$}
\author{D. Dickinson$^7$}
\author{W. Dorland$^{4,3}$}
\author{C. Giroud$^2$}
\author{D. R. Hatch$^8$}
\author{J. C. Hillesheim$^2$}
\author{J. Ruiz Ruiz$^3$}
\author{S. Saarelma$^2$}
\author{D. St-Onge$^3$}
\author{JET Contributors }
\thanks{See the author list of `Overview of JET results for optimising ITER operation' by J. Mailloux et al. to be published in Nuclear Fusion Special issue: Overview and Summary Papers from the 28th Fusion Energy Conference (Nice, France, 10-15 May 2021).}
\affiliation{$^1$Princeton Plasma Physics Laboratory, Princeton University, Princeton, NJ 08543, USA}
\affiliation{$^2$Culham Centre for Fusion Energy, Culham Science Centre, Abingdon, OX14 3DB, UK}
\affiliation{$^3$Rudolf Peierls Centre for Theoretical Physics, University of Oxford, Oxford, OX1 3PU, UK}
\affiliation{$^4$Department of Physics, University of Maryland, College Park, Maryland 20742, USA}
\affiliation{$^5$National Institutes for Quantum and Radiological Science and Technology, Rokkasho, Japan}
\affiliation{$^6$\'Ecole Polytechnique F\'ed\'erale de Lausanne (EPFL), Swiss Plasma Center (SPC), CH-1015 Lausanne, Switzerland}
\affiliation{$^7$York Plasma Institute, Department of Physics, University of York, Heslington, York. YO10 5DD, UK}
\affiliation{$^8$Institute for Fusion Studies, University of Texas at Austin, Austin, Texas, 78712, USA}

\begin{abstract}
Nonlinear multiscale gyrokinetic simulations of a Joint European Torus edge pedestal are used to show that electron-temperature-gradient (ETG) turbulence has a rich three-dimensional structure, varying strongly according to the local magnetic-field configuration. In the plane normal to the magnetic field, the steep pedestal electron temperature gradient gives rise to anisotropic turbulence with a radial (normal) wavelength much shorter than in the binormal direction. In the parallel direction, the location and parallel extent of the turbulence are determined by the variation in the magnetic drifts and finite-Larmor-radius (FLR) effects. The magnetic drift and FLR topographies have a perpendicular-wavelength dependence, which permits turbulence intensity maxima near the flux-surface top and bottom at longer binormal scales, but constrains turbulence to the outboard midplane at shorter electron-gyroradius binormal scales. Our simulations show that long-wavelength ETG turbulence does not transport heat efficiently, and significantly decreases overall ETG transport -- in our case by $\sim$40 \% -- through multiscale interactions.
\end{abstract}

\maketitle

\setlength{\parskip}{0mm}
\setlength{\textfloatsep}{5pt}

\setlength{\belowdisplayskip}{6pt} \setlength{\belowdisplayshortskip}{6pt}
\setlength{\abovedisplayskip}{6pt} \setlength{\abovedisplayshortskip}{6pt}

\section{Introduction} \label{sec:1}

\begin{figure}
        \centering
        \includegraphics[width=0.52\textwidth]{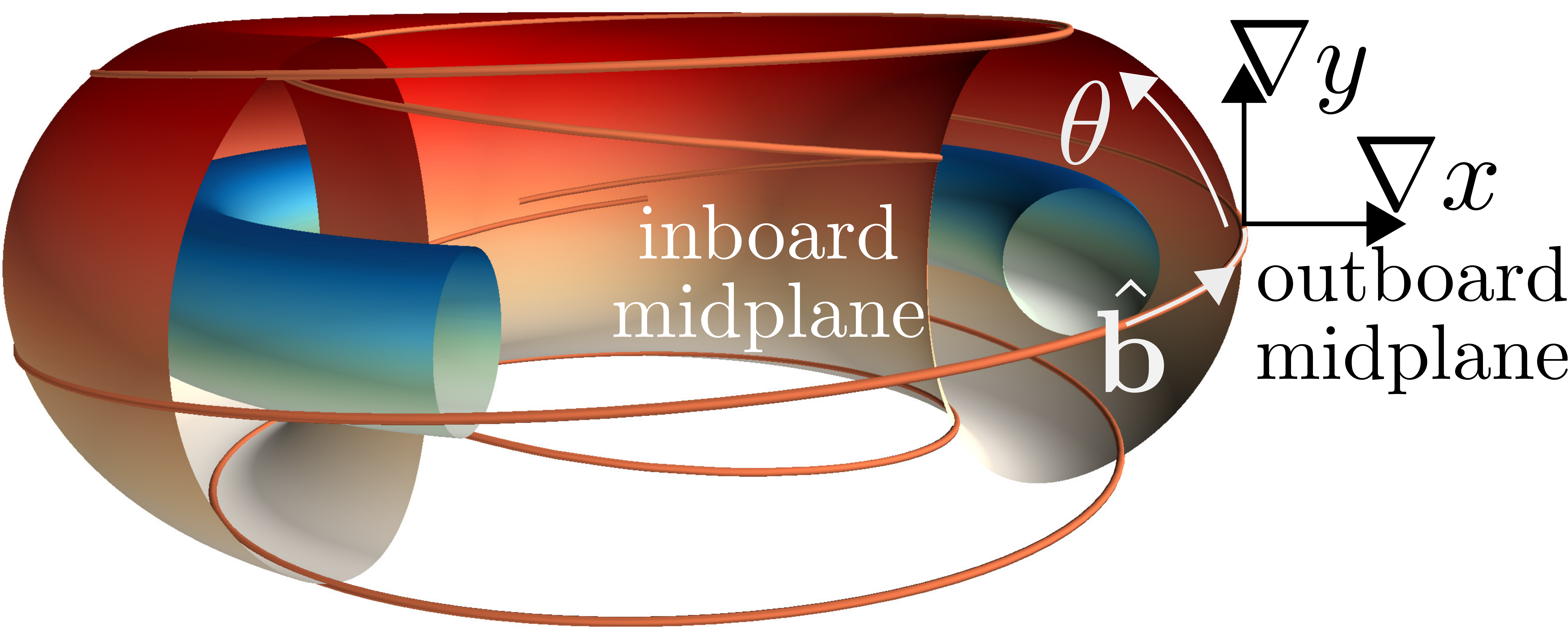}
        \caption{A circular (blue) flux surface in the tokamak core and a highly shaped (red) flux surface in the edge pedestal with a magnetic field line that completes a $2\pi$ turn in poloidal angle $\theta$. At the outboard midplane, we label the curvilinear coordinate system with radial $\nabla x$, binormal $\nabla y$, and field-line $\hat{\mathbf{b}}$ directions defined around \Cref{eq:two}. In this work, we study this highly shaped flux surface.}
        \label{fig:zero}
\end{figure}

In tokamaks, strong magnetic fields and plasma currents generate nested magnetic flux surfaces. On a flux surface, illustrated in \Cref{fig:zero}, particles move much faster parallel to the magnetic field than perpendicular to it, causing equilibrium quantities such as temperature and density to be constant within flux surfaces \cite{Hazeltine2003}. The radial gradients of equilibrium quantities drive turbulence at scales comparable to ion and electron gyroradii \cite{Mazzucato1976,Cowley1991,Jenko2000}. Such turbulence has a perpendicular eddy length that is very short compared to the perpendicular equilibrium length scale \cite{Liewer1985,Mazzucato1993}, and is radially inhomogeneous from the core to the edge \cite{Miyato2004,Wang2006,Gorler2011b}. In the tokamak core, turbulence is found to vary slowly along magnetic field lines \cite{Waltz1994,Beer1995,Barnes2011}. In these conditions, the turbulence amplitude typically peaks at the outboard midplane -- the low magnetic-field side where an interchange-like plasma instability is strongest -- and decreases in amplitude smoothly in the parallel direction away from the outboard midplane \cite{Rewoldt1990,Waltz1994}. Thus, core turbulence typically varies strongly in the plane perpendicular to the magnetic field, but has a predictable profile in the parallel direction.

In contrast, for the turbulence in the edge pedestal of tokamak plasmas -- a region of steep pressure gradients in high-performance discharges \cite{Wagner1982} -- we show that both the parallel and perpendicular physics become intriguingly complex, giving turbulence a highly inhomogeneous character. This inhomogeneous turbulence is due to steep pressure gradients and the strong parallel variation in the perpendicular physics of magnetic drifts, generating modes with very different character at different parallel (and, therefore, poloidal) locations. The flux surfaces in the pedestal are highly `shaped' \cite{Belli2008,Marinoni2009,Laggner2018,Austin2019}, unlike the more circular flux surfaces in the plasma core \cite{Kendl2006,Ball2015}. In \Cref{fig:zero}, we show both a highly shaped and circular flux surface. We find that the strong magnetic-field variation in the parallel direction and the steep temperature gradients in the edge pedestal create a non-trivial topography of regions that determines where turbulence can and cannot reside. This topography gives edge pedestal turbulence a novel three-dimensional structure not seen in the core.

Due to the steep gradients in the pedestal, we find that the parallel spatial structure of the turbulence is particularly inhomogeneous and peaked away from the outboard midplane at wavelengths as long as the ion gyroradius, $k_y \rho_i \sim 1$, where $k_y$ is the wavenumber in the binormal direction $y$ and $\rho_s$ is the Larmor radius for a species $s$. Note that the binormal length scale is still determined by electron physics, not intrinsically by $\rho_i$, but it is quantitatively comparable to $\rho_i$ at the relevant JET parameters. Thus, the ion-electron scale separation is broken in this system. In contrast, at electron-gyroradius binormal scales, $k_y \rho_e \sim 1$, turbulence becomes confined to the tokamak's low magnetic-field side.

To reveal the importance of the $k_y \rho_i \sim 1$ ETG turbulence and its unusual parallel structure, we performed multiscale \cite{Candy2007,Gorler2008,Maeyama2015, Howard2016, Hardman2019, Pueschel2020} nonlinear gyrokinetic simulations of a JET pedestal using the gyrokinetic code \texttt{stella} \cite{Barnes2019}. These multiscale simulations are novel because they resolve the spatiotemporal scales that are needed to observe the complex parallel dynamics and multiscale interactions of $k_y \rho_i \sim 1$ ETG turbulence. By means of these numerical experiments, we will show that the electron-temperature-gradient (ETG) turbulence at $k_y \rho_i \sim 1$ reduces transport due to ETG turbulence at $k_y \rho_e \sim 1$.

ETG turbulence is one of many important transport mechanisms in the edge pedestal. The pedestal, which is a key ingredient in a fusion reactor, appears once external plasma heating crosses a threshold value \cite{Ryter1996}. This steep-gradient region significantly increases a reactor's core pressure and hence fusion power \cite{Lawson1957}. The transport properties of the pedestal are determined by the nature of the turbulence, which is driven by the strong gradients. These turbulent fluxes constrain the pedestal's magnetohydrodynamic stability \cite{Keilhacker1984,Connor1998, Hill1997,Snyder2002,Diallo2014}, neoclassical transport \cite{Pusztai2016}, and scrape-off-layer processes \cite{Neuhauser2002}. Extensive experimental, numerical, and analytic results suggest that ion-temperature-gradient (ITG) \cite{Rudakov1961,Nordman1990,Cowley1991}, ETG \cite{Jenko2000,Dorland2000}, microtearing \cite{Drake1980}, kinetic-ballooning \cite{Snyder2011,Dickinson2012}, and trapped-electron modes \cite{Ernst2004} are responsible for anomalous heat losses in the pedestal \cite{Smith2013,Gao2013,Fulton2014,Holod2015,Hatch2016,Churchill2017,Kotschenreuther2019,Villard2019,Pueschel2020,Guttenfelder2021,Hatch2021,Larakers2021,Hassan2021,Nelson2021}. Pedestal instability and turbulence peaking away from the outboard midplane has been observed for ETG \cite{Told2008,Parisi2020,Parisi2020b}, ITG \cite{Kotschenreuther2017}, microtearing \cite{Hatch2015,Hatch2016,Hatch2021}, and trapped-electron modes \cite{Hatch2015}. 

The rest of this paper is organized as follows. We introduce the gyrokinetic formalism in \Cref{sec:2}. In \Cref{sec:3}, we describe the consequences of steep temperature gradients for pedestal ETG physics. Results of linear and nonlinear gyrokinetic simulations are described in \Cref{sec:4,sec:5}, respectively. In \Cref{sec:6}, we analyze a temperature-gradient scan for nonlinear simulations. \Cref{sec:7} describes the relation between geometric topography and turbulence. In \Cref{sec:8}, we use a numerical experiment to show that $k_y \rho_i \sim 1$ ETG turbulence reduces transport at $k_y \rho_e \sim 1$. We conclude in \Cref{sec:10}.

\section{Gyrokinetic Turbulence} \label{sec:2}

In the presence of a strong magnetic field, plasma perturbations are anisotropic relative to the mean magnetic field, $k_{\parallel} / k_{\perp} \sim \rho_{*s} \ll 1$, and slow relative to the Larmor frequency, $\omega / \Omega_s \sim \rho_{*s}$. Here $k_{\parallel}$ and $k_{\perp}$ are wavenumbers parallel and perpendicular to the mean magnetic field, $\rho_{*s} = \rho_s / L_p$ where $L_p$ is the pedestal width, $\omega $ is the turbulent frequency, $\Omega_s = Z_s e B / m_s c $ is the gyrofrequency, $Z_s$ is the charge number, $e$ is the proton charge, $B$ is the magnetic field strength, $m_s$ is the mass, and $c$ is the speed of light. Such plasma fluctuations are well-described by gyrokinetics \cite{Taylor1968,Catto1978,Antonsen1980,Frieman1982,Parra2008,Abel2013}. The distribution function of particles of species $s$ is split into equilibrium and turbulent components, $f_s = F_{Ms} + f _s^{tb}$, where $F_{Ms}$ is a Maxwellian and the turbulent distribution $f  _s^{tb}$ satisfies $f _s^{tb} \sim \rho_{*s} F_{Ms}$. We study turbulence governed by the gyrokinetic equation
\begin{equation}
\begin{aligned}
& \frac{\partial {h}_{s}}{\partial t} + (v_{\parallel } \hat{\mathbf{b} } + \mathbf{ v}_{Ms} + \langle \mathbf{ v}_{E} ^{tb}  \rangle_{\varphi} ) \cdot \nabla { h}_{s} \\ & = \frac{Z_s e F_{Ms}}{T_s} \frac{\partial \langle { \phi}^{tb}  \rangle_{\varphi}}{\partial t}  - \langle \mathbf{ v}_{E}^{tb}  \rangle_{\varphi} \cdot \nabla F_{Ms},
\end{aligned}
\label{eq:one}
\end{equation}
where $h_s = (Z_s e \phi ^{tb} /T_s ) F_{Ms} + f _s^{tb}$, $t$ is time, $T_s$ is the equilibrium temperature, $v_{\parallel}$ is the parallel velocity, $\hat{\mathbf{ b} } = \mathbf{B} /B$, $\mathbf{ v}_{Ms}$ is the magnetic drift velocity, $\mathbf{ v}_{E} ^{tb} = c (\hat{\mathbf{ b} } \times \nabla \phi ^{tb} )/ B$ is the $\mathbf{ E} \times \mathbf{ B}$ drift velocity, $\phi^{tb} $ is the turbulent electrostatic potential, $\langle \ldots \rangle_{A}$ is an average with respect to the variable $A$, and $\varphi$ is the gyrophase angle.

Since the turbulence is anisotropic, behaving differently in the directions perpendicular and parallel to the magnetic field, we can solve \Cref{eq:one} in a numerically efficient field-following domain called a flux tube \cite{Beer1995}, which has a narrow perpendicular extent centered on a magnetic field line, but extends far along the field line, typically performing a $2\pi$ poloidal circuit. To describe the directions perpendicular to the magnetic field, we use the flux coordinates
\begin{equation}
x = \frac{q_c}{r_c B_c} \psi, \;\;\; y = \frac{1}{B_c} \frac{\partial \psi}{\partial r} (\zeta - q \theta - \Omega_{\zeta} t - \nu),
\label{eq:two}
\end{equation}
where $q_c$ is the safety factor, $r_c$ is a minor-radial flux coordinate, both evaluated at the flux tube's center, $\psi$ is the poloidal flux divided by $2\pi$, $B_c$ is a reference magnetic field, $\zeta$ is the toroidal angle, $\theta$ is the poloidal angle, $\Omega_{\zeta}$ is the toroidal flow's angular frequency, and $\nu (r, \theta)$ is a function $2\pi$-periodic in $\theta$  \cite{Dhaeseleer1991} that is nonzero when a magnetic field line's pitch angle at a poloidal location $\theta$ differs from the mean pitch angle $\propto 1/q$ on the flux surface; $|\nu|$ is larger for highly shaped flux surfaces than for the more circular ones in the core. The quantities $q(r)$ and $\nu(r,\theta)$ are defined so that $\mathbf{B} \cdot \nabla y = 0$. The angle $\theta$ is defined so that $\theta = 0$ is the outboard midplane, $\theta = \pm \pi$ is the inboard midplane (see \Cref{fig:zero}), and $\theta = \pm \pi/2$ is approximately the flux surface's top/bottom. We Fourier transform locally in the perpendicular plane,
\begin{equation}
\widetilde{\phi} (x, y, \theta,t) = \sum_{k_{x}, k_{y}} \hat{ \phi} _{k_{x}, k_{y}}( \theta,t) \exp( i k_{x} x + i k_{y} y),
\label{eq:three}
\end{equation}
where the normalized potential is $\widetilde{\phi} = e \phi ^{tb} /T_i \rho_{*i}$ and $\hat{ \phi}_{k_{x}, k_{y}}( \theta,t)$ are its Fourier coefficients. 

We will frequently use the electron magnetic-curvature drift frequency $\omega_{\kappa e}$ and the grad-B drift frequency $\omega_{\nabla B e}$,
\begin{equation}
\begin{aligned}
&\omega_{\kappa e} = \frac{v_{te}^2 \mathbf{ k}_{\perp}}{\Omega_e} \cdot \left[ \hat{\mathbf{b} } \times \left( \nabla \ln B + \frac{4\pi}{B^2} \frac{\partial p}{\partial r} \nabla r \right) \right], \\
&\omega_{\nabla Be} = \frac{v_{te}^2 \mathbf{ k}_{\perp}}{\Omega_e} \cdot \left(\hat{\mathbf{b} } \times \nabla \ln B \right),
\label{eq:four}
\end{aligned}
\end{equation}
related to $\mathbf{v}_{Me}$ through $\mathbf{v}_{Me} \cdot \mathbf{k}_{\perp} = \omega_{\kappa e} v_{\parallel}^2/v_{te}^2 + \omega_{\nabla B e} v_{\perp}^2/2v_{te}^2 $, where $v_{\perp}$ is the perpendicular velocity, $v_{ts} = \sqrt{2 T_s / m_s}$ is the thermal speed, and $p$ is the equilibrium pressure. The perpendicular wavenumber is
\begin{equation}
\begin{aligned}
\mathbf{k}_{\perp} = & k_x \nabla x + k_y \nabla y = k_y \bigg{[}\hat{s} (\theta_0 - \theta) - \gamma_E t + \frac{r}{q} \frac{\partial \nu}{\partial r} \bigg{]} \nabla x \\ & + \frac{\partial \psi}{\partial r} \frac{1}{B_c} k_y \bigg{[} \nabla \zeta + \bigg{(}\frac{\partial \nu }{\partial \theta} - q\bigg{)} \nabla \theta  \bigg{]},
\end{aligned}
\label{eq:five}
\end{equation} 
where $\hat{s} = (r /q) (\partial q / \partial r)$ is the magnetic shear, $\theta_0 = k_x / (k_y \hat{s})$ the ballooning angle, and $\gamma_E = - (r/q)\partial_r \Omega_{\zeta}$ the radial shear of the toroidal flow. At $\theta =0$ and $t = 0$, the radial wavenumber is proportional to $\theta_0$: $\mathbf{k}_{\perp} \cdot \nabla x  = k_y \hat{s} \theta_0 |\nabla x|^2$. Another important frequency is
\begin{equation}
\omega_{*e}^T = k_y \frac{c}{B_c} \frac{T_e}{Z_e e L_{Te}},
\label{eq:six}
\end{equation}
which is the electron drift frequency associated with the temperature gradient length scale $L_{Te} \equiv -( \partial \ln T_e / \partial r)^{-1}$. This frequency appears in the linear-drive term on the right-hand side of \Cref{eq:one} and is particularly large in the pedestal due to steep temperature gradients. Typically, $\omega_{*e} ^T$ is comparable in size to the frequency of drift waves \cite{Cowley1991,Parra2019}, as has been shown for the pedestal ETG modes \cite{Parisi2020} considered in this paper. Crucially, the magnetic drift frequencies $\omega_{\kappa e}$ and $\omega_{\nabla B e}$ are proportional to $k_{\perp}$, but $\omega_{*e}^T$ is proportional only to $k_y$.

In this paper, we perform linear and nonlinear local, electrostatic, collisionless gyrokinetic simulations for JET-ILW discharge \#92174 \cite{Giroud2018} at $r/a$ = 0.974. This flux surface was chosen due to its large value of the flow shear $\gamma_E a/ v_{ti} = 0.56$, which is an important parameter for turbulence suppression \cite{Hahm1995,Barnes2011b,Yan2014}. On this surface, we use the following simulation parameters \cite{Parisi2020}: $a / L_{Te} = 42, \; a/ L_{Ti} = 11, \; a / L_{n} = 10, \; \rho_i / L_{Te} = 0.12, \; T_{e}/T_{i} = 0.56, \; T_i = 0.71 \; \mathrm{keV}, \; \hat{s} = 3.36, \; q = 5.1, \; \nu_{ee} a/vti = 0.83, \; \nu_{ii} a/vti = 0.006$, where the minor radius at the midplane is $a = 0.91$ m, the density gradient is $L_{n} \equiv -( \partial \ln n / \partial r)^{-1}$ for equilibrium density $n$, $\nu_{ss'} = \sqrt{2} \pi n_{s'} Z_s^2 Z_{s'}^2 e^4 \ln (\Lambda_{ss'}) / \sqrt{m_s} T_{s}^{3/2} $, and $\ln (\Lambda_{ss'})$ is the Coulomb logarithm. For the flux surface shape, shown in \Cref{fig:zero}, we use a Miller geometry prescription \cite{Miller1998}. For linear simulations, we use the parameters described in \cite{Parisi2020} with $\gamma_E = 0$.

\section{Pedestal ETG Characteristics} \label{sec:3}

Steep temperature gradients in the pedestal drive strong ETG instability far away from the outboard midplane, particularly at $k_y \rho_e \ll 1$. This is in stark contrast to the tokamak core, where the linear ETG growth rate and nonlinear field amplitudes peak at the outboard midplane at $k_y \rho_e \sim 1$ \cite{Dorland2000,Parisi2020b}. For the JET pedestal region investigated in this paper, two branches of ETG dominate: toroidal and slab ETG modes, which are unstable drift waves mediated by electron magnetic drifts and parallel streaming, respectively \cite{Rudakov1961,Coppi1967,Coppi1977,Horton1981,Cowley1991,Jenko2000,Dorland2000}.

For a strong toroidal ETG instability to be present, i.e., for the growth rate to be $\gamma \sim \omega_{*e}^T$, it has been shown that $\omega_{*e}^T / \omega_{\kappa e} \simeq A$ must be satisifed, where $A \simeq 3-20$ \cite{Parisi2020}. Note that $A$ is a dimensionless constant for the toroidal ETG instability, and is not specific to the discharge analyzed in this paper. We find
\begin{equation}
\frac{\omega_{*e} ^T}{\omega_{\kappa e}} \sim \frac{k_y}{k_{\perp}} \frac{R}{L_{Te}} \simeq A,
\label{eq:seven}
\end{equation}
where $R$ is the major radius. Since $R/ AL_{Te} \gg 1$ in the pedestal, for a strong toroidal ETG instability, we must have
\begin{equation}
\frac{k_{\perp}}{ k_y} \sim \frac{R}{ A L_{Te}} \gg 1.
\label{eq:eight}
\end{equation}
In the simple case where $\gamma_E = \nu = 0$, for a mode with $\theta_0 = 0$, as $\theta$ is increased away from the outboard midplane, $k_{\perp} / k_y$ becomes large due to magnetic shear, $k_{\perp} \sim k_y \hat{s} |\theta|$. For $\hat{s} \sim 1$, \Cref{eq:eight} implies
\begin{equation}
|\theta| \sim \frac{R}{ \hat{s} A L_{Te}} \gg 1,
\label{eq:nine}
\end{equation}
and so we expect linear toroidal ETG modes to be localized away from the outboard midplane. In \Cref{sec:4}, we show numerically that in our JET equilibrium, this is indeed the case for most values of $\theta_0$.

For a strong ETG instability, we also require the finite-Larmor-radius (FLR) effects not to be too strong: significant FLR damping occurs for $k_{\perp} \rho_e \gtrsim 1$ as an electron's gyromotion averages over the smaller-scale perpendicular wavelength, decreasing the linear growth rate. Thus, for a strong instability, we must have
\begin{equation}
k_{\perp} \rho_e \lesssim 1.
\label{eq:ten}
\end{equation}
Combining \Cref{eq:ten,eq:seven} we find strong toroidal ETG instability for
\begin{equation}
k_y \rho_e \lesssim \frac{A L_{Te}}{R}.
\label{eq:eleven}
\end{equation}
Since pedestal parameters often satisfy $A L_{Te}/R \lesssim \rho_e/ \rho_i$, this implies that toroidal ETG modes can be driven linearly at $k_y \rho_i \sim 1$. We stress that strong toroidal ETG instability at $k_y \rho_i \sim 1$ is a quantitative coincidence of $A L_{Te}/R \lesssim \rho_e/ \rho_i$, and is not a fundamental consequence of kinetic ion physics. However, for notational convenience, we will frequently refer to `$k_y \rho_i \sim 1$ ETG modes.'

Linear slab ETG modes dominate in the JET equilibrium used for this paper at most $k_y \rho_i$ values for $\theta_0$ = 0. One also finds strong sub-dominant slab ETG instability for $\theta_0 \neq 0$ \cite{Parisi2020}. In \Cref{sec:7}, by examining the topographies of $k_{\perp}$ and $\omega_{*e} / \omega_{\kappa, e}$ for this JET equilibrium, we show that, for $k_y \rho_e \ll 1$, both linear toroidal and slab ETG modes are expected to be unstable far away from the outboard midplane.

Nonlinearly, we also expect the ETG turbulence injection scale -- the outer scale -- at long binormal wavelengths. To estimate the wavenumber $k_y^o$ associated with this outer scale, we observe that the nonlinear decorrelation rate at the outer scale must be the same as the energy injection rate by the instability, $\omega_{*e}^T$, and then we `critically balance' this rate with the parallel streaming rate $v_{te}/ l_{\parallel}$ \cite{Goldston1995,Barnes2011,Ghim2013}, to find
\begin{equation}
k_y^o \rho_e \sim \frac{L_{Te}}{l_{\parallel}} \ll 1.
\label{eq:twelve}
\end{equation}
In the pedestal, we expect the parallel correlation length $l_{\parallel}$ to be determined by the characteristic parallel length of the local magnetic drifts and perpendicular wavenumber. In contrast, in the core, the length $l_{\parallel}$ is usually assumed to be of the order of the length of one poloidal turn along a magnetic field line, $qR$ \cite{Barnes2011}, giving $k_y^o \rho_e \sim  (1/q) (L_{Te}/R)$. 

In our equilibrium, we find that at the outer scale, $l_{\parallel} \approx qR/2$, giving $L_{Te}/l_{\parallel} \approx 1/300 \ll \rho_e / \rho_i$. Therefore, nonlinear pedestal ETG simulations that aim to capture the full ETG turbulence cascade require a wide range of binormal modes from $k_y \rho_i \sim (L_{Te}/l_{\parallel})(\rho_i/\rho_e) \lesssim 1$ to $k_y \rho_e \gtrsim 1$.

\section{Linear Simulations} \label{sec:4}

\begin{figure}
        \centering
            \begin{subfigure}[t]{0.49\linewidth}
        \includegraphics[width=1.\linewidth]{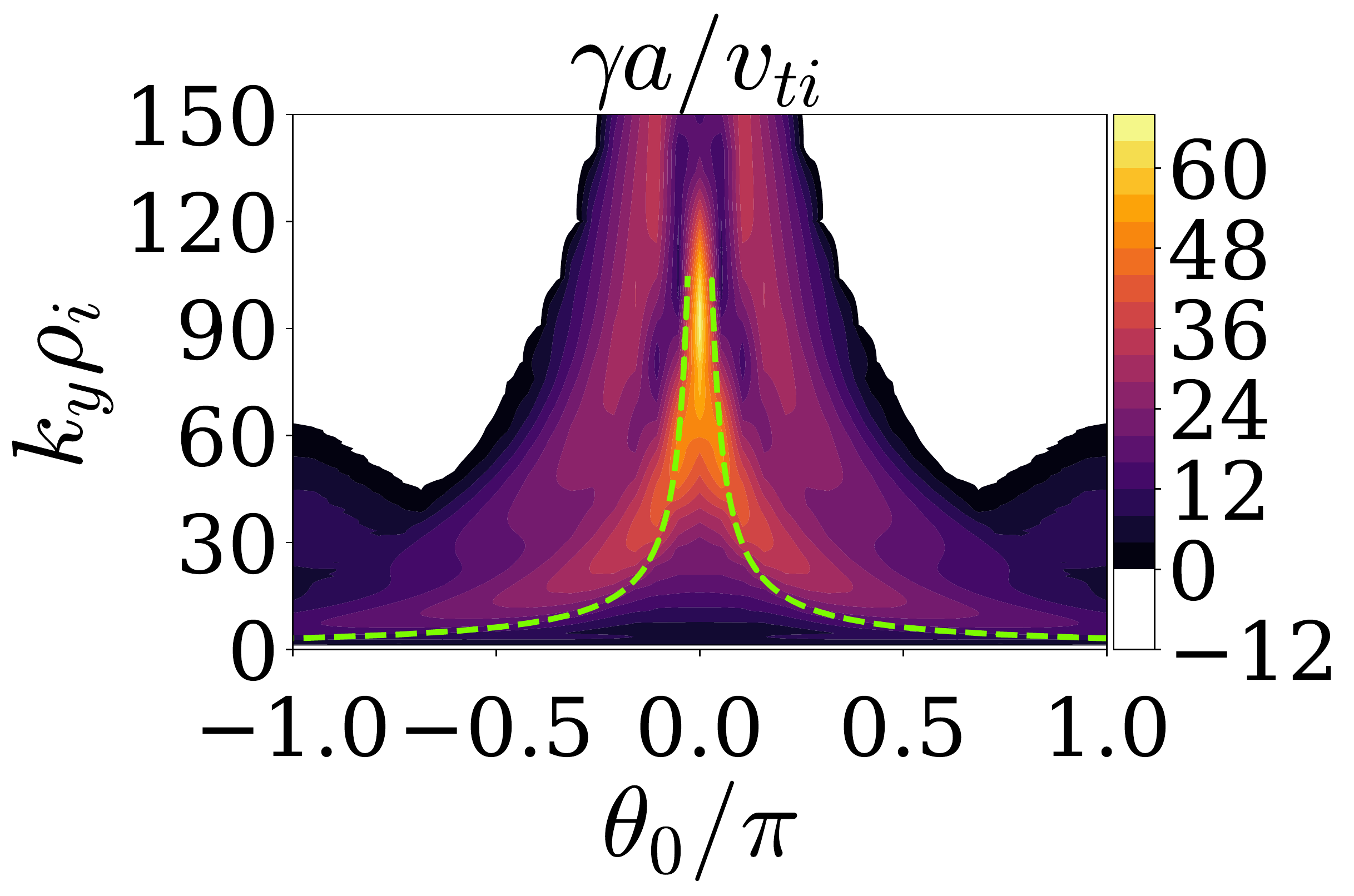}
        \caption{Linear growth rate $\gamma a / v_{ti}$.}
        \end{subfigure}
           \begin{subfigure}[t]{0.49\linewidth}
        \includegraphics[width=0.91\linewidth]{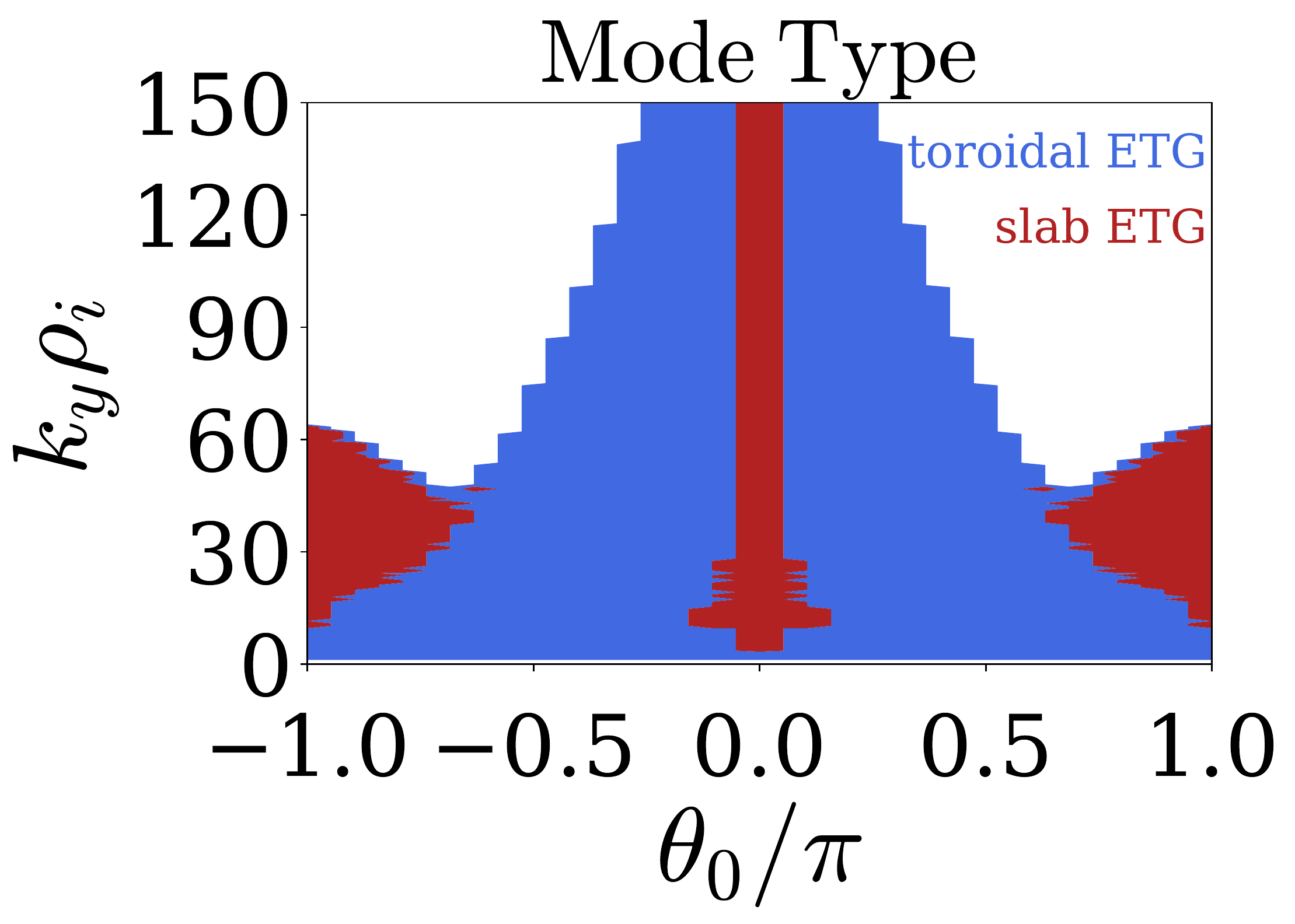}
        \caption{Fastest growing modes.}
            \end{subfigure}
        \caption{(a) The maximum linear growth rate $\gamma a / v_{ti}$ and (b) the dominant mode type, versus $k_y \rho_i$ and $\theta_0 = k_x/ (k_y \hat{s})$, for the linear \texttt{GS2} simulation described in \Cref{sec:4}. Green dashed curves in (a) denote the edge of the perpendicular $(\theta_0, k_y)$ grid for the nonlinear simulation Base150 in \Cref{sec:5}.}
        \label{fig:1}
\end{figure}

\begin{figure}
        \centering
            \begin{subfigure}[t]{0.49\linewidth}
        \includegraphics[width=1.03\linewidth]{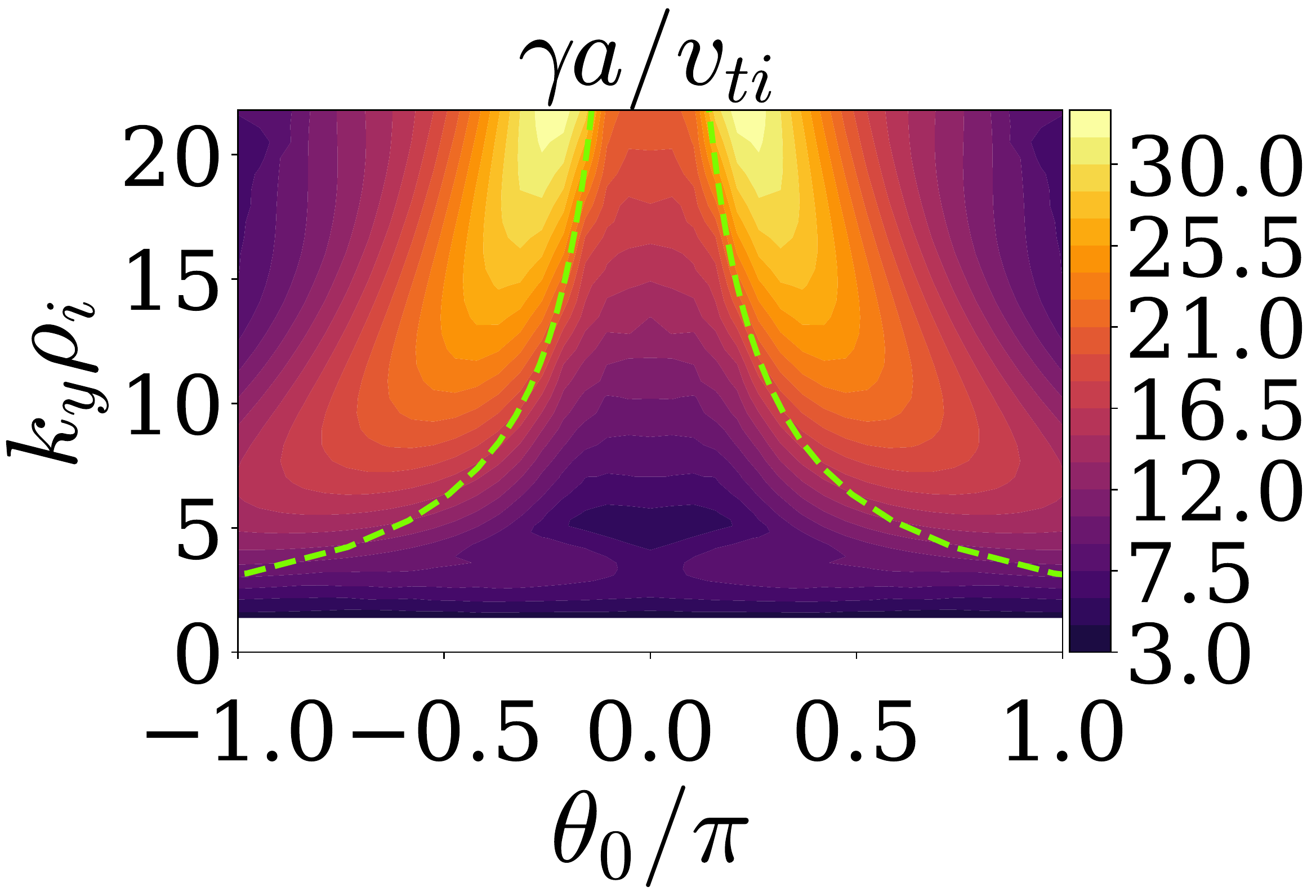}
        \caption{Linear.}
        \end{subfigure}   
           \begin{subfigure}[t]{0.49\linewidth}
        \includegraphics[width=1\linewidth]{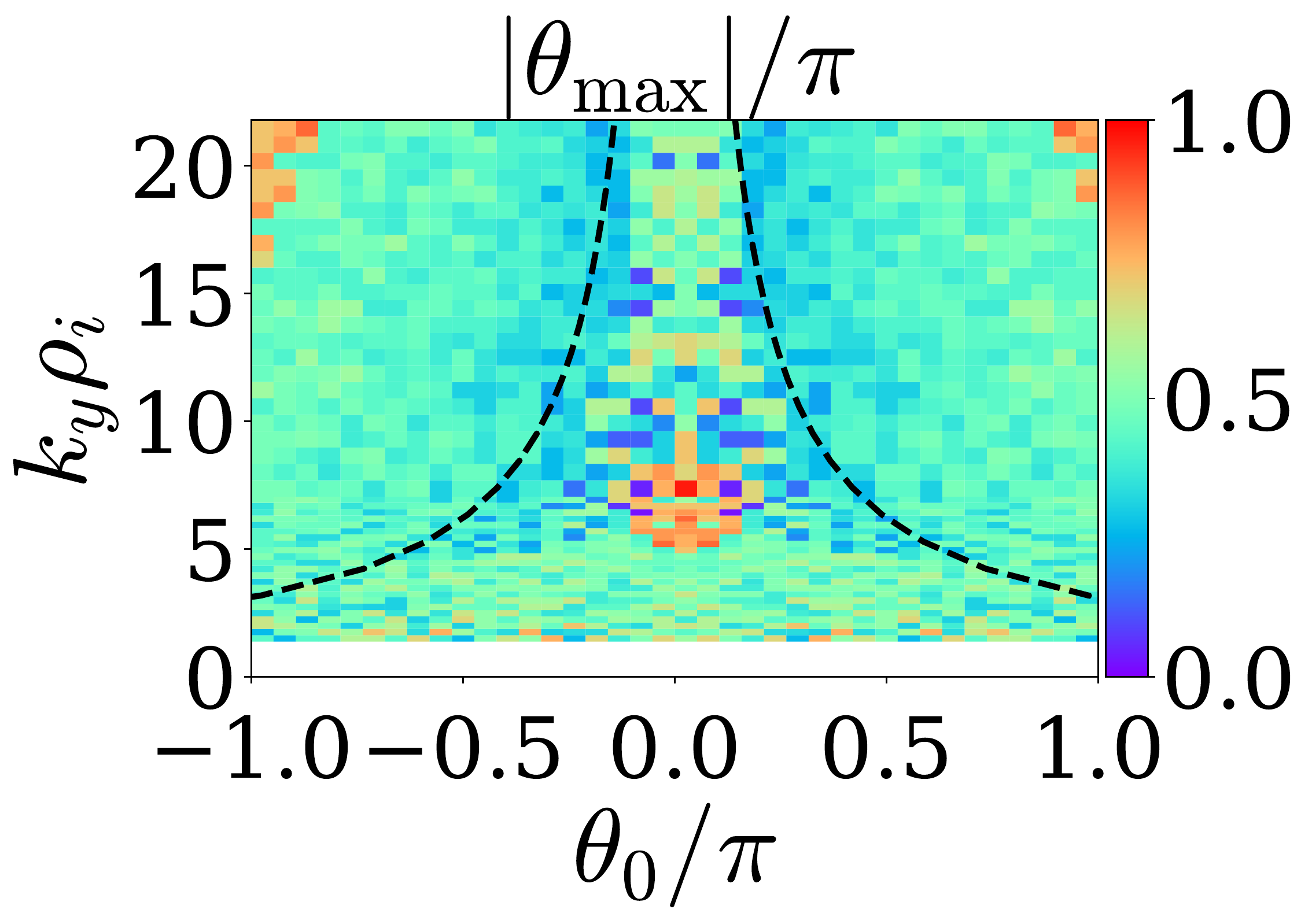}
        \caption{Linear.}
            \end{subfigure}
       \begin{subfigure}[t]{0.49\linewidth}
        \includegraphics[width=0.97\linewidth]{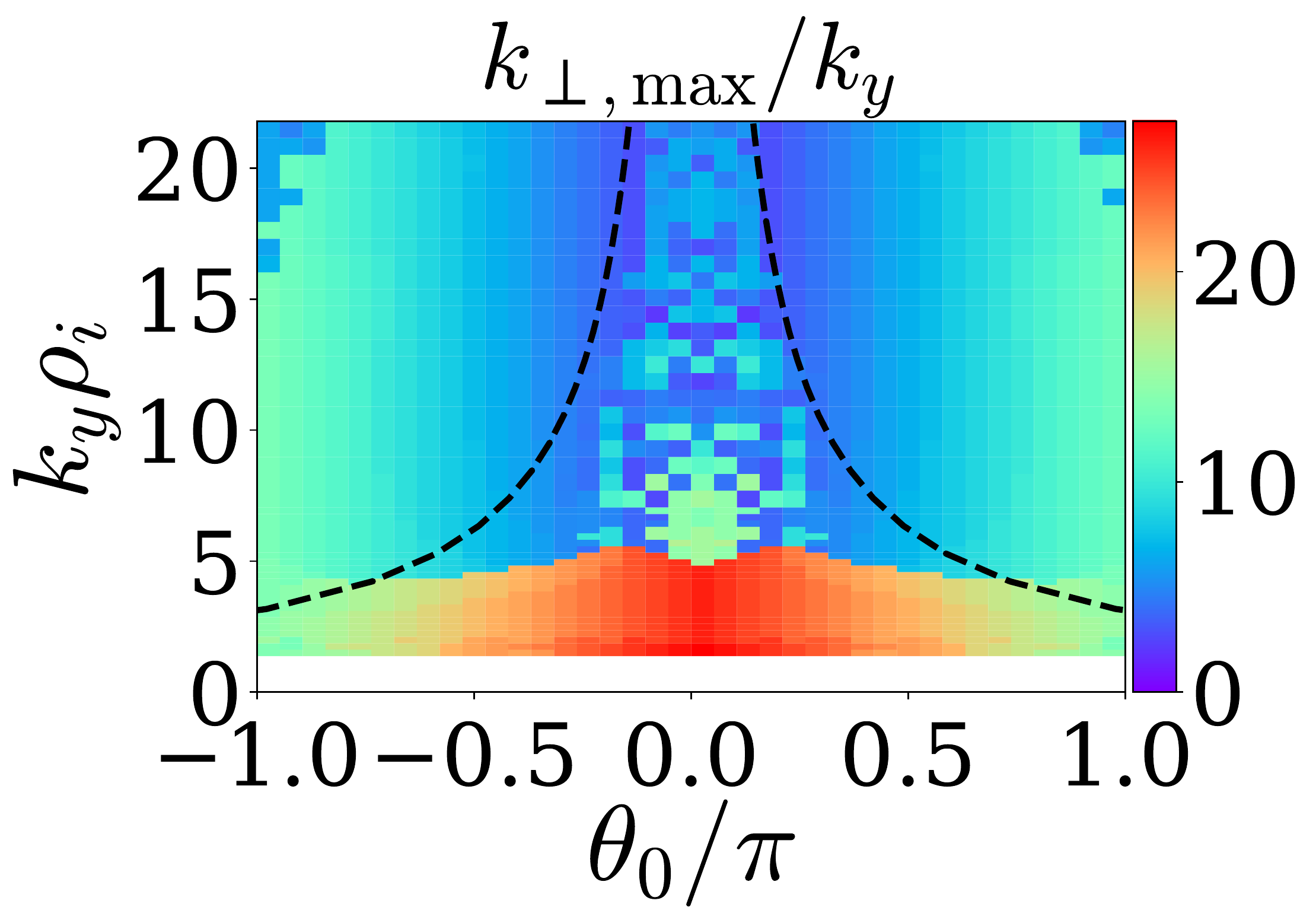}
        \caption{Linear.}
        \end{subfigure}
        \begin{subfigure}[t]{0.49\linewidth}
        \includegraphics[width=0.94\linewidth]{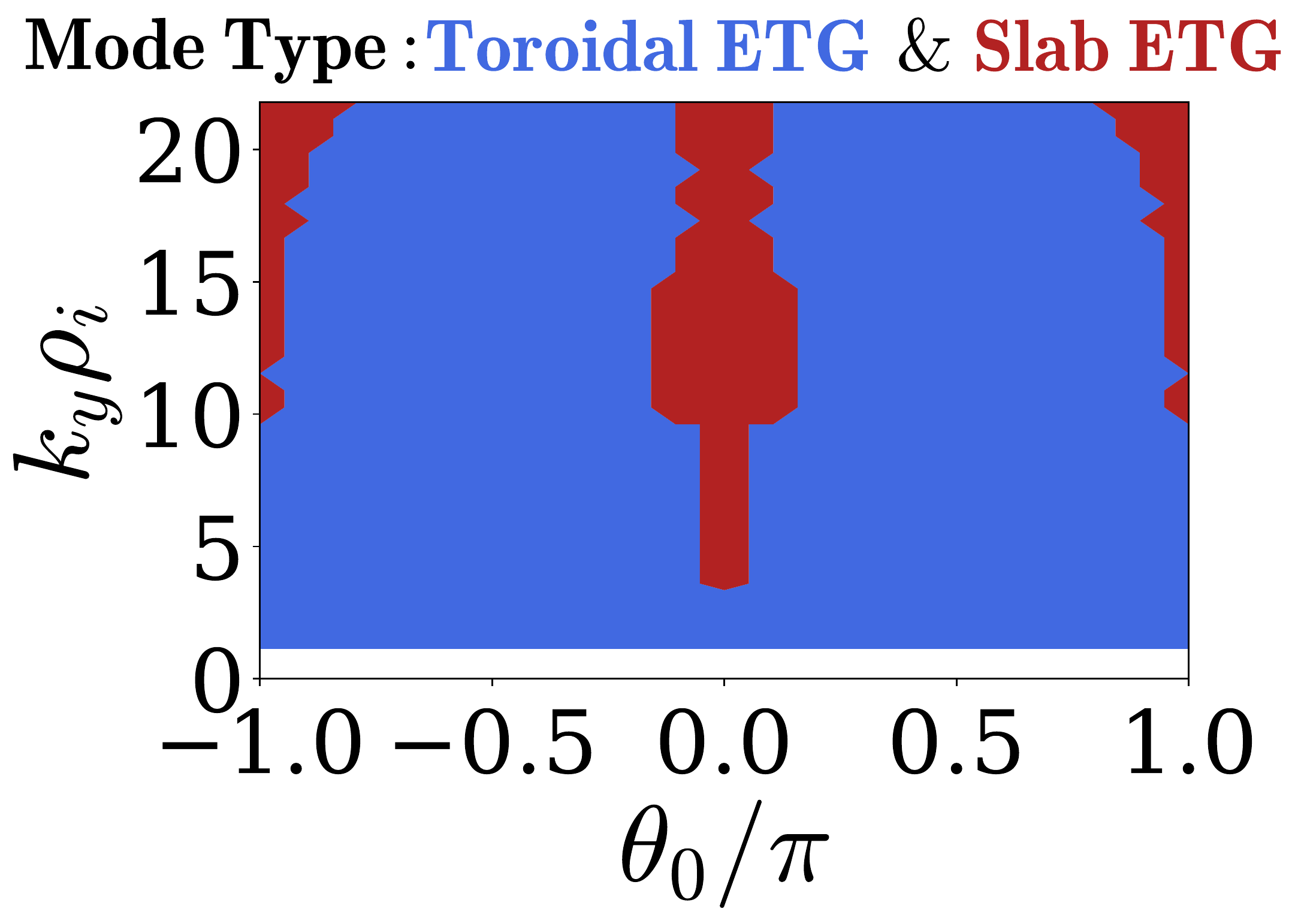}
        \caption{Linear.}
    \end{subfigure}
        \begin{subfigure}[t]{0.49\linewidth}
        \includegraphics[width=1.03\linewidth]{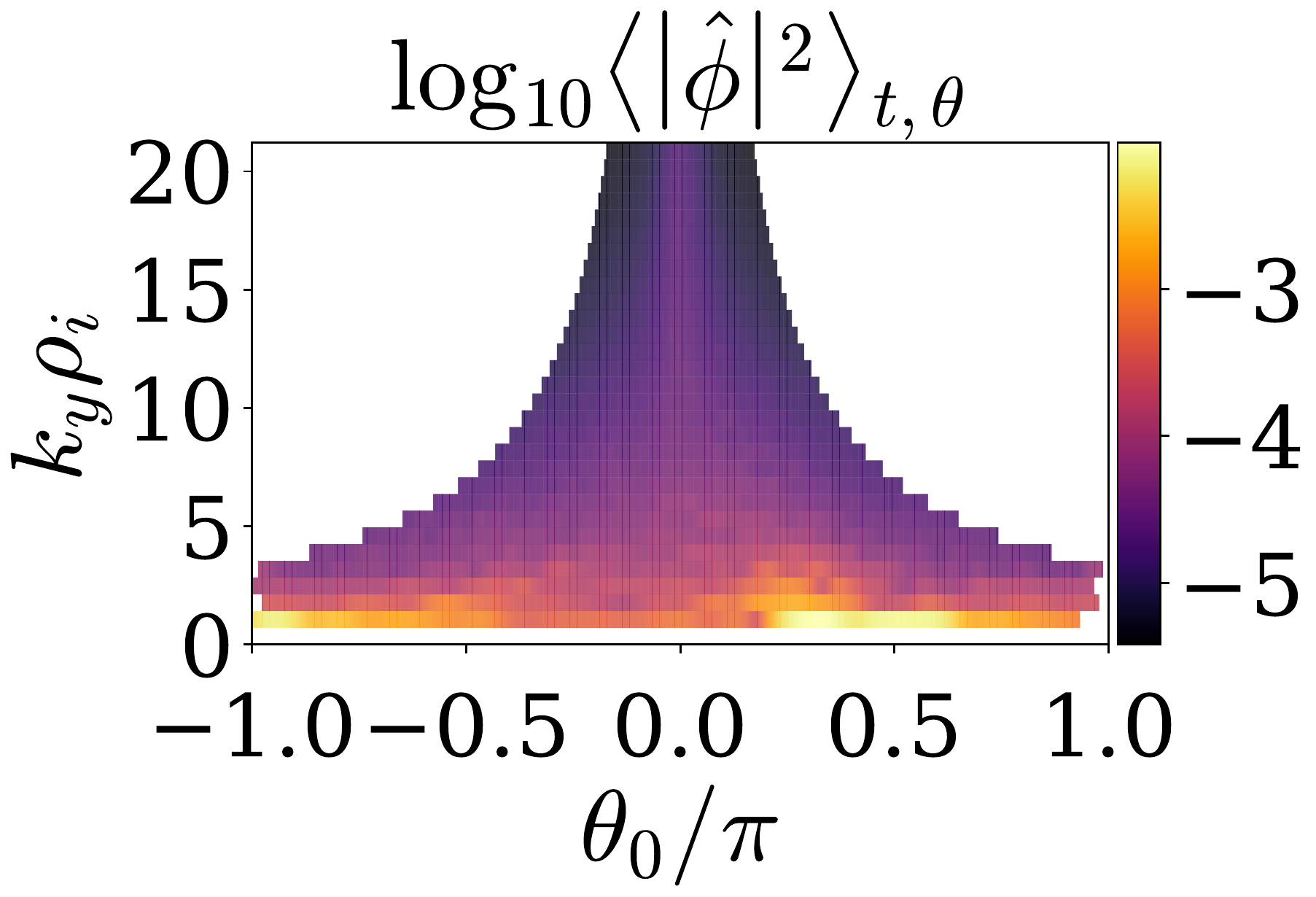}
        \caption{Nonlinear.}
        \end{subfigure}   
        \begin{subfigure}[t]{0.49\linewidth}
        \includegraphics[width=1.03\linewidth]{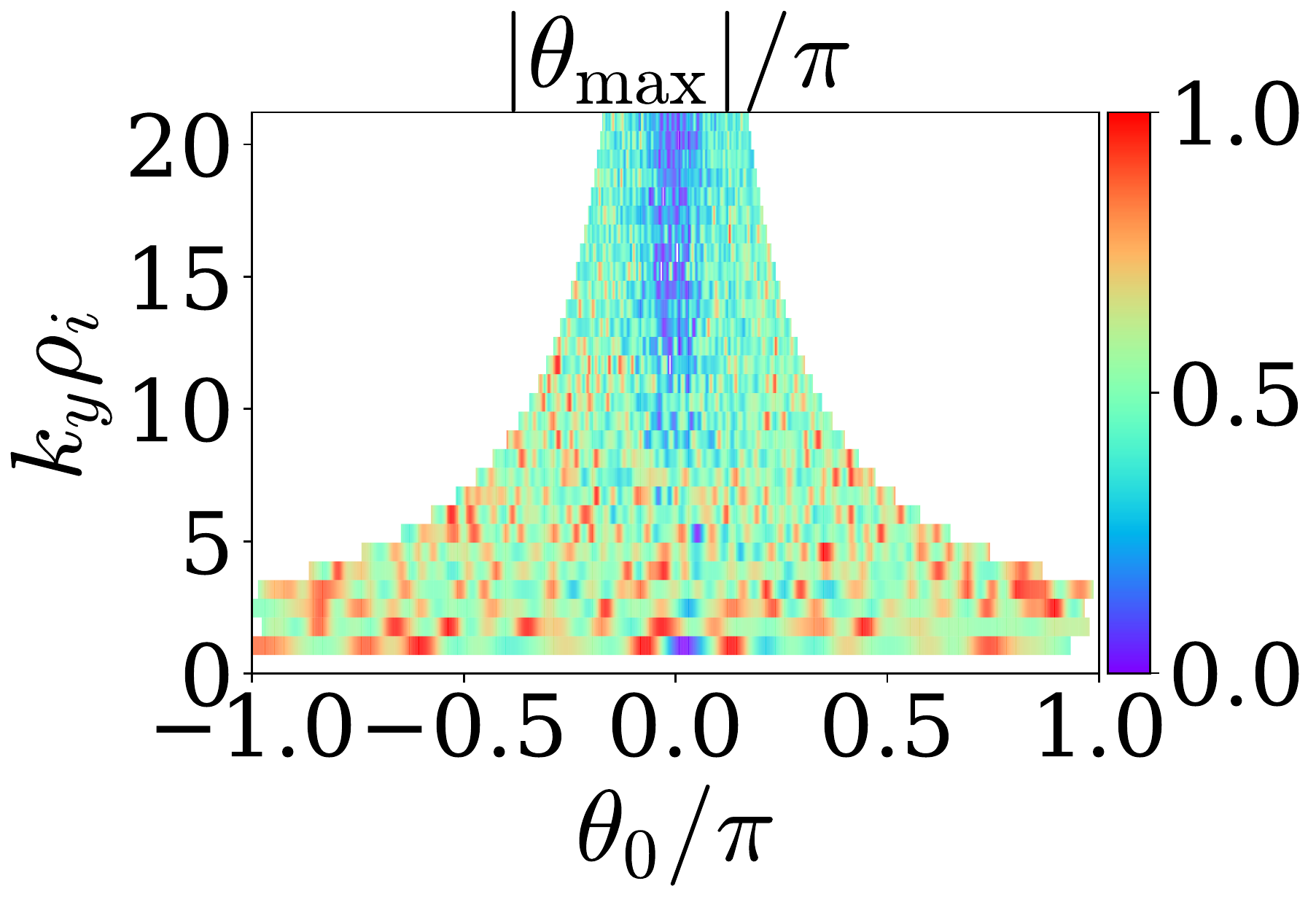}
        \caption{Nonlinear.}
        \end{subfigure}   
     \caption{Linear and nonlinear mode properties for the $k_y$ region where nonlinear potential amplitudes are highest, $k_y \rho_i \lesssim 20$. (a) Linear growth rate $\gamma$, (b) poloidal location $|\theta_{\mathrm{max}}|/\pi$ of maximum amplitude $|\hat{\phi}|$ for the linear mode, (c) perpendicular wavenumber evaluated at the linear mode's maximum and divided by the binormal wavenumber, viz., $k_{\perp, \mathrm{max} } / k_y$, and (d) the fastest growing mode's type, all versus $k_y \rho_i$ and $\theta_0$ for the linear simulation described in \Cref{sec:4}. The dashed curves in (a)-(c) denote the perpendicular grid boundary for nonlinear simulations. (e) The nonlinear amplitude $\log_{10} \langle | \hat{\phi} |^2 \rangle_{t, \theta}$ and (f) $|\theta_{\mathrm{max}}|/\pi$ versus $k_y \rho_i$ and $\theta_0$ for nonlinear simulations in \Cref{sec:5}; these are calculated using $\hat{\phi}$ averaged over $t v_{ti}/a \in [14.8,15.9]$.}
      \label{fig:2}
\end{figure}

We perform linear gyrokinetic simulations using the code \texttt{GS2} \cite{Dorland2000, Barnes2021} for $0.7 \leq k_y \rho_i \leq 150$. Due to tokamak toroidal symmetry, the linear system is $2\pi$ periodic in $\theta_0$ \cite{Connor1978,Cowley1991,Hazeltine2003}. For this reason, in \Cref{fig:1,fig:2} we plot $\theta_0$ between $-\pi$ and $\pi$ only. In \Cref{fig:1}(a), we plot the linear growth rate, $\gamma a/v_{ti}$ versus $k_y \rho_i$ and $\theta_0$. In \Cref{fig:1}(b), we indicate the fastest-growing linear mode, which for this equilibrium is associated with either toroidal or slab ETG instability. In Figures 3(a)-(d), we plot some properties of the fastest-growing modes in the $k_y \rho_i \lesssim 20$ region, where turbulent amplitudes in the nonlinear simulation described in \Cref{sec:5} are largest. In this region, toroidal ETG modes dominate except for $k_y \rho_i \gtrsim 5$ where $\theta_0 \approx 0$. In \Cref{fig:2}(a), we show that the maximum linear growth rate, $\gamma$, peaks at $\theta_0 \neq 0$. In \Cref{fig:2}(b), we plot $|\theta_{\mathrm{max}}|$, the poloidal angle at which the linear modes have maximum amplitude $|\hat{\phi}|$. Due to steep gradients, toroidal ETG modes with $k_y \rho_e \ll 1$ peak away from the outboard midplane \cite{Parisi2020}, as predicted in \Cref{eq:nine}. Toroidal ETG modes also satisfy $k_{\perp} / k_y \gg 1$ [see \Cref{eq:eight}], and $k_{\perp} \rho_e \lesssim 1$ [see \Cref{eq:ten}], shown in \Cref{fig:2}(c). In \Cref{fig:2}(d), we indicate the linear mode type at each $(k_y, \theta_0)$, showing dominant toroidal ETG instability.

Since the dominant linear instabilities found by us are ETG modes satisfying $k_{\perp} d_e \gg 1$, where $d_e$ is the electron skin depth, electromagnetic effects \cite{Horton1988, Dorland2000, Jenko2000, Holland2002, Adkins2022} are likely unimportant for these ETG modes. However, since electromagnetic modes often dominate in other pedestals \cite{Wang2012, Dickinson2013, Groebner2013, Holod2015, Hatch2016, Larakers2021}, we cannot rule out linearly subdominant electromagnetic instabilities being important nonlinearly. 

We have adopted the collisionless limit because the growth rates of the modes that we find are unaffected by collisionality \cite{Parisi2020}. {Linear parameter scans in $a/L_{Te}, a/L_{n}, \hat{s}$, and $T_e/T_i$ are performed in \cite{Parisi2020}. These scans reveal the prevalence of toroidal and slab ETG instability across a wide range of parameter values, providing confidence that our simulations capture the ETG physics of the experimental point.}

Since the dominant modes that we simulate satisfy $K_{x} L_{Te} \gg 1$, where $ K_x = \mathbf{k}_{\perp} \cdot (\nabla x) / |\nabla x|$ is the local radial wavenumber, our local flux-tube approach is justified. However, simulations of modes with longer radial wavelengths may require radially `global' approaches to better capture the physics of radial profile variation \cite{Ku2009,Gorler2011,Wan2012,Hatch2016,Candy2020,StOnge2022}.

\section{Nonlinear simulations} \label{sec:5}

\begin{figure}
        \centering
        \includegraphics[width=0.49\textwidth]{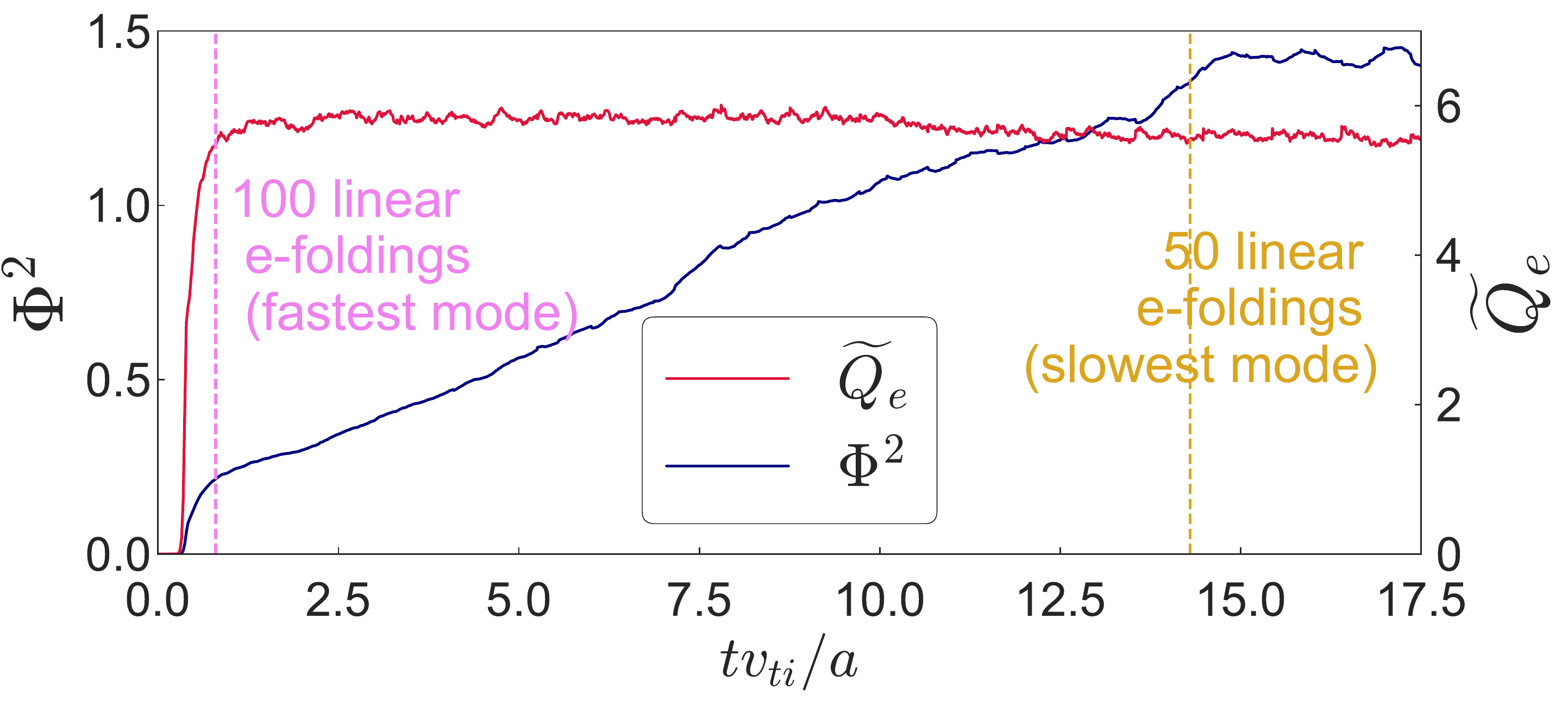}
        \caption{Time traces of the potential $\Phi^2$ and heat flux $\widetilde{Q}_e $, defined in \Cref{eq:thirteen,eq:fourteen}, respectively, for Base150 nonlinear simulation (see row one of \Cref{tab:1}).}
        \label{fig:3}
\end{figure}

\begin{figure}
        \centering
        \includegraphics[width=0.5\textwidth]{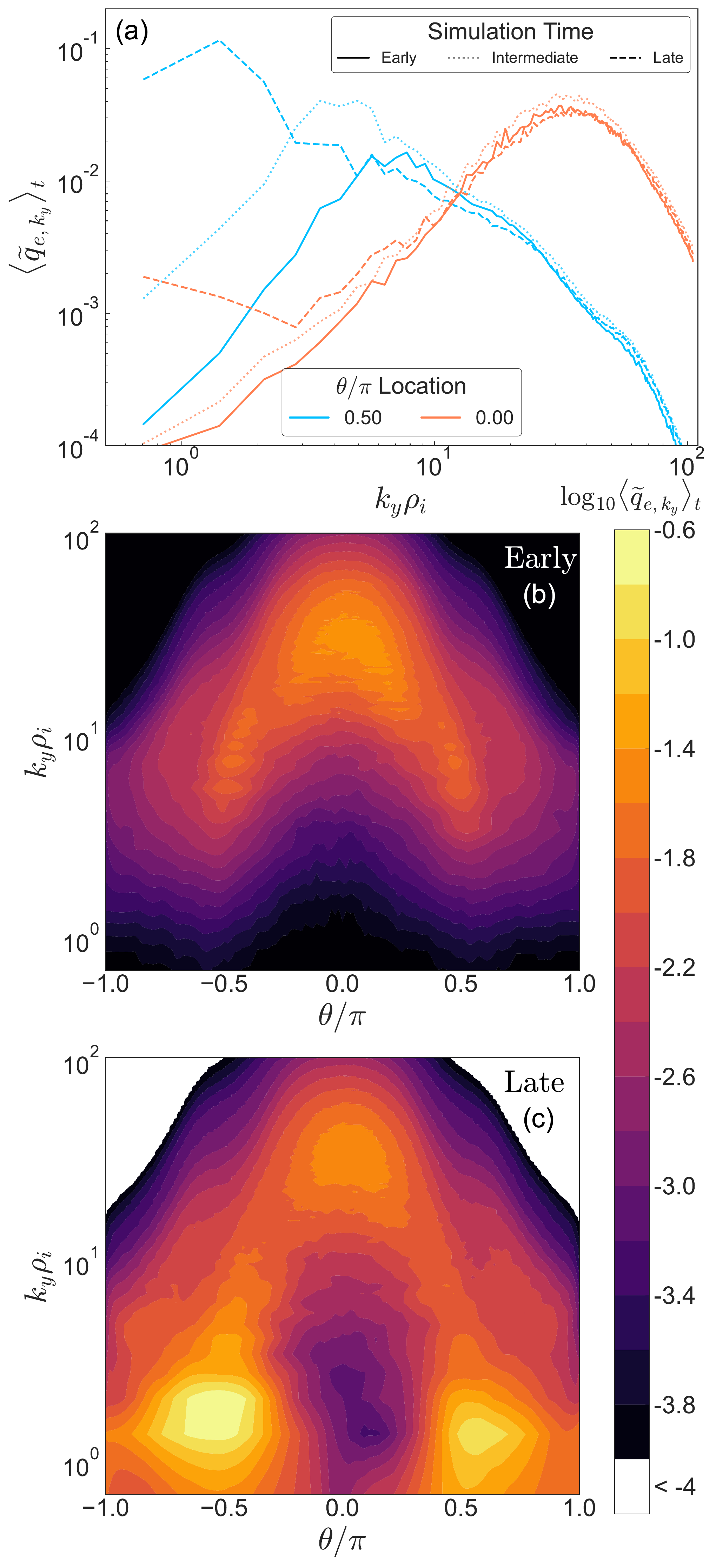}
        \caption{(a) Heat flux $\langle \widetilde{q} _{e,k_y} \rangle_t$, defined in \Cref{eq:thirteen}, versus $k_y \rho_i$ from Base150 nonlinear simulation [see \Cref{tab:1}] at two $\theta$ locations. The heat flux is averaged over $t v_{ti}/a \in [0.7, 1.7]$ (early), $t v_{ti}/a \in [7.5,8.8]$ (intermediate), and $t v_{ti}/a \in [14.8,15.9]$ (late). (b) $\langle \widetilde{q} _{e,k_y} \rangle_t$ versus $k_y \rho_i$ and $\theta/ \pi$ at early times. (c) $\langle \widetilde{q} _{e,k_y} \rangle_t$ versus $k_y \rho_i$ and $\theta/ \pi$ at late times.}
        \label{fig:4}
\end{figure}

\begin{figure*}
        \centering
        \includegraphics[width=0.98\textwidth]{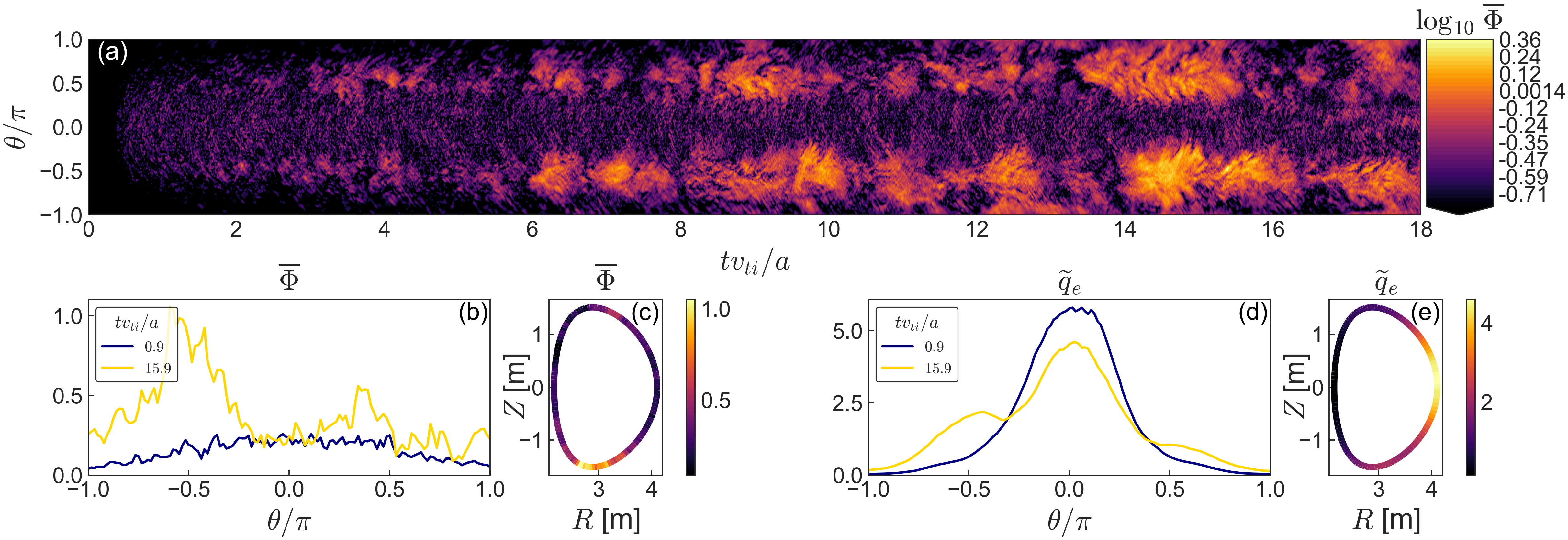}
        \caption{(a): $\log_{10} \overline{ \Phi} $ versus time and $\theta/\pi$. (b) $\overline{ \Phi}$ and (d) $ \widetilde{ q}_e $ versus $\theta/\pi$ evaluated at two times: $t v_{ti}/ a = 0.9$ and $t v_{ti}/ a = 15.9$. (c) $\overline{ \Phi}  $ and (e) $\widetilde{ q}_e $ projection onto the flux surface, both also evaluated at $t v_{ti}/ a = 15.9$. The quantities $\widetilde{q}_e $ and $\overline{ \Phi} $ are defined in \Cref{eq:fourteen,eq:thirteen}, respectively. Data from simulation Base150 [see \Cref{tab:1}].}
        \label{fig:5}
\end{figure*}

\begin{figure}
        \centering
    \begin{subfigure}[c]{0.22\textwidth}
        \includegraphics[width=1.1\textwidth]{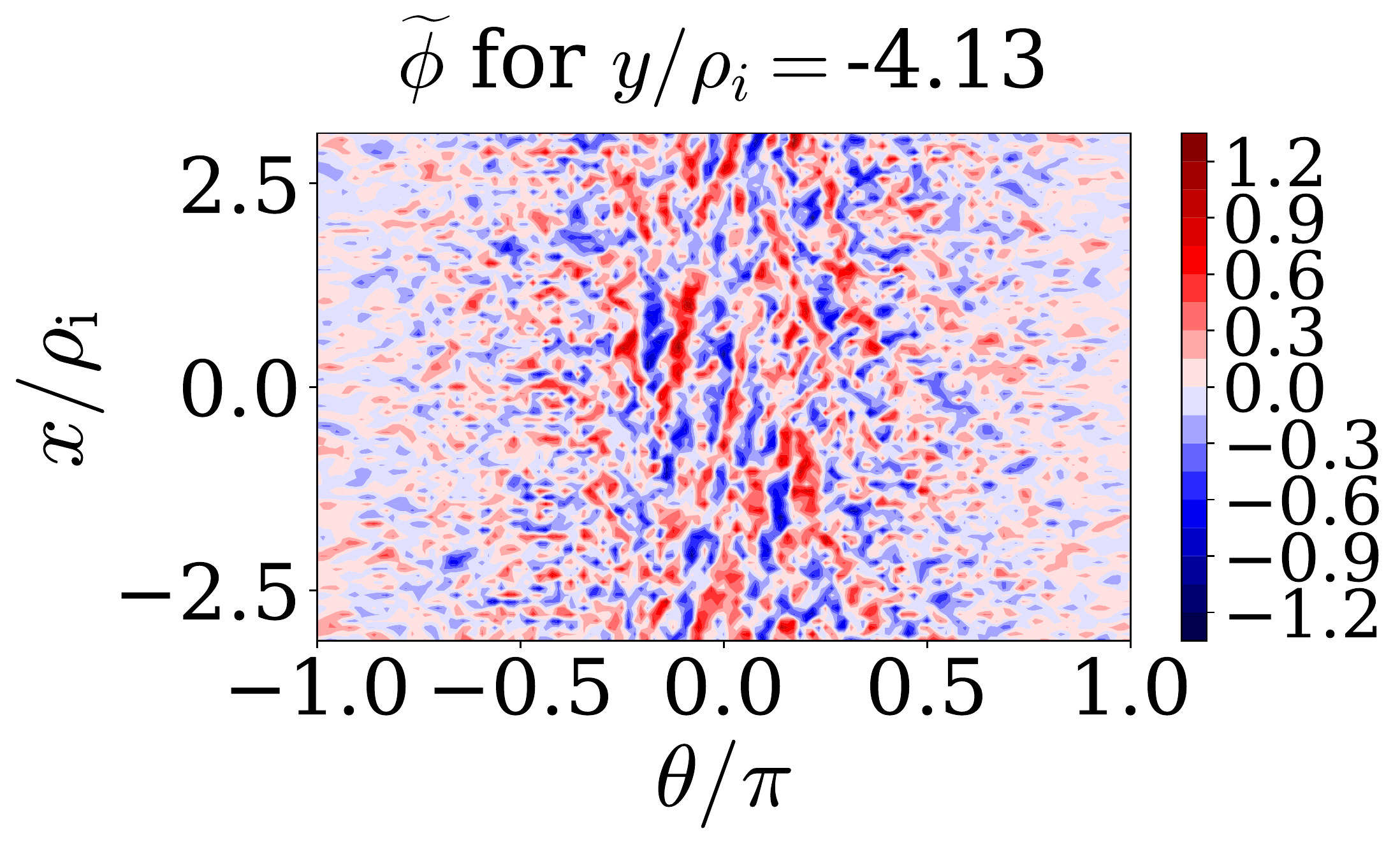}
        \caption{Early times (dominant $k_y \rho_e \sim 1$ ETG).}
    \end{subfigure} \hspace{0.7mm}
    \hspace{0.1mm} \begin{subfigure}[c]{0.22\textwidth}
        \includegraphics[width=1.1\textwidth]{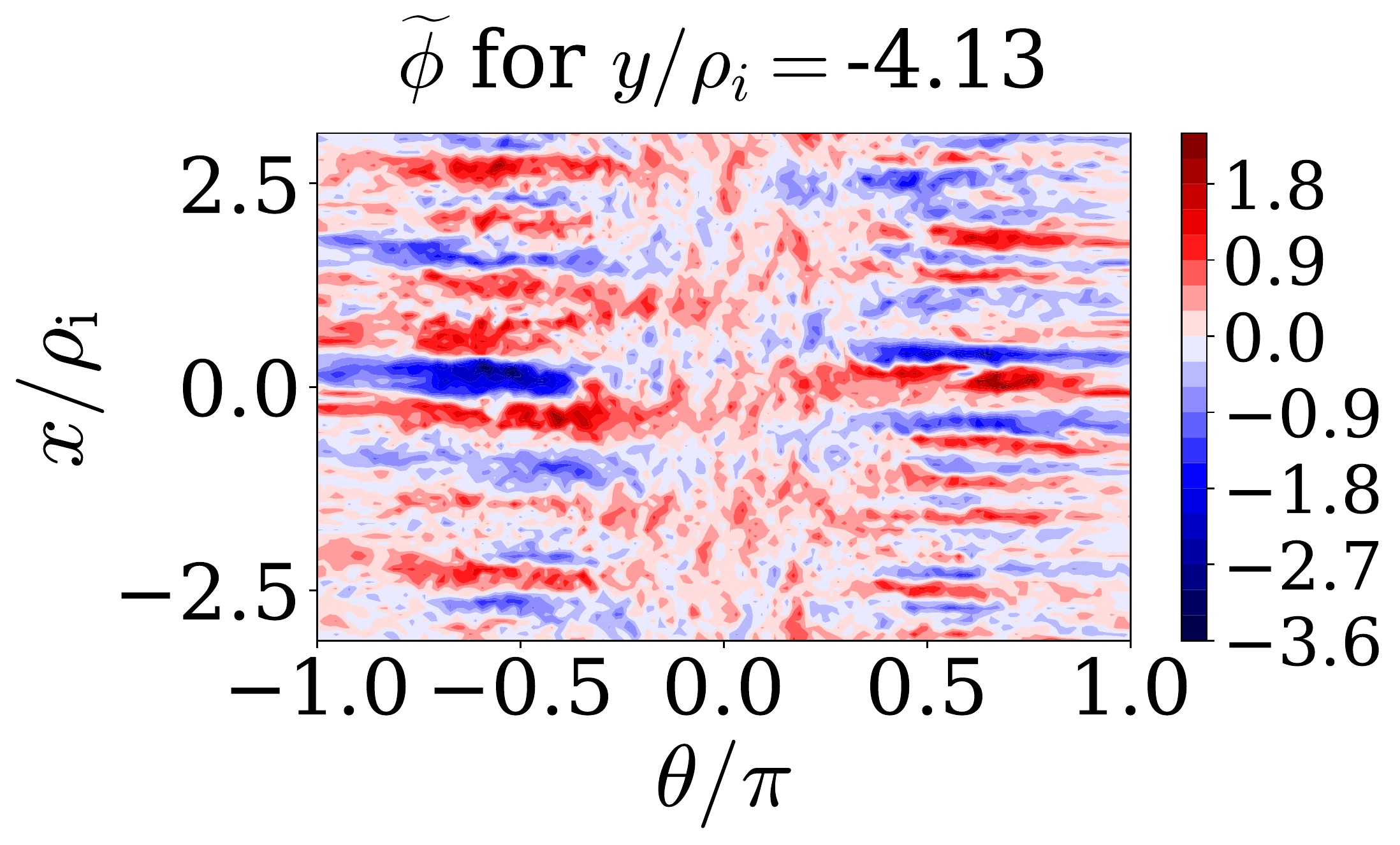}
        \caption{Late times (dominant $k_y \rho_i \sim 1$ ETG).}
    \end{subfigure}
        \begin{subfigure}[t]{0.22\textwidth}
        \includegraphics[width=1.1\textwidth]{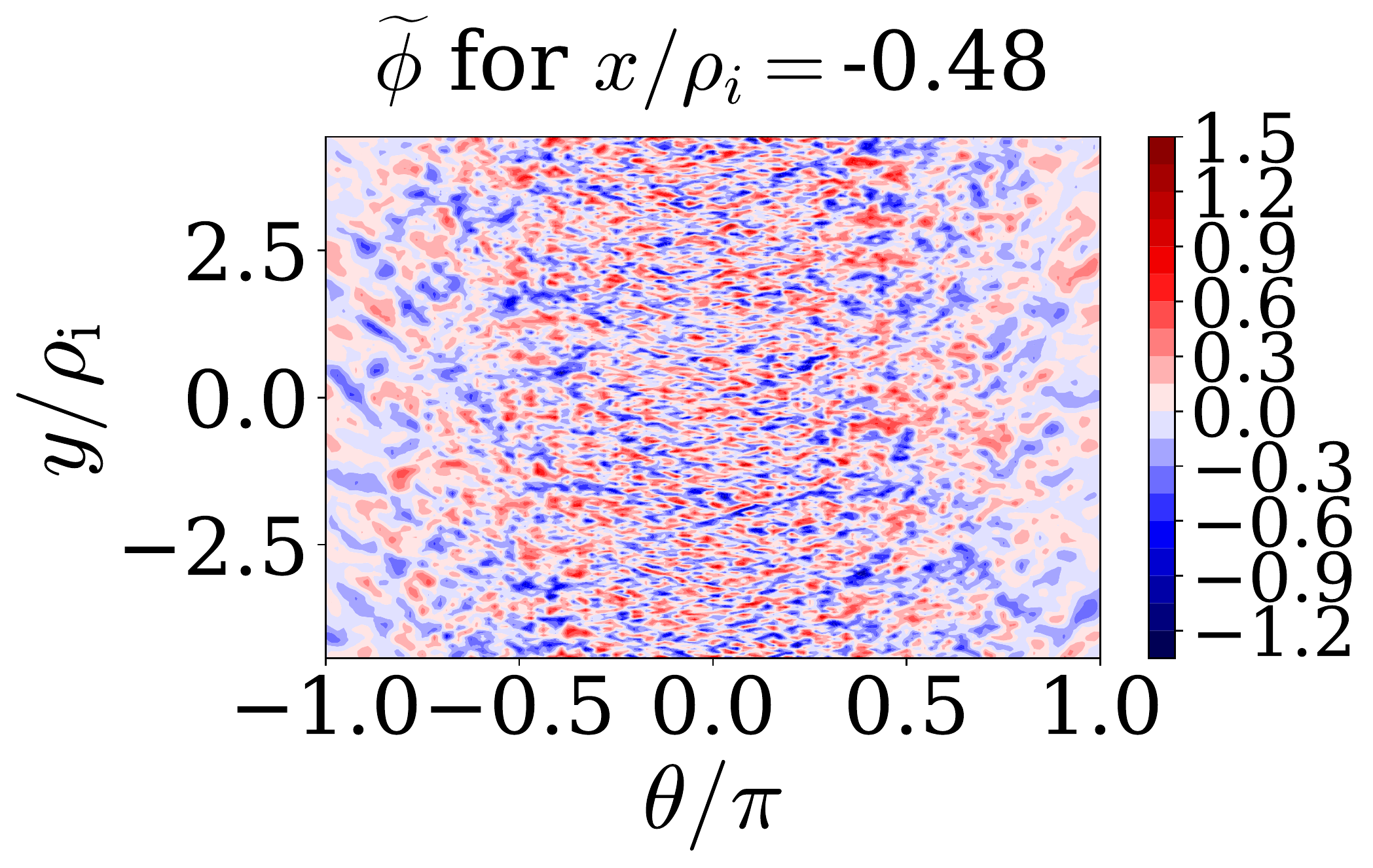}
         \caption{Early times (dominant $k_y \rho_e \sim 1$ ETG).}
    \end{subfigure} \hspace{0.7mm}
    \hspace{0.1mm} \begin{subfigure}[t]{0.22\textwidth}
        \includegraphics[width=1.1\textwidth]{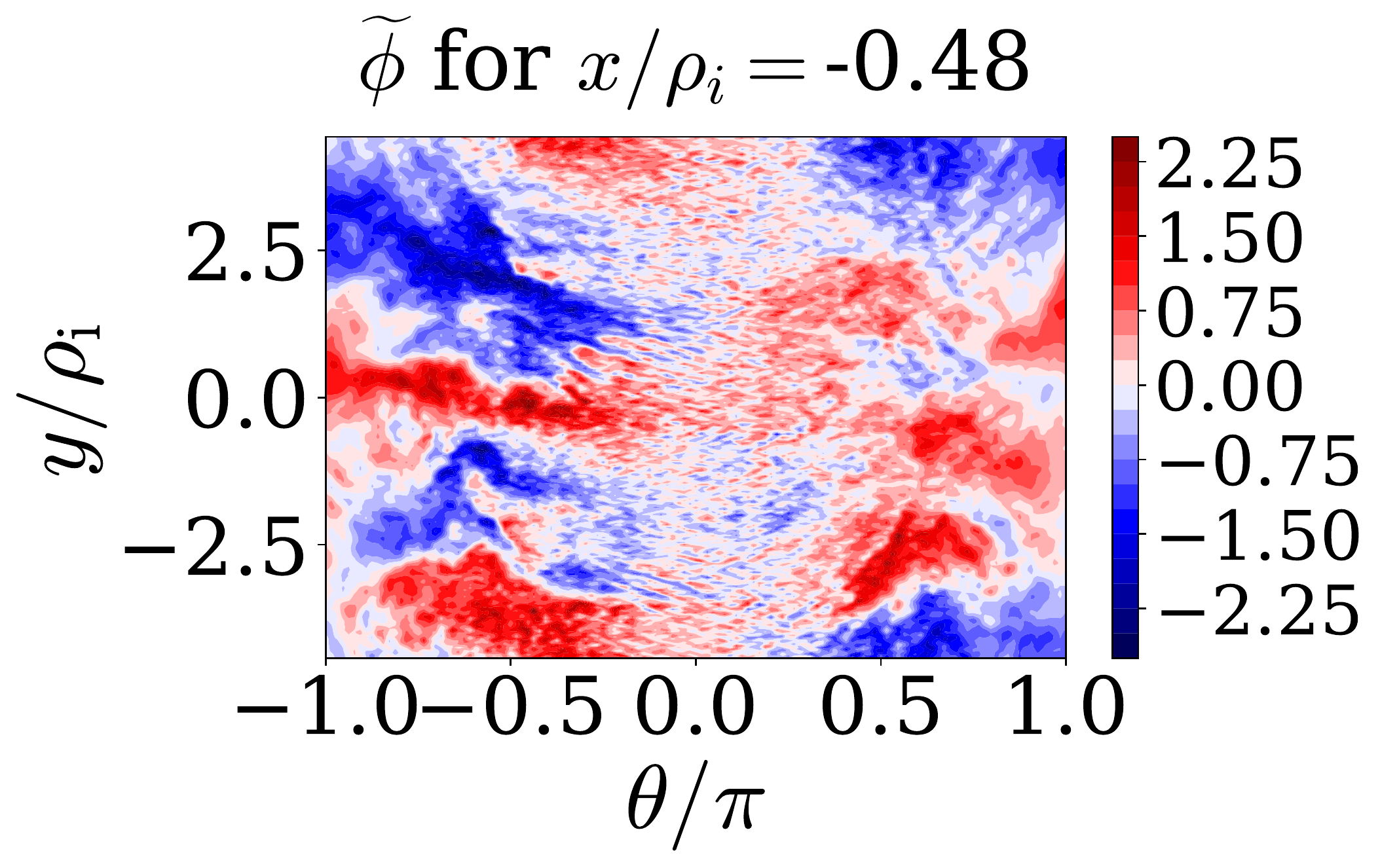}
        \caption{Late times (dominant $k_y \rho_i \sim 1$ ETG).}
    \end{subfigure}
        \begin{subfigure}[t]{0.23\textwidth}
        \includegraphics[width=1\textwidth]{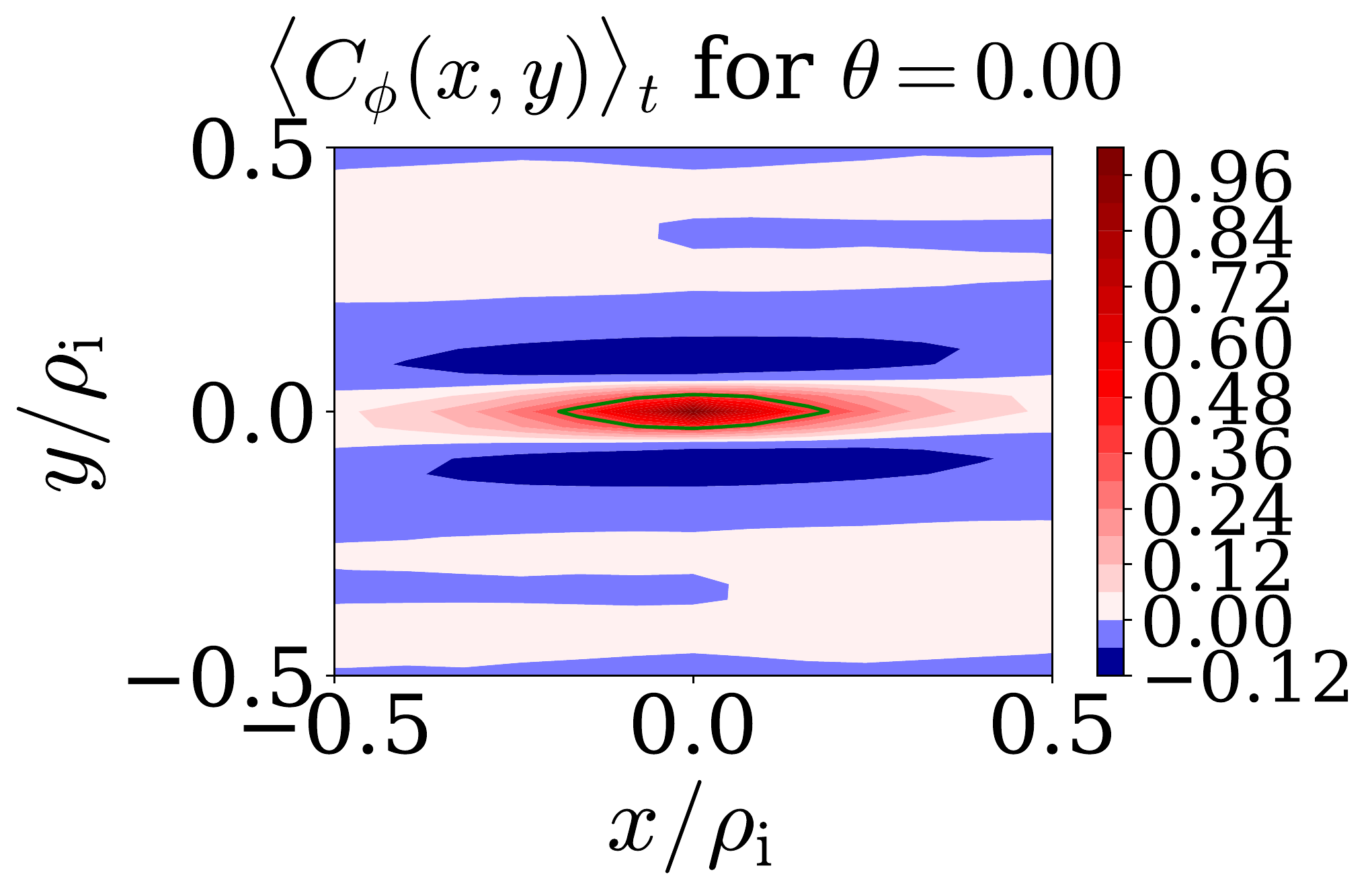}
        \caption{Early times (dominant $k_y \rho_e \sim 1$ ETG).}
    \end{subfigure} 
    \begin{subfigure}[t]{0.23\textwidth}
        \includegraphics[width=1\textwidth]{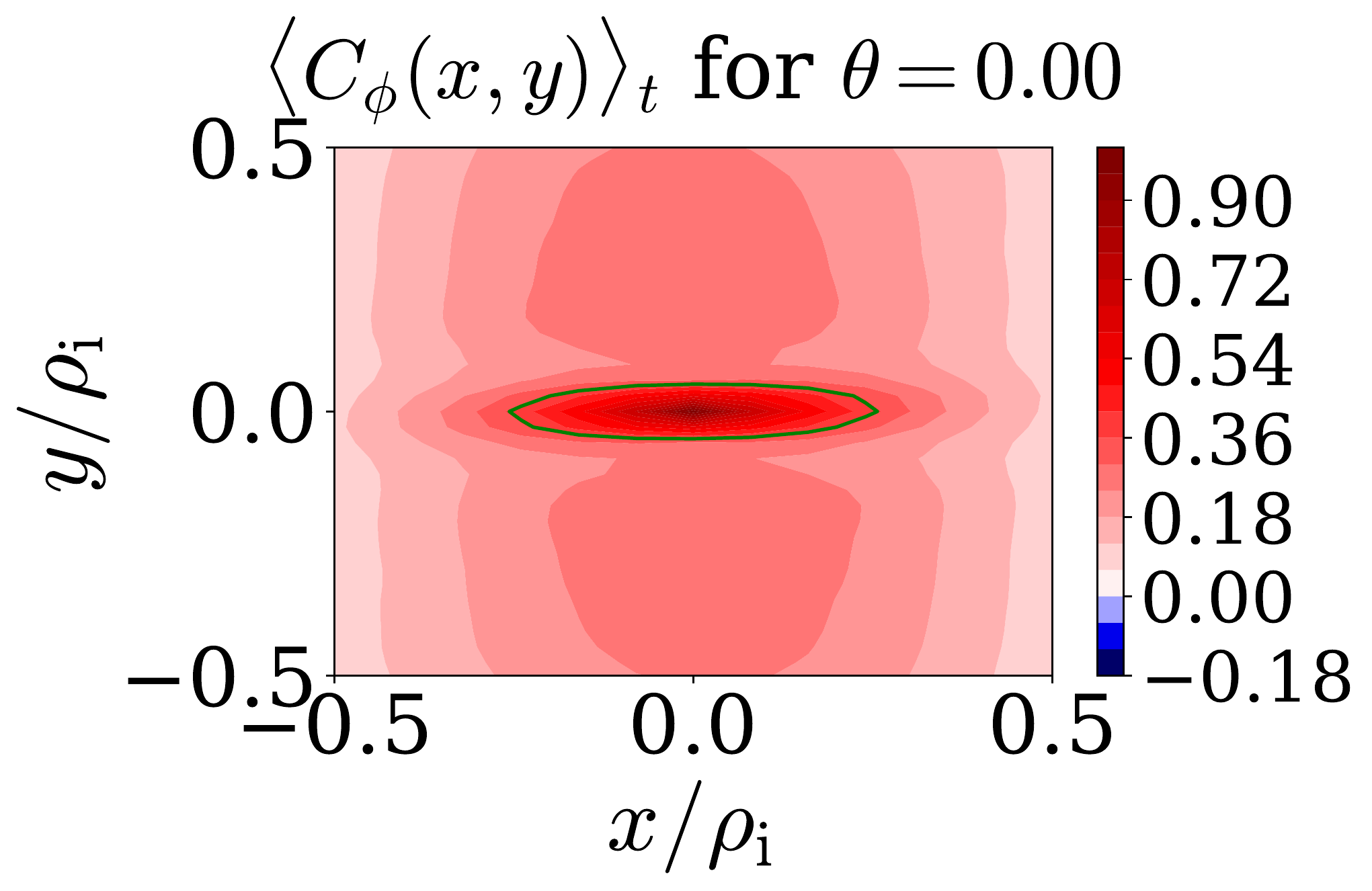}
        \caption{Late times (dominant $k_y \rho_i \sim 1$ ETG).}
    \end{subfigure}
            \begin{subfigure}[t]{0.23\textwidth}
        \includegraphics[width=1\textwidth]{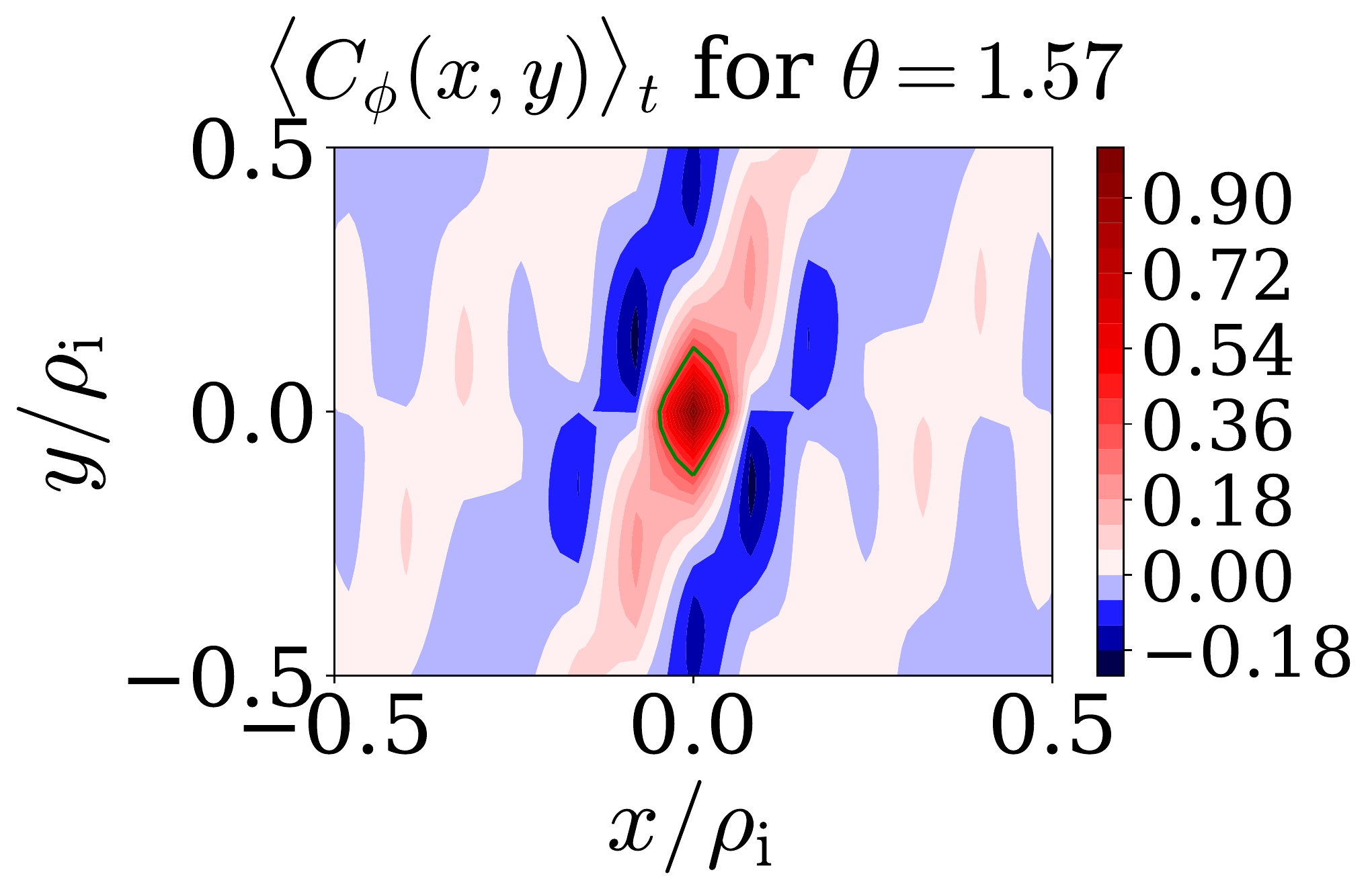}
        \caption{Early times (dominant $k_y \rho_e \sim 1$ ETG).}
    \end{subfigure} 
    \begin{subfigure}[t]{0.23\textwidth}
        \includegraphics[width=1\textwidth]{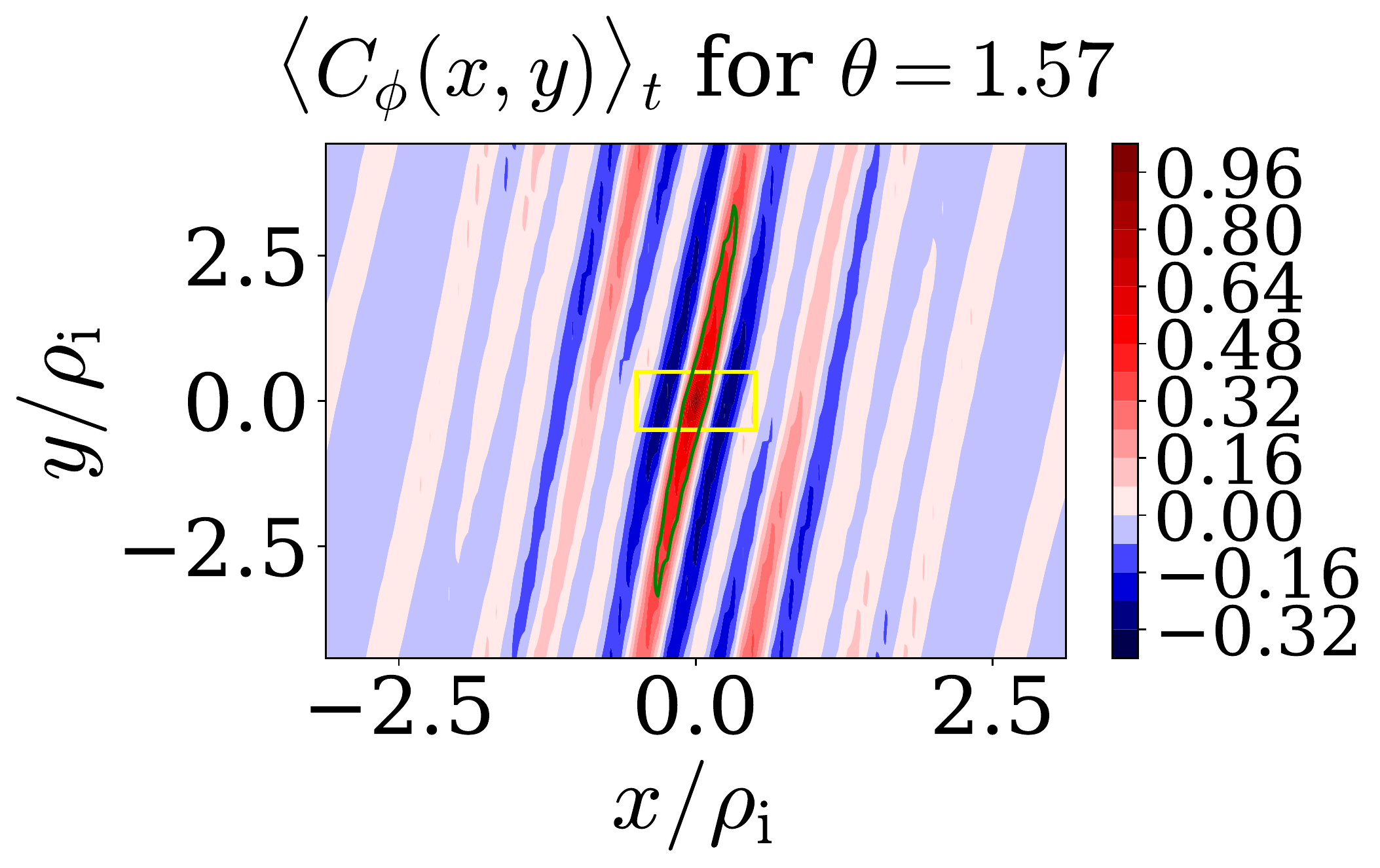}
        \caption{Late times (dominant $k_y \rho_i \sim 1$ ETG).}
    \end{subfigure}
    \caption{(a) - (d) electrostatic potential $\widetilde{\phi}$ at early ($t v_{ti}/a  = 1.1$, left column) and late ($t v_{ti}/a  = 15.9$, right column) times. (e) - (h) Correlation functions at early ($t v_{ti}/a  = 1.1$, left column) and late ($t v_{ti}/a  = 15.9$, right column) times. The contour of value $1/e$ is plotted in green to indicate a correlation length. Note that the plot ranges in (h) are much larger than in (e), (f), and (g). In (h), a yellow box shows the plot ranges of (g). Data from simulation Base150 [see \Cref{tab:1}].}
    \label{fig:6}
\end{figure}
Simulating turbulence away from the outboard midplane imposes demanding radial-resolution requirements, necessitating large numbers of radial grid points \cite{Parisi2020b}. Additionally, capturing the fine structure in the linear spectra in \Cref{fig:1,fig:2} requires narrow perpendicular grid spacing in $\theta_0$ and $k_y$. Given that there is strong linear instability for $k_y \rho_e \gg 1$, we also need a large maximum $k_y \rho_i$.

\begin{table}
\centering
\begin{tabular}{|p{1.45cm}|p{1.09cm}|p{0.8cm}|p{0.47cm}|p{0.35cm}|p{0.46cm}|p{0.93cm}|p{0.62cm}|p{0.62cm}|p{0.61cm}|}
 \hline
 Simulation & $k_{y, _{\rm{max}}} \rho_i$ & $ \Delta k_y \rho_i$ & $\frac{a}{L_{Te}}$ & $\frac{a}{L_n}$ & $\frac{a}{L_{Ti}}$ & \makecell{$D_{hy}$ \\ $(10^{-7})$} & $\texttt{nakx}$ & $\texttt{naky}$ & \texttt{nzed} \\
 \hline
 \makecell{Base150} & 105.8 & 0.71 & 42 & 10 & 11 & $ 10$ & 67 & 150 & 128 \\
  \hline
  \makecell{Radial100} & 70.5 & 0.71 & 42 & 10 & 11 & $ 30$ & 134 & 100 & 128 \\
  \hline
   \makecell{Scan100a \\ ($L_n$ fixed)} & $87.6$ & 0.88 & 34 & 10 & 11 & $17$ & 67 & 100 & 128 \\
    \hline
 \makecell{Scan100b \\ ($\eta_e$ fixed)} & $87.6$ & 0.88 & 34 & 8 & 11 & $17$ & 67 & 100 & 128 \\
   \hline
 \makecell{Scan100c \\ ($L_n$ fixed)} & $141.0$ & 1.41 & 21 & 10 & 11 & $4.6$ & 67 & 100 & 128 \\
 \hline
 \makecell{Scan100d \\ ($\eta_e$ fixed)} & $141.0$ & 1.41 & 21 & 5 & 11 & $4.6$ & 67 & 100 & 128 \\
 \hline
 \makecell{Scan100e \\ (circle, \\ $\eta_e$ fixed)} & $176.3$ & 1.76 & 4 & 1 & 1 & $2.6$ & 67 & 100 & 32 \\
  \hline
 \makecell{Scan200f \\ (circle)} & $70.5$ & 0.35 & 42 & 10 & 11 & $30$ & 67 & 200 & 64 \\
 \hline
\end{tabular}
\caption{Nonlinear simulations with varying box sizes. Base150 is the main simulation used throughout the text, Scan100 and Scan200 are used for $a/L_{Te}$ scans in \Cref{sec:6}, with $\Delta k_y \propto a/L_{Te}$, except for cases with circular flux surfaces, where we retain relatively small $\Delta k_y$ to resolve possible $k_y \rho_i \sim 1$ ETG turbulence. Radial100, performed to test radial resolution, has double \texttt{nakx} of other simulations. Quantities \texttt{nakx}, \texttt{naky}, and \texttt{nzed} are the number of $k_x$ and $k_y$ wavenumbers and parallel grid points, respectively.}
\label{tab:1}
\end{table}

We perform nonlinear simulations that attempt to satisfy these demanding resolution requirements using the gyrokinetic code \texttt{stella}. We simulate the non-adiabatic response of both ions and electrons by evolving $h_s$ for both species according to \Cref{eq:one}. Simulations have $\Delta k_{x} \rho_i = 1.38$, $\Delta k_{y} \rho_i = 0.71$, 150 $k_y$ modes, 67 $k_x$ modes, 128 parallel grid points, 12 $\mu =  m_s v_{\perp}^2/(2B)$ grid points, and 48 $v_{\parallel}$ grid points. These simulation parameters are referred to as the `Base150' simulation. Simulations are performed with an experimentally relevant flow-shear value $\gamma_E a/ v_{ti} = 0.65$ \cite{Parisi2020}, without which ETG streamers \cite{Dorland2000} at $k_y \rho_i \sim 1$ appear at long times (we found them at $t v_{ti}/a \simeq 12$ in a simulation with $\gamma_E = 0$). Hyperviscosity in $k_y$ prevents spectral pile-up at $k_y \rho_e \gtrsim 1$, and is discussed further around \Cref{eq:hyperviscousDh}.

In \Cref{fig:1}(a), the green dashed curves denote the edge of the $k_y \rho_i$ and $\theta_0$ grids for our nonlinear simulations. The variable $\theta_0$ is not periodic in $2\pi$ in nonlinear simulations, but we can ignore $|\theta_0| > \pi$ because we find very low turbulent amplitudes for these higher values of $k_x$. For $k_y \rho_i \gg 1$, the $|\theta_0|$ gridpoints have small values since $\theta_0 \sim 1 / k_y$, limiting the resolution of turbulence at $k_y \rho_i \gg 1$ away from $\theta_0 = 0$. This limitation occurs because the radial grid in nonlinear simulations is evenly spaced in $k_x$, but not in $\theta_0$. We checked this limitation in $\theta_0$ for our nonlinear simulations by doubling the number of radial modes in a cheaper simulation (referred to as Radial100 in \Cref{tab:1}) using 100 $k_y$ modes (rather than 150 $k_y$ modes in Base150). This doubles the maximum $|\theta_0|$ value included in the simulation at any given $k_y \rho_i$, allowing us to resolve more structure in $\theta_0$. We found that doubling the number of radial modes did not qualitatively change the nature of the turbulence.

In \Cref{fig:2}(e), we plot the turbulence amplitude $\log_{10} \langle | \hat{\phi} |^2 \rangle_{t, \theta}$ averaged over $t v_{ti}/a \in [14.8,15.9]$ from our nonlinear simulations versus $k_y \rho_i$ and $\theta_0$, zoomed in to the $k_y \rho_i \lesssim 20$ region where $\log_{10} \langle | \hat{\phi} |^2 \rangle_{t, \theta}$ is the largest. In \Cref{fig:2}(f), we plot the $\theta$ location where $\log_{10} \langle | \hat{\phi} |^2 \rangle_{t}$ has a maximum for each $(k_y \rho_i, \theta_0)$ value. For lower $k_y \rho_i$ modes, the mode amplitudes peak far away from $\theta = 0$.

In our simulation, the fastest-growing modes at $k_y \rho_i \approx 1$ and $k_y \rho_i \approx 90$ have linear growth rates $\gamma a/v_{ti} \simeq 1$ and $\gamma a/v_{ti} \simeq 70$, respectively. To resolve these modes and their nonlinear interactions, the simulation must satisfy $t \gg 1/ \gamma_{\mathrm{slowest} }$, where $\gamma_{\mathrm{slowest} } a/ v_{ti} \simeq 3.5$ is the slowest linear growth rate over all dominant modes with $k_y \rho_i \lesssim 20$ in our simulation domain. This is demonstrated by the simulation time traces in \Cref{fig:3}. For $0 < t v_{ti}/a \lesssim 2$, the heat flux is dominated by faster high-$k_y \rho_i$ slab ETG modes similar to `conventional' ETG, with the heat flux peaked at $k_y \rho_e \sim 1$ at $\theta = 0$, shown by the `early' curves of the heat flux in \Cref{fig:4}(a) and the heat flux contours in \Cref{fig:4}(b). Figures 5(a) and (b) show local contributions in $\theta$ to the electron heat flux, $\widetilde{q}_{e,k_y}$, due to the turbulence at particular $k_y \rho_i$ values. The total turbulent electron heat flux through the flux surface is then
\begin{equation}
\widetilde{ Q}_e (t) = \int \widetilde{q}_e d \theta, \;\;\;\;\; \widetilde{q}_{e} (\theta,t) = \sum_{k_y} \widetilde{q}_{e,k_y} (k_y, \theta, t).
\label{eq:thirteen}
\end{equation}
Here, $\widetilde{Q}_e $ is normalized to ion gyroBohm units, $Q_{gB} = (\rho_i/a)^2 p_i v_{ti}$, where $p_i$ is the equilibrium ion pressure. At the early times $0 < t v_{ti}/a \lesssim 2$, while the heat flux appears steady and one might erroneously believe that saturation has been reached, the total electrostatic potential
\begin{equation}
\Phi^2(t)  = \int \overline{ \Phi}^2 d \theta, \;\; \overline{ \Phi} ^2(\theta, t) = \sum_{k_x,k_y} \left| \hat{\phi}_{k_x,k_y}(\theta,t) \right|^2,
\label{eq:fourteen}
\end{equation}
is, in fact, still increasing (see \Cref{fig:3}). We plot the parallel structure of $\overline{ \Phi}$ and $\widetilde{q}_e$ in Figures 6(b) and 6(d), showing their maximum amplitudes near the outboard midplane at early times.

At later times ($t v_{ti}/a \gtrsim 2$), the slower-growing ETG modes increase in amplitude, causing turbulence to peak away from the outboard midplane. \Cref{fig:4}(a) shows non-negligible heat transport at lower $k_y \rho_i$ from modes near the flux surface's top/bottom at `intermediate' and `late' times. `Intermediate' times are averaged over $t v_{ti}/a \in [7.5,8.8]$ and `late' times are averaged over the saturated state for $t v_{ti}/a \in [14.8,15.9]$. \Cref{fig:4}(c) further demonstrates that the largest individual $\widetilde{q}_{e, k_y}$ contributions are at $\theta \simeq \pm \pi/2$ and $k_y \rho_i \simeq 1$. However, while the individual heat flux contribution $\widetilde{q}_{e, k_y}$ per $k_y$ from such modes is very high, in \Cref{sec:8} we will show that summed over a small $k_y$ interval $0.7 < k_y \rho_i \leq 4.3$, these modes not only transport little heat, but \textit{decrease} the total heat flux in the simulation substantially through multiscale interactions.

These slower-growing $k_y \rho_i \sim 1$ ETG modes also have maximum $\overline{\Phi}$ amplitudes at $\theta \simeq \pm \pi/2$; \Cref{fig:5}(a) shows how fluctuations at $\theta \simeq \pm \pi/2$ grow to be largest at $t v_{ti}/a \gtrsim 3$, which suppress $\widetilde{ q}_e$ at smaller $|\theta|$. This is evidenced in Figures 6(b) and (d), which show $\overline{ \Phi}$ and $\widetilde{q}_e$ becoming less peaked at $\theta = 0$ and, in the case of $\overline{ \Phi}$, reaching their highest values away from the outboard midplane due to low $k_y \rho_i$ ETG modes. In Figures 6(c) and (e), we project $\overline{ \Phi}$ and $\widetilde{q}_e $ onto the flux surface at later times, revealing the maximum turbulent amplitudes (but not heat flux) far away from $\theta = 0$. 

Despite $\overline{\Phi}$ having substantial magnitude near the inboard midplane ($\theta = \pm \pi$), evidenced in \Cref{fig:5}(b), the heat flux at the inboard midplane, shown in \Cref{fig:5}(d), is strikingly small. This is due to a combination of $\widetilde{ \phi}$, the turbulent electron temperature $\mathcal{T}_e$, and the cross-phase angle between $\widetilde{ \phi}$ and $\mathcal{T}_e$ all being small near the inboard midplane \cite{Parisi2020b}. Numerical constraints also limit the maximum value of $|\theta_0|$ at high $k_y \rho_i$, which may artificially suppress some turbulence near the inboard midplane. As discussed further in \Cref{sec:7}, the magnetic geometry allows slab ETG turbulence to be driven strongly near the inboard midplane, even though the heat flux there is small. The non-twisting-flux-tube approach \cite{Ball2021} might do better at resolving this high-$|\theta_0|$ turbulence near the inboard midplane and determine its importance definitively.

In \Cref{fig:6}, we compare snapshots of $\widetilde{ \phi}$ and its correlation functions at the early (left column) and late (right column) times. In panels (a) and (b), we plot $\widetilde{\phi}$ versus $\theta$ and $x$ at fixed $y$. At late times, radially narrow eddies that are extended in $\theta$ emerge. These are responsible for reducing overall heat transport in the outboard midplane (see \Cref{fig:5}(d)). In the $(y, \theta)$ cross-sections shown in panels 7(c) and (d), the fluctuations that emerge at later times are seen to have $k_y \rho_i \sim 1$. In panels 7(e)-(h), we plot the (time averaged) correlation functions
\begin{equation}
\langle C_{\phi} (x,y) \rangle_t = \left\langle \frac{ \sum\limits_{k_x, k_y} |\hat{\phi}_{k_x, k_y}|^2 e^{ i k_x x + i k_y y} }{ \sum\limits_{k_x, k_y} |\hat{\phi}_{k_x, k_y}|^2} \right\rangle_t,
\label{eq:corrfun}
\end{equation}
at $\theta = 0$ and $\theta = 1.57$ at early and later times. At $\theta = 1.57$, the correlation length in $y$ increases significantly with time, indicating the importance of the slower growing, low-$k_y \rho_i$ modes away from $\theta = 0$, as anticipated in \Cref{sec:3}.

To test for convergence, we performed scans in the number of parallel and perpendicular velocity and spatial grid points, and in hyperviscosity. We are unable to resolve turbulence at $k_y \rho_e \gtrsim 1$ fully, due to computational resource constraints, which currently prevent us from substantially increasing the maximum value of $k_y \rho_i$ in the simulation at fixed $\Delta k_y \rho_i < 1$. However, we believe that our results are close to reality thanks to our use of hyperviscosity.

We used dimensionless hyperviscous coefficents $D_{hy} = 10^{-6}$ and $D_{hx} = 3.5 \times 10^{-7}$, with the hyperviscous damping rate $\gamma_h$ given by \cite{Parisi2020b}
\begin{equation}
\gamma_h \frac{a}{v_{ti}} = - D_{hx} (k_x \rho_i)^{4} - D_{hy} \left(k_{y} \rho_i\right)^{4}.
\label{eq:hyperviscousDh}
\end{equation}
This was the weakest hyperviscosity possible that still admitted a well-converged simulation at high $k_y \rho_i$. To determine whether this value of hyperviscous damping is physically acceptable, we performed nonlinear simulations with a 60\% smaller perpendicular box size ($\Delta k_y \rho_i = 1.75$) at a fixed number of binormal wavenumbers (and so a much larger maximum $k_y \rho_i$ value), and used much smaller hyperviscous coefficients $D_{hx} = D_{hy} = 10^{-9}$. We found that the heat-flux peak for $\theta = 0$ in \Cref{fig:4}(a) remained at $k_y \rho_i \simeq 30$. We do not show these simulations with a larger maximum $k_y \rho_i$ and smaller hyperviscosity because they fail to capture low-$k_y \rho_i$ physics that regulates the heat flux at higher $k_y \rho_i$ values, discussed in \Cref{sec:8}. We refer the reader to Refs. \cite{Hatch2019,Guttenfelder2021,Chapman2022} for thorough investigations of $k_y \rho_e \gtrsim 1$ pedestal ETG turbulence.

To determine whether the low-$k_y \rho_i$ heat flux is wholly due to toroidal ETG turbulence, we performed simulations with no magnetic drifts, $\mathbf{v}_{Ms} = 0$. We found large heat flux and turbulent amplitudes driven by slab ETG modes at $k_y \rho_i \sim 1$ away from the outboard midplane and at $k_y \rho_e \sim 1$ at the outboard midplane. This shows that at $k_y \rho_i \sim 1$, both toroidal and slab ETG modes are driven strongly away from the outboard midplane. Even so, the toroidal ETG modes are important because, when $\mathbf{v}_{Me} \neq 0$, we observe the poloidally extended radial structures, shown in \Cref{fig:6}(b), which are absent when $\mathbf{v}_{Me} = 0$.

\begin{figure}
        \centering
        \includegraphics[width=0.4\textwidth]{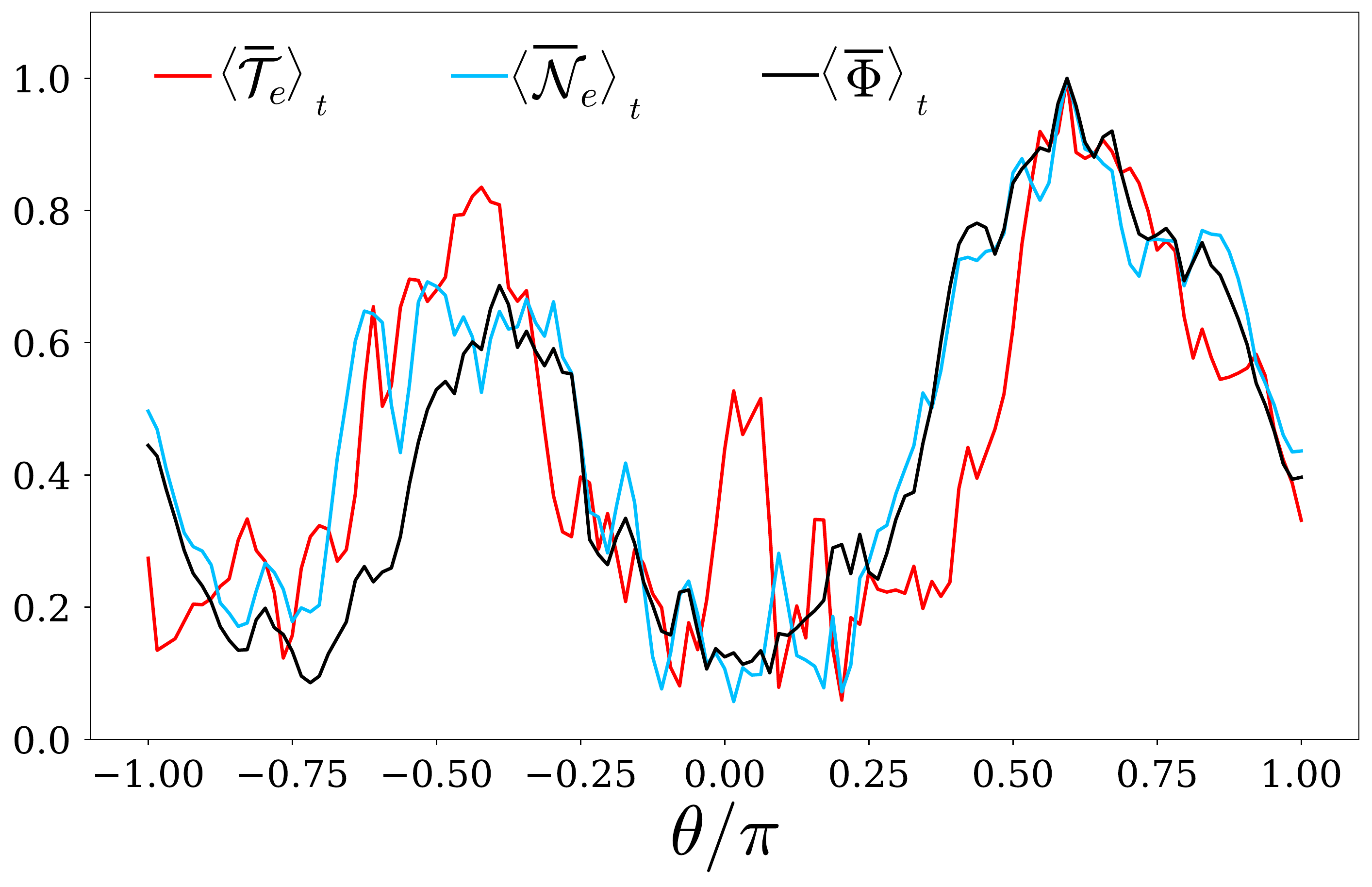}
        \caption{Root-mean-square perturbations of temperature $\langle \overline{\mathcal{T}}_e \rangle_t$, density $\langle \overline{\mathcal{N}}_e \rangle_t$, and potential $\langle \overline{\Phi} \rangle_t$, averaged over $tv_{ti}/a \in [16.7-17.8]$ for Base150 and normalized so that the maximum value for each is one.}
        \label{fig:13}
\end{figure}

We now show density and temperature fluctuations for ETG turbulence. We define the perturbed density $n_s ^{tb}$ and temperature $T_s ^{tb}$ as
\begin{equation}
\begin{aligned}
& n ^{tb}_s = \int d^3 v \langle f_s ^{tb} \rangle_{\mathbf{r}} = \int d^3 v \left( \langle h_s \rangle_{\mathbf{r}} -  \frac{Z_s e}{T_s} \phi ^{tb} F_{Ms} \right) ,\\
& T_s ^{tb} = \frac{1}{n_s} \int d^3 v \left( \frac{m_s{v^2}}{3} - T_s \right) \langle f_s ^{tb} \rangle_{\mathbf{r}},
\end{aligned}
\label{eq:17}
\end{equation}
where $\langle \ldots \rangle_{\mathbf{r}}$ is a gyrophase average performed at fixed $\mathbf{r}$. Normalizations for the density and temperature are
\begin{equation}
\begin{aligned}
& \widetilde{n}_s = \frac{n ^{tb} _s}{n_i \rho_{*i}}, \;\;\;\;\overline{ \mathcal{N}}_s^2(\theta, t) = \sum_{k_x,k_y} \left| \hat{n}_{s,k_x,k_y}(\theta,t) \right|^2, \\
& \widetilde{T}_s = \frac{T ^{tb} _s}{T_i \rho_{*i}}, \;\;\;\;\overline{ \mathcal{T}}_s^2(\theta, t) = \sum_{k_x,k_y} \left| \hat{T}_{s,k_x,k_y}(\theta,t) \right|^2,
\end{aligned}
\label{eq:18}
\end{equation}
where $\hat{T}_{k_x,k_y,s}$ and $\hat{n}_{k_x,k_y,s}$ are the Fourier coefficients of  $\widetilde{T}_s$ and $\widetilde{n}_s$, respectively. In \Cref{fig:13}, we plot potential, density, and temperature fluctuations averaged over $tv_{ti}/a \in [16.7-17.8]$ for Base150, showing that while $\overline{\Phi}(\theta)$, $\overline{\mathcal{N}}_e(\theta)$, and $\overline{\mathcal{T}}_e (\theta)$ are largest away from $\theta = 0$, $\overline{\mathcal{T}}_e (\theta)$ also has a high amplitude at $\theta \approx 0$, which gives rise to high electron heat transport around $\theta \approx 0$.

\begin{figure}
        \centering
       \begin{subfigure}[t]{0.241\textwidth}
        \includegraphics[width=\textwidth]{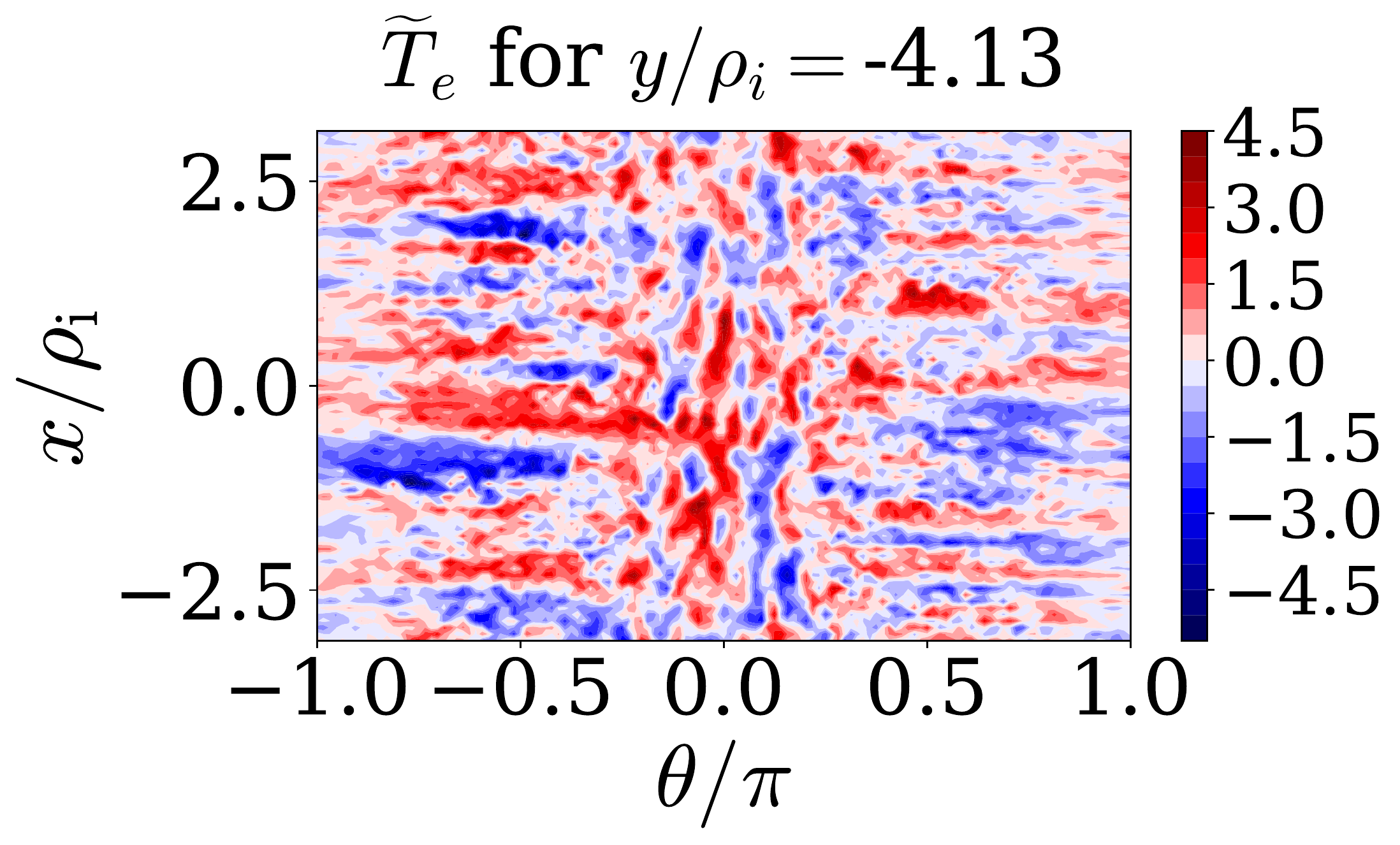}
        \caption{Temperature $\widetilde{T}_e$.}
    \end{subfigure} 
    \begin{subfigure}[t]{0.23\textwidth}
        \includegraphics[width=\textwidth]{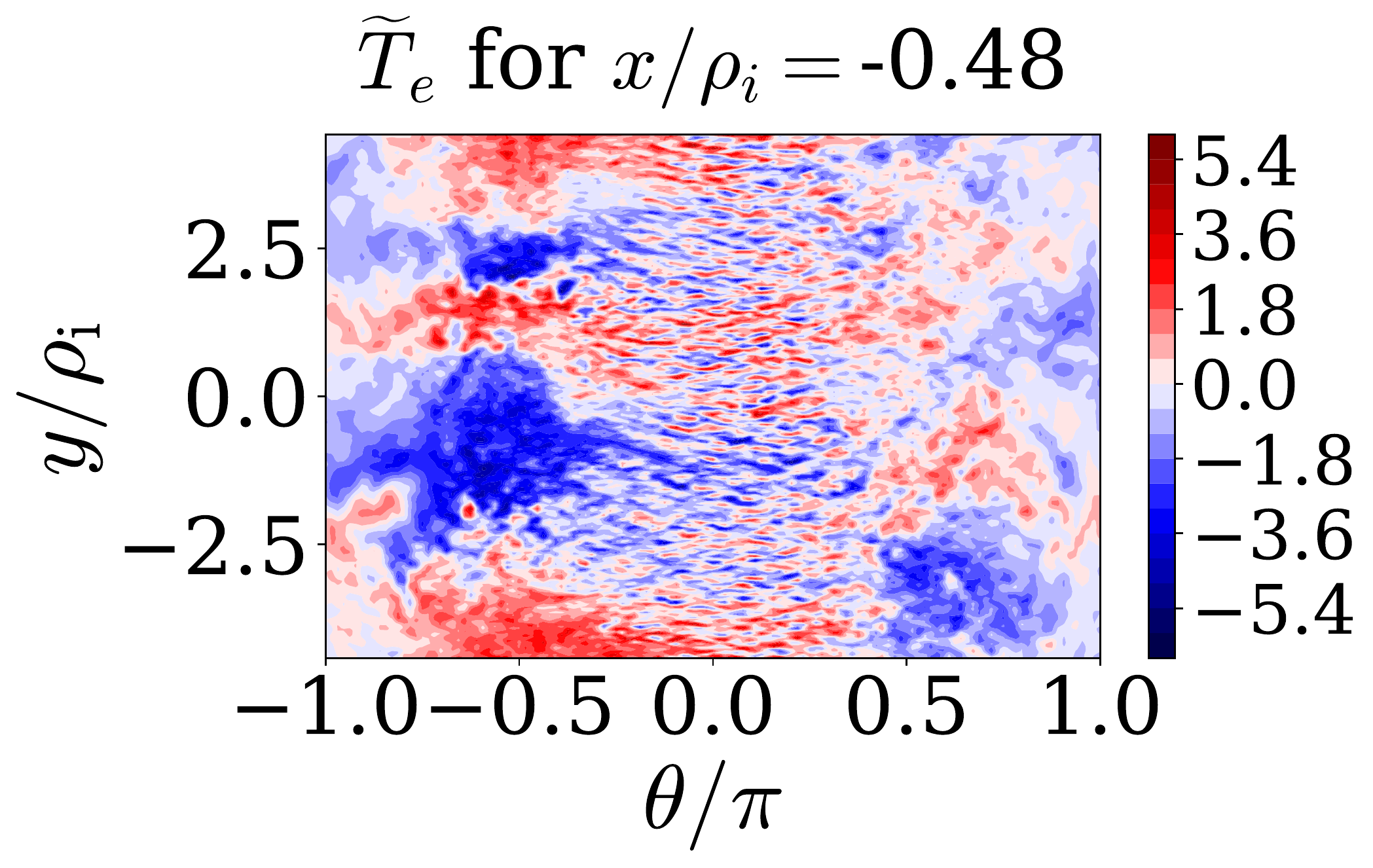}
        \caption{Temperature $\widetilde{T}_e$.}
    \end{subfigure}
    \begin{subfigure}[c]{0.23\textwidth}
        \includegraphics[width=1\textwidth]{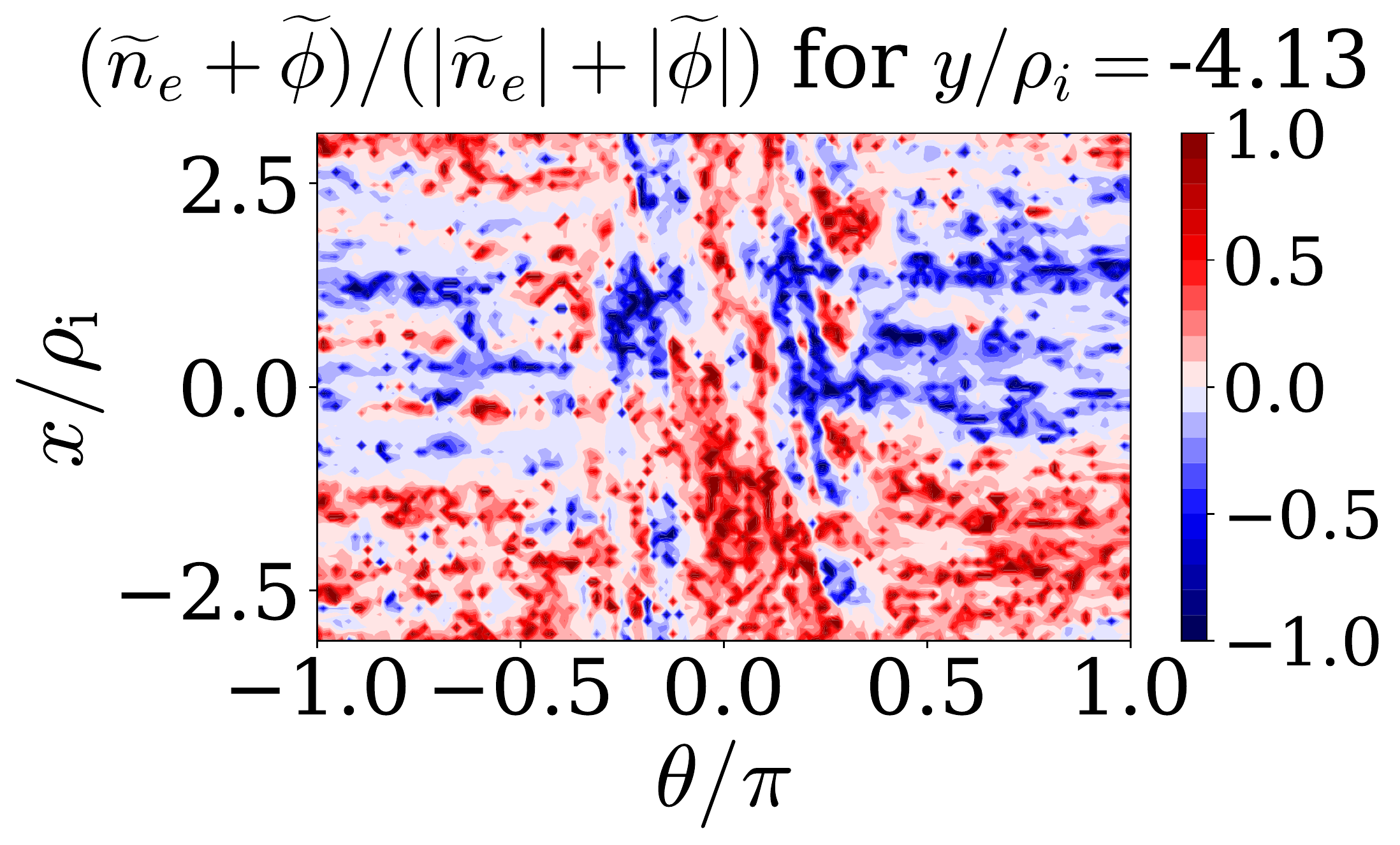}
        \caption{$(\widetilde{n}_e +\widetilde{\phi})/(|\widetilde{n}_e| + |\widetilde{\phi}|)$.}
    \end{subfigure} \hspace{0.7mm}
    \hspace{0.1mm} \begin{subfigure}[c]{0.23\textwidth}
        \includegraphics[width=1\textwidth]{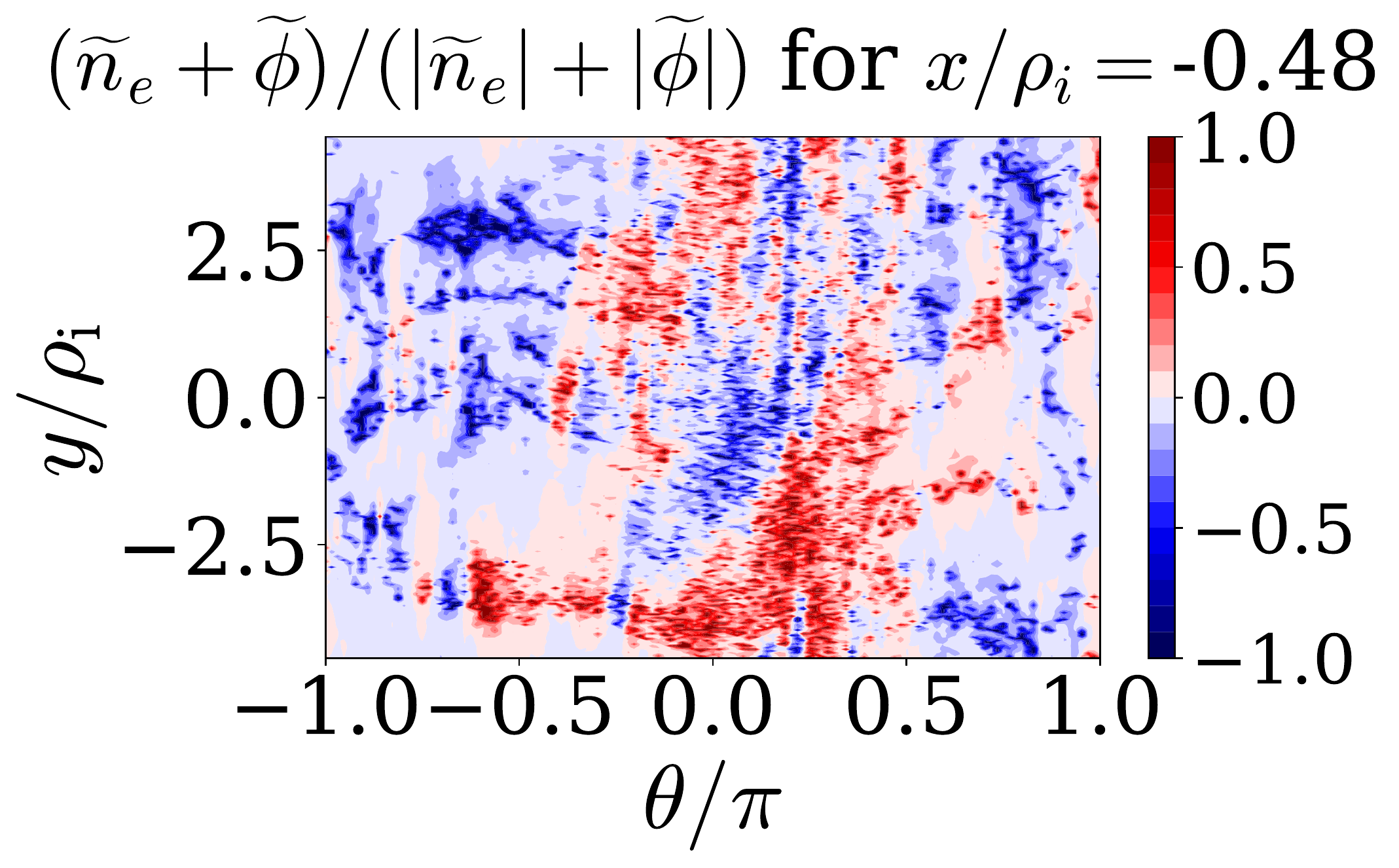}
        \caption{$(\widetilde{n}_e +\widetilde{\phi})/(|\widetilde{n}_e| + |\widetilde{\phi}|)$.}
    \end{subfigure}
    \caption{(a) and (b) temperature fluctuations, (c) and (d) potential plus density fluctuations, showing a near cancellation at many locations, particularly where $|\widetilde{ n}_e|$ is largest. Definitions for $\widetilde{T}_e$ and $\widetilde{n}_e$ are given in \Cref{eq:17,eq:18}.}
    \label{fig:14}
\end{figure}

\Cref{fig:14}(a) shows that $\widetilde{T}_e$ features thin radial layers similar to $\widetilde{\phi}$ [see Figures 7(b) and (d)], but unlike $\widetilde{\phi}$, has prominent radially elongated structures around $\theta \approx 0$ that cause heat transport. \Cref{fig:14}(b) shows that like $\widetilde{\phi}$, the temperature $\widetilde{T}_e$ also has low-$k_y \rho_i$ structure away from $\theta = 0$ and fine-scale $k_y \rho_i$ structure around $\theta \approx 0$. The perturbed density $\widetilde{n}_e$ often has similar amplitudes to $\widetilde{\phi}$ [see Figures 7(b) and (d)], but with the opposite sign, demonstrated by plots of $(\widetilde{n}_e + \widetilde{\phi} ) / (|\widetilde{n}_e| + |\widetilde{\phi}|)$ in panels 9(c) and (d). We find that $(\widetilde{n}_e + \widetilde{\phi} ) / (|\widetilde{n}_e| + |\widetilde{\phi}|)$ is smallest in regions of highest amplitudes of $\widetilde{n}_e$ and $\widetilde{\phi}$, where toroidal ETG turbulence satisfying $k_{\perp} \rho_i \gg 1$ resides. The property $\widetilde{ n}_e \approx - \widetilde{\phi}$ is expected for ETG turbulence with adiabatic ions ($k_{\perp} \rho_i \gg 1$), using $|\langle h_i \rangle_{\mathbf{r}}| \ll |(e \phi ^{tb}/T_i) F_{Mi}|$ [see \Cref{eq:17}] and quasineutrality $\widetilde{n}_e = \widetilde{ n}_i$. Given that $\widetilde{n}_e \simeq - \widetilde{\phi}$, we find density fluctuations to cause little electron heat transport \cite{Evensen1998,White2008,Parisi2020b}, while temperature fluctuations cause $\sim$99\% of heat transport.

\section{Temperature-gradient scan} \label{sec:6}

\begin{figure}
        \centering
        \includegraphics[width=0.49\textwidth]{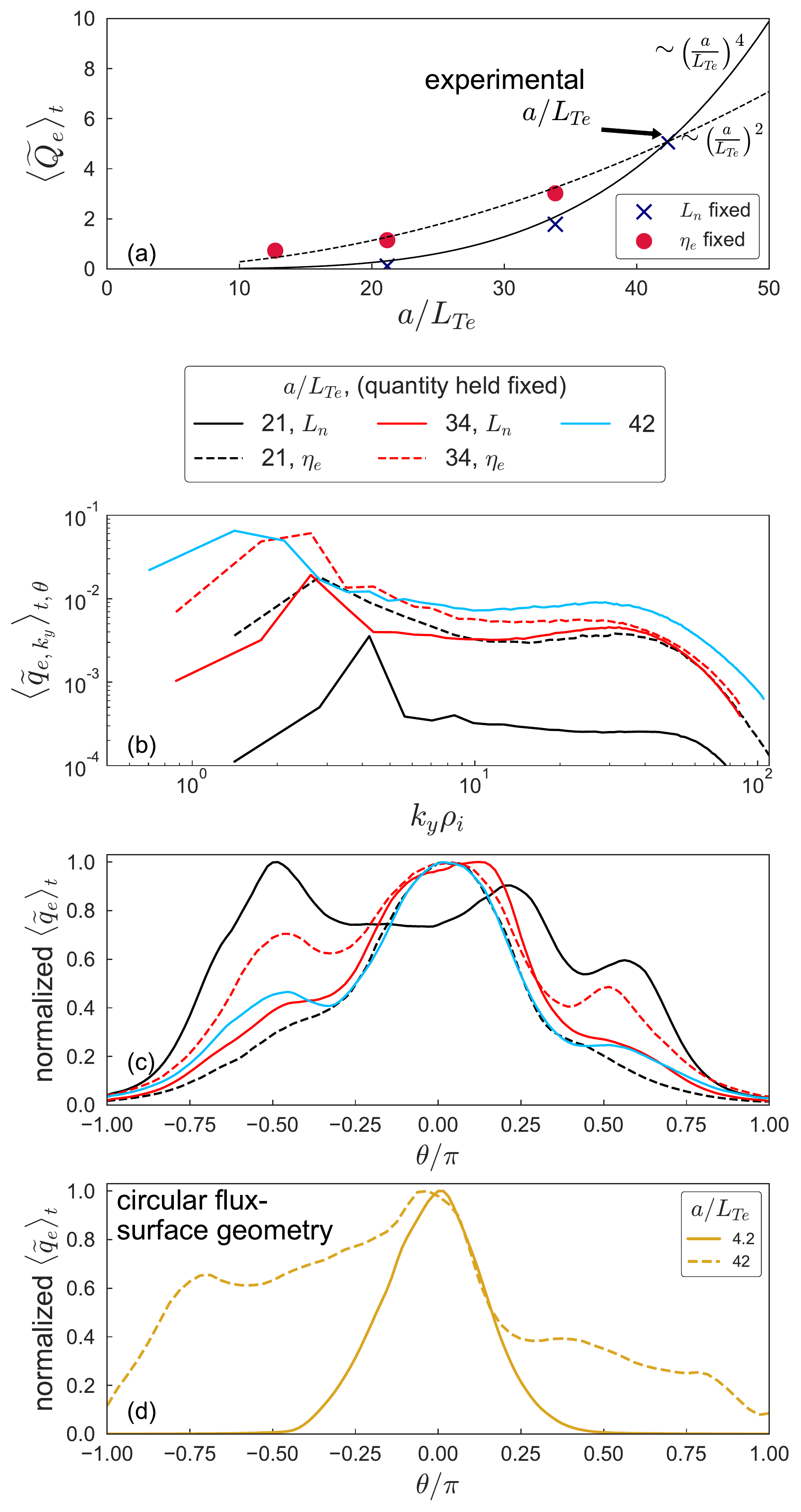}
        \caption{Results from the temperature gradient scan in \Cref{sec:6}. (a) Heat flux versus $a/L_{Te}$ with either $L_n$ or $\eta_e$ fixed for pedestal geometry; (b) $\widetilde{q} _{e,k_y}$ versus $k_y \rho_i$ for selected $a/L_{Te}$ values for pedestal geometry; (c) normalized $\widetilde{q} _e$ versus $\theta/\pi$ for selected $a/L_{Te}$ values for pedestal geometry; (d) normalized heat flux $\widetilde{ q} _e$ versus $\theta/\pi$ for circular flux-surface geometry with two $a/L_{Te}$ values. The fluxes are time-averaged over the saturated state. See \Cref{eq:thirteen} for definitions. The up-down asymmetry in (c) and (d) is explained in \Cref{sec:7}.}
        \label{fig:7}
\end{figure}

We now study the effect of the electron temperature gradient and flux-surface shape on pedestal ETG turbulence. The heat flux's scaling with the temperature gradient is found by performing a scan in $a/L_{Te}$ with values less than the experimental $a/L_{Te} = 42$. To save computational resources, we performed this scan using simulations with 100 binormal modes, rather than 150 modes. For these simulations (denoted by Scan100a, Scan100b, etc, in \Cref{tab:1}), we used a hyperviscosity determined by the maximum $k_y \rho_i$ value in the simulation. We changed $\Delta k_x$ and $\Delta k_y$, scaling them with $L_{Te}$ \cite{Barnes2011}. We were unable to resolve satisfactorily simulations with values of $a/L_{Te}$ above the experimental value. The scan in $a/L_{Te}$ was performed in two ways: (1) with $\eta_e = L_n / L_{Te}$ fixed, and (2) with $L_n$ fixed. As we changed $L_{Te}$ and $L_n$, the pressure gradient $\partial p / \partial r$ was changed consistently. A relatively large value of $\partial p / \partial r$, which appears in \Cref{eq:four} and implicitly in other geometrical coefficients, is known to stabilize turbulence \cite{Bourdelle2003}.  We kept all parameters not shown in \Cref{tab:1} constant.

The heat flux's scalings with $a/L_{Te}$ are shown in \Cref{fig:7}(a). With $L_n$ fixed, $\widetilde{Q} _e \propto (a/L_{Te})^4$, which is a steeper scaling than in strongly driven core toroidal ITG turbulence: $\widetilde{Q}  _i \propto (a/L_{Ti})^3$ \cite{Barnes2011}. With $\eta_e$ fixed, the scaling much is shallower, $\widetilde{Q}_e \propto (a/L_{Te})^2$. Both scalings are consistent with previous findings in \cite{Guttenfelder2021,Chapman2022}. In \Cref{fig:7}(b), we plot $\widetilde{ q}_{e,k_y}$ versus $k_y$ for several $a/L_{Te}$ values keeping either $L_n$ fixed or $\eta_e$ fixed. The peak in $\widetilde{ q}_{e,k_y}$ in \Cref{fig:7}(b) exhibits the trend $k_y^o \propto L_{Te}$ \cite{Barnes2011} that is consistent with \Cref{eq:twelve} if $l_{\parallel}$ is fixed; we verified that $l_{\parallel}$ is indeed fixed at $l_{\parallel} \approx qR / 2$ at the ETG turbulence outer scale for all scans shown in \Cref{fig:7}(b).

In \Cref{fig:7}(c), we plot $\widetilde{q}  _e$ versus $\theta$. With both $\eta_e$ fixed and $L_n$ fixed, the dependence of $\widetilde{q}  _e$ on $L_{Te}$ is complicated. With $L_n$ fixed, the shape of $\widetilde{q}  _e$ is similar for $a/L_{Te} = 42$ and $a/L_{Te}= 34$, but has the maximum value of $\widetilde{q}  _e$ away from the outboard midplane for $a/L_{Te} = 21$. With $\eta_e$ fixed, the intermediate temperature gradient $a/L_{Te} = 34$ has the highest relative off-midplane transport. The reasons behind this dependence of the heat flux's poloidal profile on the temperature gradient are beyond the scope of this work, but the different sensitivity of the growth rates and stability boundaries of toroidal and slab ETG modes to $\eta_e$ may be playing a role \cite{Jenko2001,Parisi2020}.

We also performed two simulations with the circular flux-surface geometry (by setting Miller shaping parameters to `circular' values), one with the experimental gradient for the pedestal ($a/L_{Te} = 42$) and a second with core-like gradients ($a/L_{Te} = 4.2$, $\eta_e$ fixed, and $a/L_{Ti}$ decreased by a factor of ten to $a/L_{Ti} = 1.1$). These two simulations are denoted by Scan200f and Scan100e in \Cref{tab:1}, respectively. The simulation with $a/L_{Te} = 4.2$ had $\Delta k_y \rho_i = 1.76$. We chose a relatively small $\Delta k_y$ for this simulation because we wished to determine whether significant $k_y \rho_i \sim 1$ ETG turbulence would appear, which it did not. The circular-geometry simulation with $a/L_{Te} = 42$ had $\Delta k_y \rho_i = 0.35$ --- we required this very small $\Delta k_y \rho_i$ to resolve significant $k_y \rho_i \sim 1$ ETG turbulence. To make the $a/L_{Te} = 42$ simulation affordable, we used $64$ parallel gridpoints.

For the $a/L_{Te} = 4.2$ simulation, the heat flux's profile versus $\theta$, shown in \Cref{fig:7}(d), resembles that observed in core ETG/ITG turbulence simulations, peaked at the outboard midplane and decaying smoothly in the parallel direction \cite{Waltz1994,Beer1995}. In contrast, for the $a/L_{Te} = 42$ simulation, the heat flux had substantial off-midplane contributions, shown in \Cref{fig:7}(d), due to ETG modes away from the outboard midplane. This demonstrates that even in circular flux-surface geometry, steep gradients can produce turbulence with a novel parallel structure.  

\section{Topography of turbulence} \label{sec:7}

\begin{figure}
        \centering
         \begin{subfigure}[t]{0.49\textwidth}
    \includegraphics[width=\textwidth]{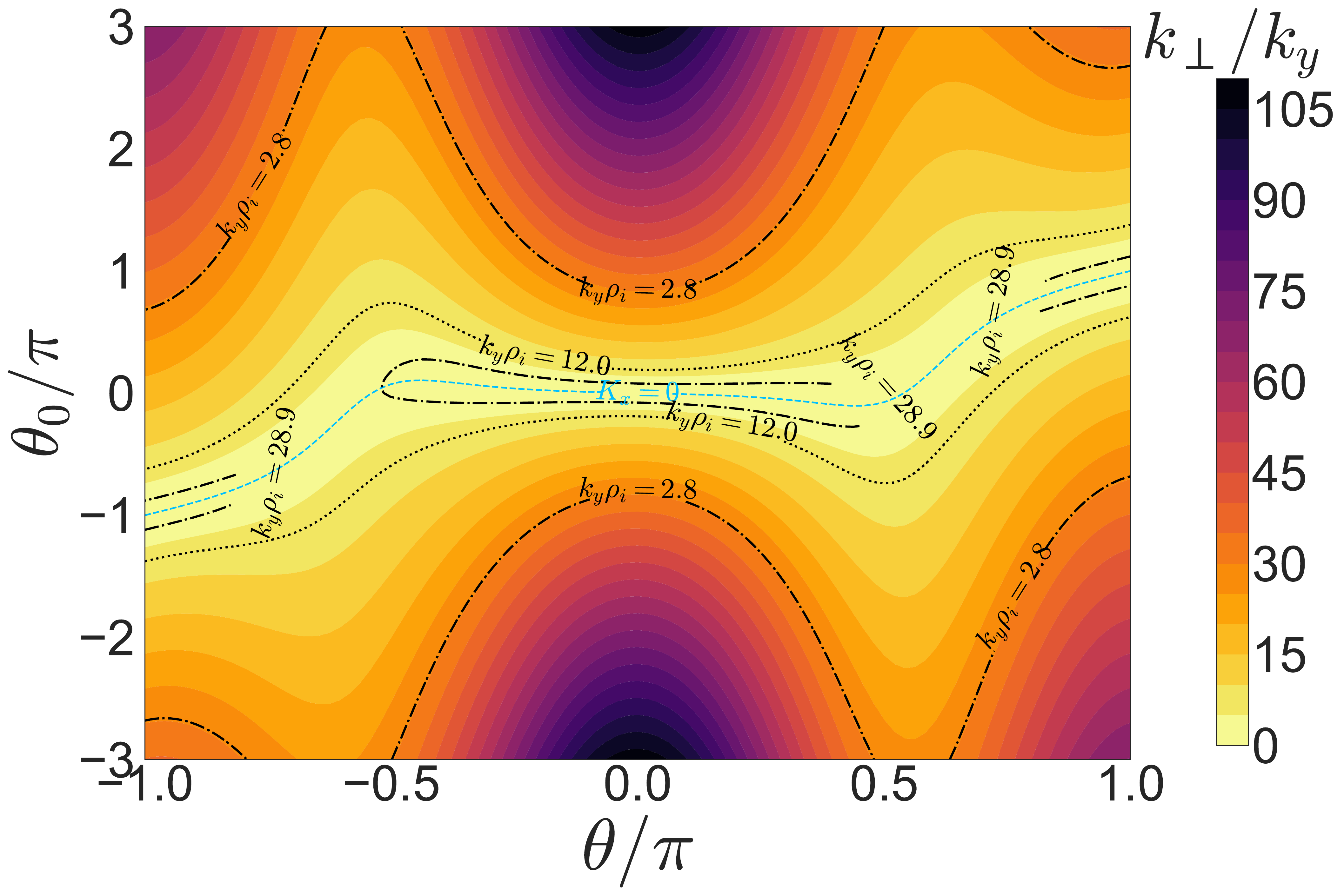}
        \caption{Pedestal geometry: $k_{\perp}/ k_y$ versus $\theta$ and $\theta_0$. On the $K_x = \mathbf{k}_{\perp} \cdot \nabla x / |\nabla x| = 0$ curve, $k_{\perp} = k_y$. Black dashed and dotted curves indicate $k_{\perp} \rho_e =1$ for the $k_y \rho_i$ values in (b), (c), (d), as labelled on the contours.}
    \end{subfigure}
               \begin{subfigure}[t]{0.49\textwidth}
        \includegraphics[width=\textwidth]{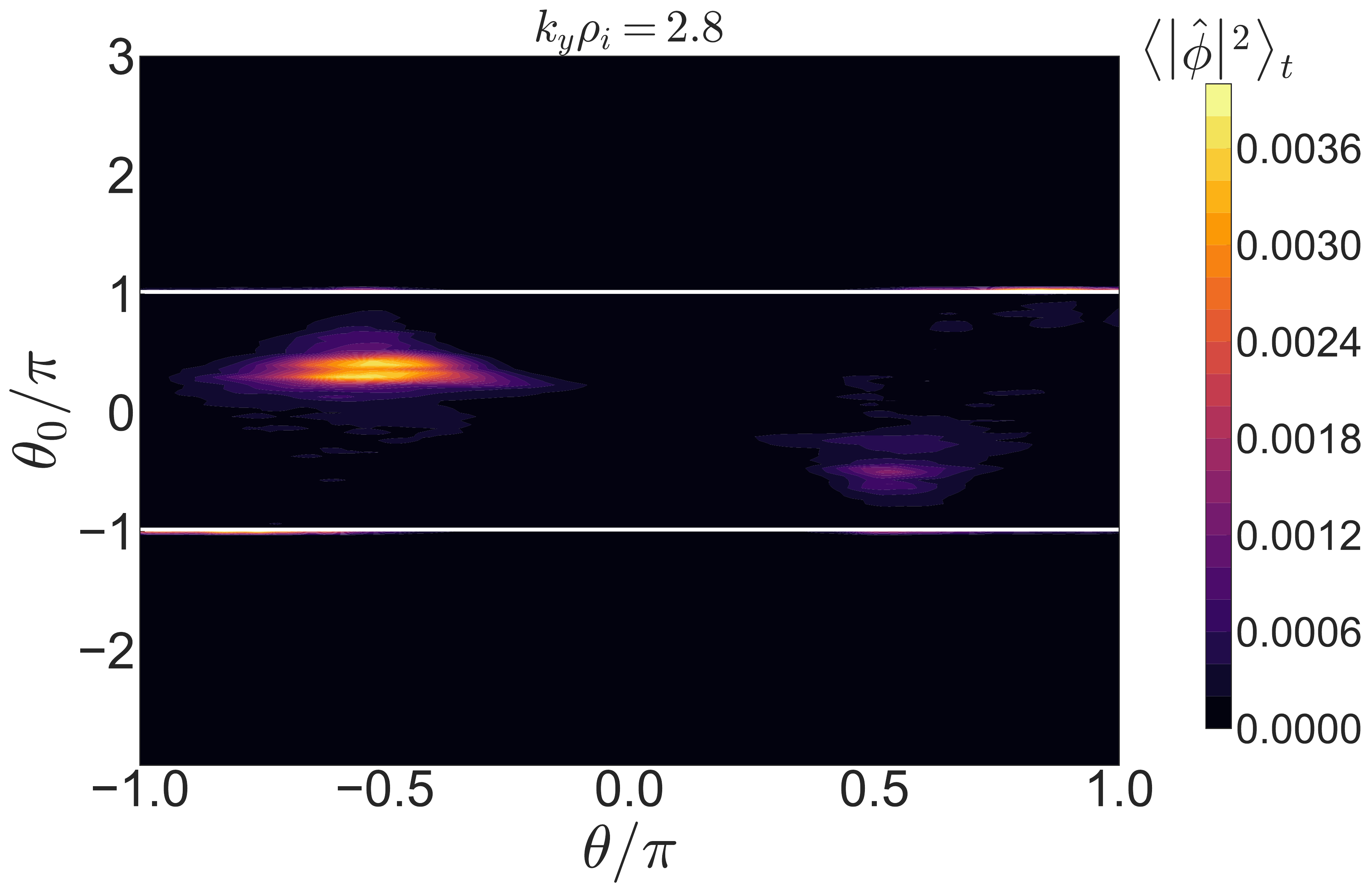}
        \caption{$\langle | \hat{\phi} |^2 \rangle_t$ versus $\theta$ and $\theta_0$ for $k_y \rho_i = 2.8$, for simulation Base150 [see \Cref{tab:1}]}
    \end{subfigure} \hspace{-0.7mm}
        \begin{subfigure}[c]{0.48\textwidth}
        \includegraphics[width=\textwidth]{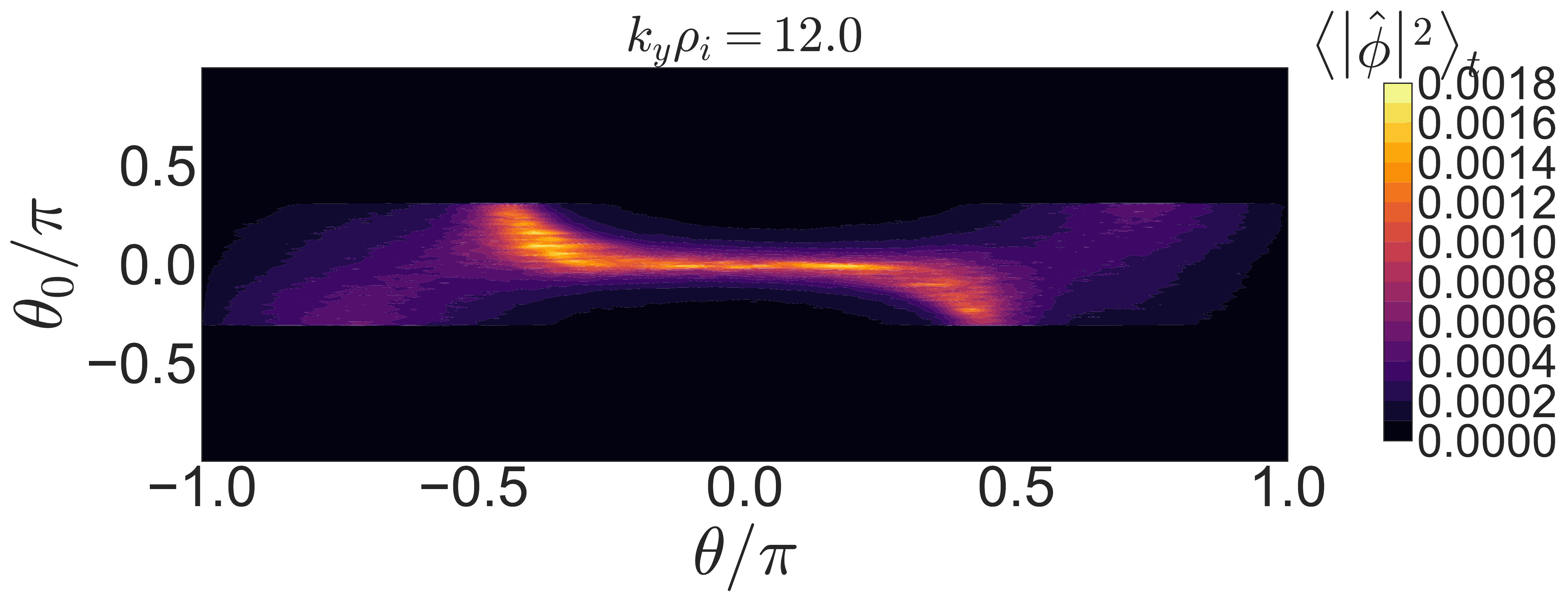}
        \caption{$\langle | \hat{\phi} |^2 \rangle_t$ versus $\theta$ and $\theta_0$ for $k_y \rho_i = 12.0$, for simulation Base150.}
    \end{subfigure} \hspace{-0.7mm}
    \hspace{0.1mm} \begin{subfigure}[c]{0.48\textwidth}
        \includegraphics[width=\textwidth]{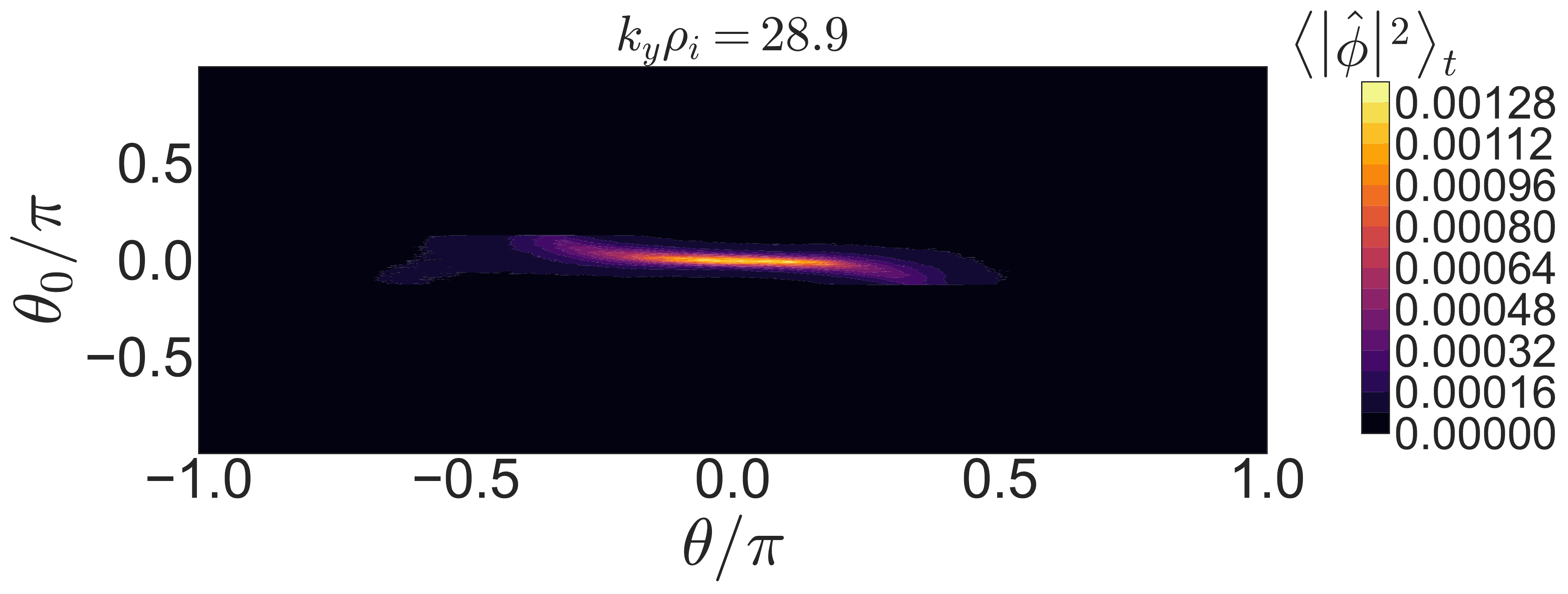}
        \caption{$\langle | \hat{\phi} |^2 \rangle_t$ versus $\theta$ and $\theta_0$ for $k_y \rho_i = 28.9$, for simulation Base150.}
    \end{subfigure}
     \caption{Effect of FLR ($k_{\perp} \rho_e$) topography (a) on the spatial distribution of the turbulent amplitudes $|\hat{ \phi}|$ in (b), (c), and (d). Contours of $k_{\perp} \rho_e = 1$ in (a) are strongly correlated with (c) and (d), indicating the importance of FLR effects for the distribution of slab ETG turbulence at higher $k_y \rho_i $ values. The amplitudes are averaged over $t v_{ti}/a \in [14.8 -16.8]$. The ETG turbulence is driven mainly by toroidal instability in (b), but is driven mainly by slab instability in (c) and (d).}
      \label{fig:8}
\end{figure}

In this section, we show how FLR effects and magnetic-drift profiles determine the parallel distribution of turbulence. These influences act at different scales: while the magnetic-drift profiles $\omega_{*e}^T/ \omega_{\kappa e}$ are independent of $k_y$ (at fixed $\theta_0$), the strength of electron FLR damping, measured by the reduction in the linear instability's growth rate and the resulting turbulence amplitude, is almost always greater at higher $k_{\perp} \rho_e$ values \cite{Parisi2020}. The $\omega_{*e}^T/ \omega_{\kappa e}$ topography is mostly relevant for toroidal ETG modes, whereas the $k_{\perp} \rho_e$ topography is important for both toroidal and slab ETG modes.

In order for ETG turbulence and transport to be strong, the FLR damping, occurring when $k_{\perp} \rho_e \gtrsim 1$, cannot be too large. Therefore, we expect $|\hat{\phi}|$ to be higher in regions of the flux surface where $k_{\perp} \rho_e$ is lower. In \Cref{fig:8}(a), we plot the ratio $k_{\perp} / k_y$ for our flux surface as a function of $(\theta,\theta_0)$. It is important to note that $k_{\perp} / k_y$ is independent of $k_y$ at fixed $\theta_0$. Due to strong magnetic shaping in the pedestal, the quantity $k_{\perp}$ varies more strongly in $\theta$ and $\theta_0$ than for flux-surface shapes characteristic of the core. Therefore, we expect turbulence and transport in the pedestal to have a stronger dependence on $\theta$ than in the core. To map out the regions of weaker FLR damping at different $k_y \rho_i$ values, in \Cref{fig:8}(a), we plot the curves of $k_{\perp} \rho_e = 1$ for different $k_y \rho_i$ values. In the areas bounded by these curves, $k_{\perp} \rho_e \leq 1$, so, heuristically, we expect weaker FLR damping there, and hence stronger turbulence.

For slab ETG turbulence, amplitudes are inversely correlated with $k_{\perp} \rho_e$. This can be seen by comparing values of $k_{\perp}/k_y$ at given $(\theta, \theta_0)$ in \Cref{fig:8}(a) with $|\hat{\phi}|^2$ in (c) and (d) at the same $(\theta, \theta_0)$. Figures 11(c) and (d) show $|\hat{\phi}|^2$ for $k_y \rho_i = 12.0$ and $k_y \rho_i = 28.9$, respectively. We see that $|\hat{\phi}|^2$ becomes narrower in $\theta$ and $\theta_0$ at higher $k_y \rho_i$. This is because, in \Cref{fig:8}(a), there are fewer regions of $k_{\perp}/k_y$ satisfying the weak FLR-damping constraint, $k_{\perp} \rho_e \lesssim 1$, at higher values of $k_y \rho_i$. At lower values of $k_y \rho_i$, $k_{\perp} \rho_e \lesssim 1$ is satisfied at more values of $\theta_0$ and $\theta$. Therefore, at lower $k_y \rho_i$, we expect slab ETG turbulence to be present over more of the $(\theta, \theta_0)$ plane.

However, slab ETG turbulence does not dominate for all values of $k_y \rho_i$. In \Cref{fig:8}(b), we plot $|\hat{\phi}|^2$ for a relatively small $k_y \rho_i = 2.8$: clearly, $|\hat{\phi}|^2$ does not occupy the lowest $k_{\perp} \rho_e$ values from \Cref{fig:8}(a). This is because the turbulence at $k_y \rho_i = 2.8$ is primarily toroidal ETG turbulence. While the parallel extent of slab ETG turbulence is constrained primarily by $k_{\perp} \rho_e$ increasing along the field line, the parallel extent and location of toroidal ETG turbulence is subject to two constraints. Namely, for strong toroidal ETG turbulence to exist at a given parallel location, not only must FLR damping be relatively weak $(k_{\perp} \rho_e \lesssim 1)$, but also the value of $\omega_{*e}^T / \omega_{\kappa e}$ must allow strong toroidal ETG instability, requiring $\omega_{*e}^T / \omega_{\kappa e} \approx A \simeq 3 - 20$ [see discussion around \Cref{eq:seven}]. 

\begin{figure}
        \centering
        \begin{subfigure}[t]{0.49\textwidth}
        \includegraphics[width=\textwidth]{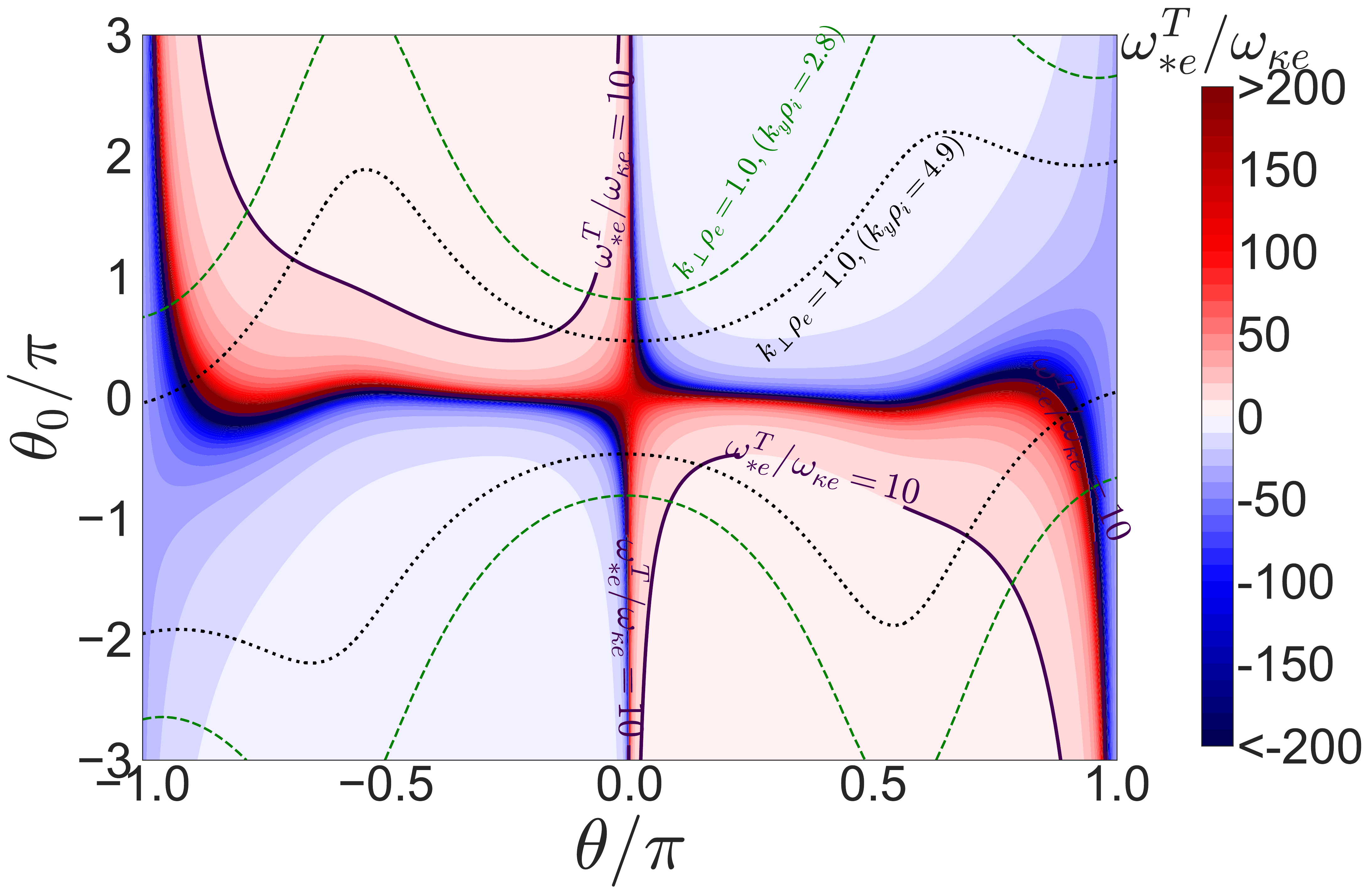}
        \caption{Pedestal geometry: $\omega_{*e}^T / \omega_{\kappa e} $ versus $\theta$ and $\theta_0$. Dashed and dotted lines show $k_{\perp} \rho_e = 1$ for different $k_y \rho_i$ values.}
    \end{subfigure}
           \begin{subfigure}[t]{0.49\textwidth}
        \includegraphics[width=\textwidth]{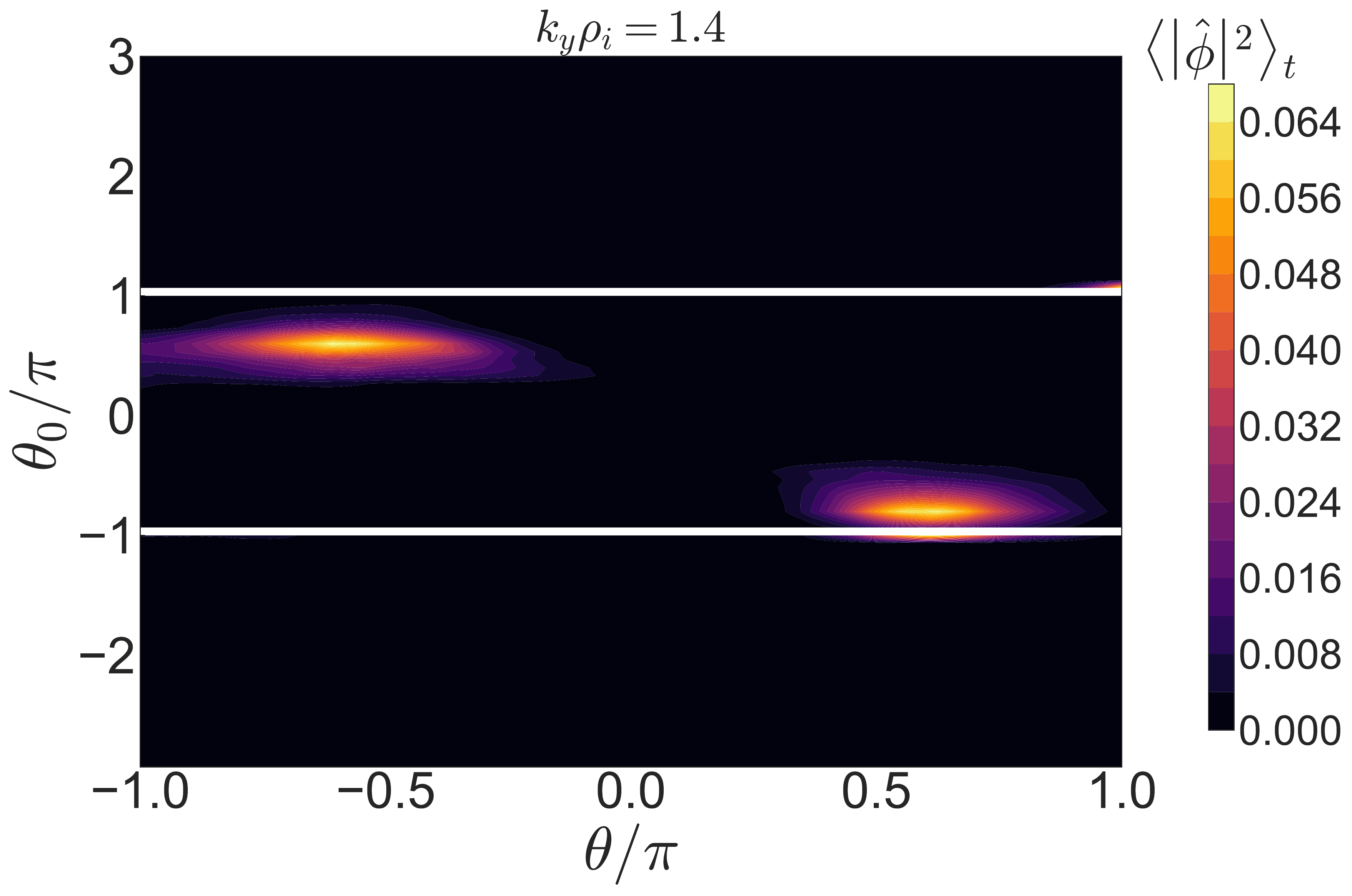}
        \caption{$\langle \left| \hat{\phi}  \right|^2 \rangle_t$ versus $\theta$ and $\theta_0$ for $k_y \rho_i = 1.4$ averaged for $t v_{ti}/a \in [14.8 -16.8]$, for simulation Base150 [see \Cref{tab:1}]. Mainly toroidal ETG turbulence.}
    \end{subfigure}
     \caption{The turbulent amplitudes (b) for $k_y \rho_i = 1.4$ are localized in regions where $\omega_{*e}^T / \omega_{\kappa e} \sim 10$ (a), demonstrating the importance of magnetic drifts for the spatial distribution of toroidal ETG turbulence.}
      \label{fig:9}
\end{figure}

In \Cref{fig:9}(a), we plot $\omega_{*e}^T / \omega_{\kappa e}$ for our flux surface, which shows a topography very different to the FLR constraints in \Cref{fig:8}(a). While the FLR damping in \Cref{fig:8}(a) tends to be weakest around $\theta_0 \simeq 0$ and $\theta \simeq 0$, the magnetic drifts are most favorable to the excitation of turbulence at $\theta_0 / \pi \approx \pm 1$ and $\theta \neq 0$. Since the ratio $\omega_{*e}^T / \omega_{\kappa e}$ is independent of $k_y$, but $k_{\perp} \rho_e$ is not, at lower $k_y \rho_i$ values where FLR damping is weaker, we expect toroidal ETG modes to be freer to occupy $\theta$ locations where $\omega_{*e}^T / \omega_{\kappa e}$ is optimal. For example, in \Cref{fig:9}(a), the dashed green line shows the rough FLR damping boundary, $k_{\perp} \rho_e = 1$, for $k_y \rho_i  = 2.8$. Within this region, $\omega_{*e}^T / \omega_{\kappa e}$ has optimal values for strong toroidal ETG instability, and hence we expect strong toroidal ETG turbulence at $k_y \rho_i = 2.8$. Indeed, the turbulent amplitude $|\hat{\phi}|^2$ for $k_y \rho_i = 2.8$ and $1.4$ in \Cref{fig:8}(b) and \Cref{fig:9}(b), respectively, has maxima in the optimal regions of $\omega_{*e}^T / \omega_{\kappa e}$ of \Cref{fig:9}(a), demonstrating that at these binormal scales, the turbulence has a strong toroidal ETG character. 

\begin{figure}
        \centering
        \begin{subfigure}[t]{0.49\textwidth}
        \includegraphics[width=\textwidth]{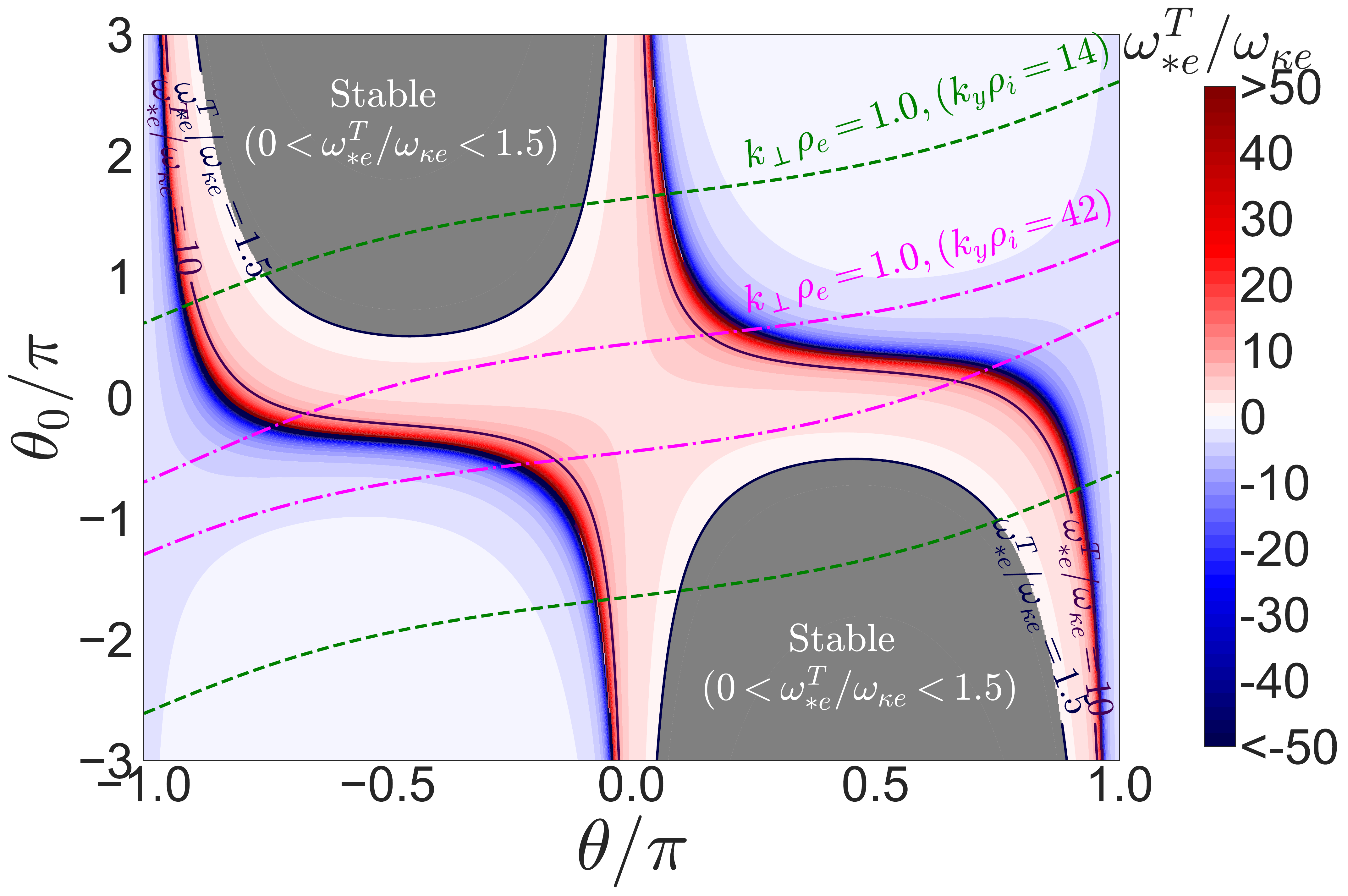}
        \caption{Circular flux-surface geometry with $a/L_{Te} = 4.2$: $\omega_{*e}^T / \omega_{\kappa e} $ versus $\theta$ and $\theta_0$. Dashed and dash-dotted lines show $k_{\perp} \rho_e = 1$ for different $k_y \rho_i$ values.}
    \end{subfigure}
        \begin{subfigure}[t]{0.49\textwidth}
        \includegraphics[width=\textwidth]{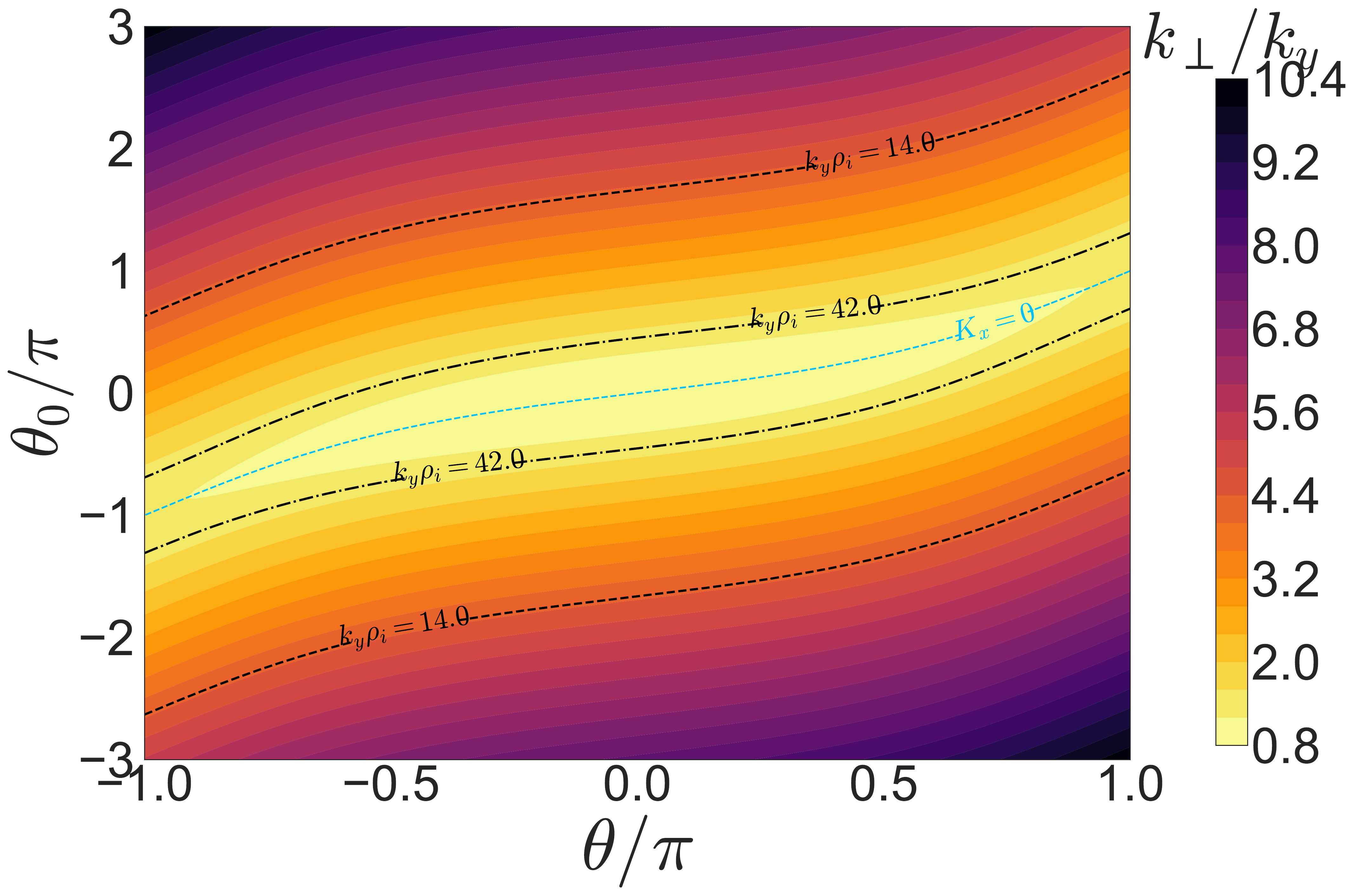}
        \caption{Circular flux-surface geometry: $k_{\perp}/ k_y$ versus $\theta$ and $\theta_0$. On the $K_x = \mathbf{k}_{\perp} \cdot \nabla x / |\nabla x| = 0$ curve, $k_{\perp} = k_y$. Black dashed and dash-dotted curves indicate $k_{\perp} \rho_e =1$ for the $k_y \rho_i$ values of $k_y \rho_i = 14, 42$.}
    \end{subfigure}
        \begin{subfigure}[t]{0.49\textwidth}
        \includegraphics[width=\textwidth]{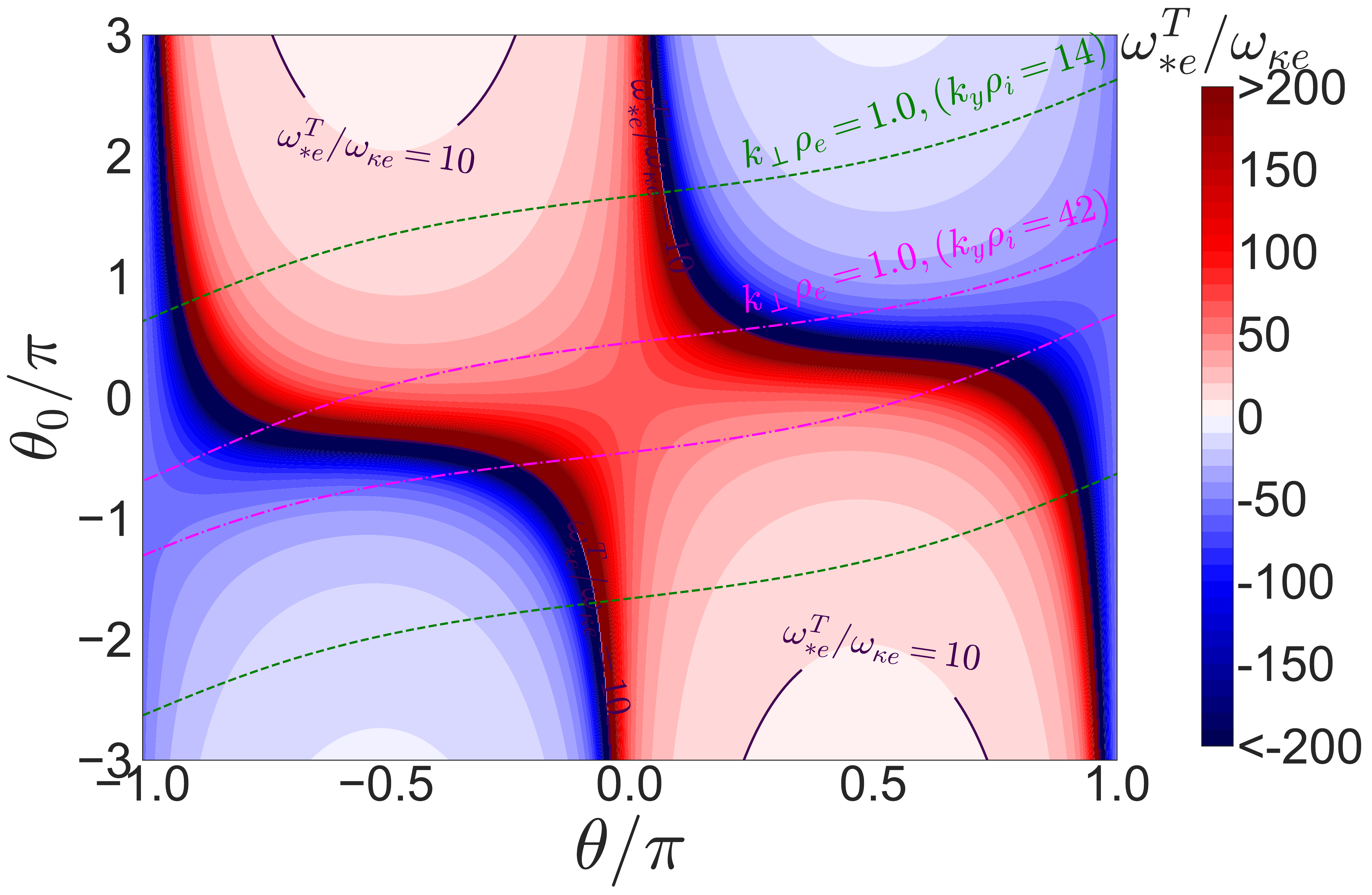}
        \caption{Circular flux-surface geometry with $a/L_{Te} = 42$: $\omega_{*e}^T / \omega_{\kappa e} $ versus $\theta$ and $\theta_0$. Dashed and dash-dotted lines show $k_{\perp} \rho_e = 1$ for different $k_y \rho_i$.}
    \end{subfigure}
     \caption{Profiles of $\omega_{*e}^T / \omega_{\kappa e}$ in (a) and (c), and $k_{\perp}/ k_y$ in (b) for the circular flux-surface geometry. In (a), $a/L_{Te} = 4.2$; in (c), $a/L_{Te} = 42$.}
      \label{fig:10}
\end{figure}

At higher values of $k_y \rho_i$, FLR damping becomes stronger in regions where $\omega_{*e}^T / \omega_{\kappa e}$ has optimal values for the excitation of toroidal ETG modes and so toroidal ETG turbulence must occupy regions with less favorable, in this case higher, values of $\omega_{*e}^T / \omega_{\kappa e}$. For example, at $k_y \rho_i = 12.0$ and $28.9$ in Figures 11(c) and (d), respectively, the turbulence has a stronger slab ETG character, as suggested by the fact that the amplitudes $|\hat{\phi}|^2$ are inversely correlated with $k_{\perp}/k_y$. The stronger competition between magnetic drifts and FLR damping at higher values of $k_y \rho_i$ causes toroidal ETG turbulence to be less virulent than slab ETG turbulence at these scales. Note that, as discussed earlier, the decrease in the maximum $|\theta_0|$ value with increasing $k_y \rho_i$ due to grid and computational resource constraints may also artificially suppress toroidal ETG turbulence at higher $k_y \rho_i$.

In the core, the effect of magnetic drifts and FLR damping on toroidal ETG instability is qualitatively different from the one in the pedestal. In the pedestal, the toroidal ETG instability at the outer scale [given by \Cref{eq:twelve}] is strongest away from the outboard midplane, whereas in the core, it occurs at higher $k_y \rho_i$ due to gentler gradients and is strongest at $\theta \approx 0$. In \Cref{fig:10}(a), we plot $\omega_{*e}^T / \omega_{\kappa e}$ for the circular flux-surface geometry with $a/L_{Te} =4.2$ [see Scan100e in \Cref{sec:7,tab:1}]. This confirms that the most favorable values of $\omega_{*e}^T / \omega_{\kappa e}$ for toroidal ETG instability in the core are at $\theta \approx 0$ and $\theta_0 = 0$. The grey regions indicate parallel locations where $\omega_{\kappa e}$ is too large ($0< \omega_{*e}^T / \omega_{\kappa e} \lesssim 2$) for instability, even in bad-curvature regions \cite{Parisi2020}. In \Cref{fig:10}(b), we plot $k_{\perp} / k_y$ for the tokamak core geometry. As in the pedestal, FLR effects in the core typically favor the outboard midplane as the preferred location for unstable modes with higher $k_y \rho_i$. Thus, while toroidal ETG instability in the pedestal is favored at low $k_y \rho_i$ because FLR effects damp the modes at higher $k_y \rho_i$, toroidal ETG instability in the core can be strong at higher $k_y \rho_i$. This is because, in the core, unlike in the pedestal, there is an alignment of favorable FLR effects and values of $\omega_{*e}^T / \omega_{\kappa e}$ at higher $k_y \rho_i$. In \Cref{fig:10}(a), we also plot contours of $k_{\perp} \rho_e = 1$ for $k_{y} \rho_i = 42$, near the approximate outer scale for the core turbulence, showing that strong toroidal ETG instability is driven at $\theta \approx 0$. Recall that unlike the pedestal gradients, core gradients cannot support $k_y \rho_i \sim 1$ ETG turbulence because $a/L_{Te}$ is too small, according to the outer scale estimate in \Cref{eq:twelve} for the core, $k_y^o \rho_i \sim (\rho_i / \rho_e) (L_{Te} / qR) \gg 1$, where we used $l_{\parallel} \sim qR$.

If we keep the circular flux-surface geometry but increase the gradient to the pedestal value $a/L_{Te} = 42$, strong toroidal ETG turbulence is pushed away from $\theta \approx 0$ [see Scan200f in \Cref{fig:7}(d) and \Cref{tab:1}]. Increasing $a/L_{Te}$ from $4.2$ to $42$ increases $\omega_{*e}^T / \omega_{\kappa e}$ in \Cref{fig:10}(a) by a scalar factor of 10, resulting in \Cref{fig:10}(c). Notably, this transformation leaves the $k_{\perp}/k_y$ profile unchanged because for the circular flux-surface geometry, we set $\partial p/\partial r$ in \Cref{eq:five} to zero. \Cref{fig:10}(c) reveals that regions where linear toroidal ETG instability is most virulent, viz., $3 \lesssim \omega_{*e}^T / \omega_{\kappa e} \lesssim 20$, are now located away from the outboard midplane. The emergence of favorable $\omega_{*e}^T / \omega_{\kappa e}$ regions away from the outboard midplane, as well as a decrease in the outer scale $k_y^o \rho_i $ due to steeper $a/L_{Te}$, explains why the set-up with circular flux-surface geometry and $a/L_{Te} =42$ in \Cref{fig:7}(d) exhibits significant contributions to the heat flux from off-midplane turbulence, whereas the case with $a/L_{Te} =4.2$ does not: in circular flux-surface geometry with $a/L_{Te} =42$, both slab and toroidal ETG turbulence at lower $k_y \rho_i$ values are supported, and can be driven away from the outboard midplane.

It is important to recall that our Miller geometry is up-down symmetric \cite{Peeters2005,Camenen2010,Parra2011b,Ball2014,Zhu2018,Janhunen2022}. Accordingly, so are the perpendicular-wavenumber and magnetic-drift topographies in \Cref{fig:8,fig:9,fig:10}, viz., they are invariant under the transformation $(\theta, \theta_0) \to (-\theta,-\theta_0)$. In contrast, inspection of the poloidal dependence of $\widetilde{q} _e$ and $\hat{\phi}$ in Figures 6(d), 10(c), 10(d), 11(b), 11(c), and 12(b) reveals an up-down \textit{asymmetry} in the parallel spatial distribution of turbulence. Thus, in Figures 6(d), 10(c), 10(d), 11(b), 11(c), and 12(b), near $\theta \approx 0$, $\widetilde{ q} _e$ is larger for $\theta > 0$, whereas away from $\theta \approx 0$, $\widetilde{ q} _e$ is larger for $\theta < 0$. Averaging over longer time periods confirms this up-down asymmetry. It is caused by flow shear, with opposite asymmetry for toroidal and slab ETG turbulence. We have verified numerically that the asymmetry is reversed when the sign of $\gamma_E$ is reversed. The asymmetry occurs because toroidal ETG modes prefer $\mathrm{sign}(\theta_0) = - \mathrm{sign}(\theta)$, as is seen by inspecting regions of $\omega_{*e}^T / \omega_{\kappa e} > 0$ in \Cref{fig:9}(a). In contrast, examination of \Cref{fig:8}(a) shows that slab ETG modes usually prefer regions of $\mathrm{sign}(\theta_0) = \mathrm{sign}(\theta)$ where $k_{\perp}$ is lower. For $\gamma_E > 0$, the effective $\theta_0$ decreases with time [see \Cref{eq:five}] \cite{Hammett2006,Christen2021} and, as a result, turbulence amplitudes peak at negative values of $\theta_0$. In turn, toroidal ETG moves to $\theta > 0$ and slab ETG to $\theta < 0$. Thus, the effect of flow shear on the relative up-down poloidal distribution of slab and toroidal ETG transport can be predicted qualitatively by inspecting the $k_{\perp}$ and $\omega_{*e}^T / \omega_{\kappa e}$ topographies.

\section{Multiscale ETG-ETG interactions} \label{sec:8}

\begin{figure}
        \centering
        \includegraphics[width=0.49\textwidth]{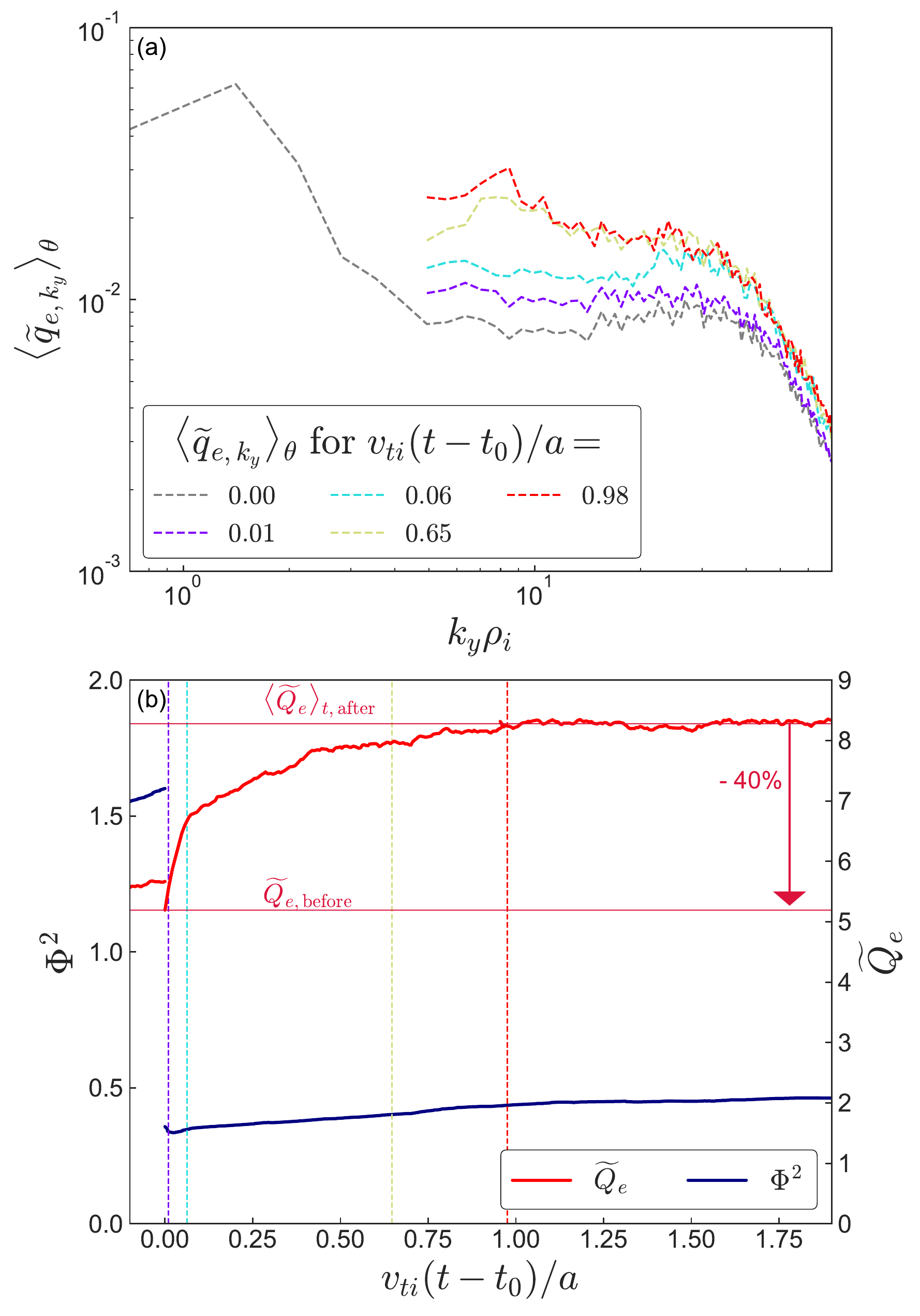}
        \caption{Numerical experiment where modes with $k_y \rho_i \leq 4.3$ are artifically damped starting at time $t_0$. (a) Heat flux $\widetilde{q}_{e,k_y} $ [see \Cref{eq:thirteen}] versus $k_y \rho_i$ for different times. (b) Heat flux $\widetilde{ Q}  _e$ and potential $\Phi^2$ versus time, with vertical lines denoting the instances for which $\widetilde{q}_{e,k_y} $ is plotted in (a), in matching colors. In (b), $ \widetilde{Q}_{e, \mathrm{before}}$ is the heat flux at the time immediately after the $k_y \rho_i \leq 4.3$ modes are damped, and $\langle \widetilde{Q}_e \rangle_{t, \mathrm{after}}$ is the time-averaged heat flux in the saturated state after the $k_y \rho_i \leq 4.3$ modes are damped. Low-$k_y \rho_i$ modes reduce turbulent heat transport by higher-$k_y \rho_i$ modes. This numerical experiment was performed with Base150-like parameters [see \Cref{tab:1}].}
        \label{fig:11}
\end{figure}

We now demonstrate that $k_y \rho_i \sim 1$ ETG turbulence decreases overall ETG transport substantially, in our case by $\sim$40\%. We show this by introducing artificial damping for low $k_y \rho_i$ modes from an initial condition corresponding to the saturated state of our Base150 calculation.

We damp modes with $k_y \rho_i \leq k_{y, \mathrm{cutoff} } \rho_i = 4.3$ to test whether $k_y \rho_i \sim 1$ ETG turbulence affects $k_y \rho_e \sim 1$ ETG turbulence and transport. To damp $k_y \rho_i \sim 1$ modes, we multiply the perturbed distribution function $f _s ^{tb}$ for these modes by $10^{-4}$ at each timestep. At $t = t_0$, just before these modes are damped, there is a significant heat flux contribution from each low $k_y \rho_i$ value. In \Cref{fig:11}(a), we show how $\widetilde{ q}_{e, k_y} $ evolves after time $t_0$ when we begin damping them.

At the time immediately after these modes are damped, the heat flux drops by roughly 5\%. This instantaneous decrease in the heat flux represents the loss of heat flux carried by the now-damped $k_y \rho_i \leq k_{y, \mathrm{cutoff} } \rho_i = 4.3$ modes. At this time, modes with $k_y \rho_i > k_{y, \mathrm{cutoff} } \rho_i$ still carry information about multiscale interactions with the $k_y \rho_i \leq k_{y, \mathrm{cutoff} } \rho_i$ modes. Therefore, we will call the heat flux at this time $\widetilde{Q}_{e, \mathrm{before}}$ and use it as the point of comparison with the heat flux in the new saturated state at later times, $\langle \widetilde{Q}_e \rangle_{t, \mathrm{after}}$.

As $t$ increases, there is negligible heat flux from $k_y \rho_i \leq k_{y, \mathrm{cutoff} } \rho_i$ modes and there is a significant increase in $\widetilde{ q}_{e, k_y} $ at larger values of $k_y \rho_i$. \Cref{fig:11}(b) shows how, over a period of several linear times of the slowest undamped linear modes, the total electron heat flux $\widetilde{Q}_e$ increases from $\widetilde{Q}_{e, \mathrm{before}} \simeq 5.2$ to $\langle \widetilde{Q}_e \rangle_{t, \mathrm{after}} \simeq 8.4$ in the new steady state. Thus, there is a $\sim$40 \% reduction in $\widetilde{Q} _e$ when the lower-$k_y \rho_i$ modes are allowed to play a role. We have demonstrated that $k_y \rho_i \sim 1$ ETG turbulence suppresses higher-$k_y \rho_i$ ETG transport. 

We have shown that retaining $k_y \rho_i \sim 1$ ETG modes is crucial to capture correctly the $k_y \rho_e \sim 1$ electron heat flux. This is relevant for tokamak turbulence modeling \cite{Waltz1998,Bourdelle2015,Meneghini2015,Staebler2020,Guttenfelder2021,Staebler2021,Hatch2022} that aims to predict experimental fluxes accurately. While this result is not the first to show $k_y \rho_i \sim 1$ turbulence suppressing $k_y \rho_e \sim 1$ turbulence \cite{Candy2007, Maeyama2015, Howard2016, Hardman2020}, it is the first to show ETG turbulence at $k_y \rho_i \sim 1$ suppressing ETG turbulence and transport at $k_y \rho_e \sim 1$. It is important to re-emphasize that the ETG turbulence at $k_y \rho_i \sim 1$ is strongly-driven not because of kinetic-ion physics, but because the pedestal temperature gradients are so steep [see discussion surrounding \Cref{eq:twelve}].

The multiscale mechanism for the suppression of electron-scale transport in the pedestal remains to be investigated in future work. In the core, cross-scale interactions between electron-scale turbulence (driven by ETG instability) and ion-scale turbulence (driven by ITG and other instabilities) can suppress electron-scale and enhance ion-scale transport \cite{Gorler2008, Maeyama2015, Howard2016, Hardman2020}. In contrast, because steep temperature gradients in the pedestal break electron-ion scale separation, interactions between electron-scale turbulence and ion-scale turbulence, where turbulence at \textit{both} scales is driven by ETG instability, is possible.

\section{Discussion} \label{sec:10}

The main result of this paper is that electron-temperature-gradient turbulence in a typical JET pedestal has a rich three-dimensional spatial structure in directions both parallel and perpendicular to the magnetic field. This structure arises due to the steep temperature gradient and the highly shaped magnetic geometry. Steep temperature gradients enable strong ETG turbulence to be driven at much longer binormal wavelengths than in core tokamak plasmas, often at wavelengths numerically comparable to the ion gyroradius, $k_y \rho_i \sim 1$. The $k_y \rho_i \sim 1$ ETG turbulence has the highest fluctuation amplitudes but produces modest heat transport due its short radial correlation length, and also reduces the overall turbulent heat transport through multiscale interactions.

Experimental measurements of off-midplane potential fluctuations are needed to test our predictions, but could prove challenging due to turbulence diagnostics conventionally being located at the outboard midplane, with some exceptions \cite{Mazzucato1976,Brower1987}. Our results might be consistent with Beam-Emission-Spectroscopy measurements of ion-gyroradius scale turbulence in MAST, showing correlation lengths that are longer in the binormal direction than in the radial direction \cite{Ghim2013,VanWyk2017}, hinting at experimental signatures of $k_{\perp} \gg k_y$ anisotropic turbulence of a nature described in this paper.

The parallel spatial distribution of toroidal and slab ETG turbulence at all scales can be qualitatively predicted from the perpendicular-wavenumber and magnetic-drift profiles [see Figures 11(a) and 12(a)]. Both have complex topography due to strong magnetic shaping in the pedestal. Due to finite-Larmor-radius damping, turbulence and transport are highest in the outboard midplane for $k_y \rho_e \sim 1$, but for $k_y \rho_i \sim 1$, electrostatic-potential fluctuations are largest near the flux surface's top and bottom [see \Cref{fig:8,fig:9}]. The adiabatic ion nature of toroidal ETG turbulence prevents large heat transport arising from large density fluctuations away from the outboard midplane.

The results of \Cref{sec:7,sec:8} suggest using magnetic shaping to optimize transport in the pedestal and internal transport barriers. This could be achieved by modifying parallel correlation lengths and hence the outer scale of the turbulence [see \Cref{eq:twelve}] using FLR effects and magnetic-drift profiles, and by maneuvering toroidal and slab ETG turbulence into similar poloidal locations, so that their multiscale interactions could suppress $k_y \rho_e \sim 1$ transport. 

\section{Code and data availability}

The data used for the material in this paper are available at the following dataset archive \cite{Parisi2022b}.

\section{Acknowledgements}

We are grateful for stimulating conversations with T. Adkins, N. Christen, A. Field, W. Guttenfelder, G. W. Hammett, and M. J. Pueschel.

JFP was supported by a Culham Fusion Research Fellowship. FIP, MRH, MB, AAS, DS, and DD were supported in part by the TDoTP project funded by EPSRC (grant number EP/R034737/1). This work was supported by the U.S. Department of Energy under contract number DE-AC02-09CH11466. The United States Government retains a non-exclusive, paid-up, irrevocable, world-wide license to publish or reproduce the published form of this manuscript, or allow others to do so, for United States Government purposes. This work has been carried out within the framework of the EUROfusion Consortium, funded by the European Union via the Euratom Research and Training Programme (Grant Agreement No 101052200 — EUROfusion). Views and opinions expressed are however those of the author(s) only and do not necessarily reflect those of the European Union or the European Commission. Neither the European Union nor the European Commission can be held responsible for them. This work has been carried out within the framework of the Contract for the Operation of the JET Facilities and has received funding from the European Union’s Horizon 2020 research and innovation programme. This work was performed using the Cambridge Service for Data Driven Discovery (CSD3), operated by the University of Cambridge Research Computing on behalf of the STFC DiRAC HPC Facility. The DiRAC component of CSD3 was funded by BEIS capital funding via STFC capital grants ST/P002307/1 and ST/R002452/1 and STFC operations grant ST/R00689X/1. This work was supported by the US Department of Energy through grant DE-SC0018429. This work was carried out using the JFRS-1 supercomputer system at Computational Simulation Centre of International Fusion Energy Research Centre (IFERC-CSC) in Rokkasho Fusion Institute of QST (Aomori, Japan).

\bibliographystyle{apsrev4-2}
\bibliography{EverythingPlasmaBib}

\end{document}